%% file: thesis.tex
\RequirePackage{ifpdf}

\documentclass[11pt]{uopthesis}
\usepackage{amsmath}
\usepackage{amssymb}
\usepackage{epsfig}
\usepackage{latexsym}
\usepackage{color}
\usepackage{pstricks}

\newcommand{\da}{\dot{A}}
\newcommand{\daa}{\frac{\dot{A}}{A}}

\newcommand{\ddaa}{\frac{\ddot{A}}{A}}
\newcommand{\paa}{\frac{A'}{A}}
\newcommand{\ppaa}{\frac{A''}{A}}
\newcommand{\dbb}{\frac{\dot{B}}{B}}
\newcommand{\ddbb}{\frac{\ddot{B}}{B}}
\newcommand{\pbb}{\frac{B'}{B}}

\newcommand{\dnn}{\frac{\dot{N}}{N}}

\newcommand{\pnn}{\frac{N'}{N}}
\newcommand{\ppnn}{\frac{N''}{N}}
\newcommand{\oa}{\Omega_{\alpha}}
\newcommand{\op}{\Omega_{\phi}}
\newcommand{\ob}{\Omega_{\lambda}}
\newcommand{\om}{\Omega_{\rm m}}
\newcommand{\orc}{\Omega_{\rm r_c}}

\ifpdf%
  \hypersetup{
      pdftitle={Brane world cosmology with Gauss-Bonnet and induced gravity terms},
      pdfauthor={Richard A. Brown (richard.brown@port.ac.uk)},
      pdfsubject={PhD, Thesis},
      pdfkeywords={Brane worlds,Cosmology,Gauss-Bonnet,Induced gravity},
      pdfpagemode=UseOutlines,
      pdfstartpage=1,
      pdfstartview=FitV,
      pdfview=FitV
  }
\fi

\title{Brane world cosmology with Gauss-Bonnet and induced gravity terms}

\ifpdf
  \author{\href{mailto:richard.brown@port.ac.uk}{Richard A.~Brown}}
  \department{\href{http://www.tech.port.ac.uk/icg/}{Institute of Cosmology & Gravitation}}
  \university{\href{http://www.port.ac.uk}{University of Portsmouth, UK}}
  \principaladvisor{\href{http://www.tech.port.ac.uk/icg/people/staff/maartens.html}{Prof.~R.~Maartens}}
  \firstreader{First Reader Name}
  \secondreader{Second Reader Name}
\else
  \author{Richard A. Brown}
  \department{Institute of Cosmology & Gravitation}
  \university{University of Portsmouth, UK}
  \principaladvisor{Prof.~R.~Maartens}
  \firstreader{First Reader Name}
  \secondreader{Second Reader Name}
\fi

\submitdate{November, 2006} \copyrightyear{2006}

\tablespagefalse

\begin{document}
\begin{preliminary}

\begin{dedication}
  To my family and friends.
\end{dedication}

\newpage
\mbox{} \vspace{50pt}

\noindent Edgar Allan Poe (1809-1849) ponders the subject of matter
undergoing collapse: \vspace{12pt}
\begin{center}
\begin{quote}
\textit{\Large In sinking into Unity, it will sink at once into that
Nothingness which, to all Finite Perception, Unity must be--into
that Material Nihility from which alone we can conceive it to have
been evoked--to have been created by the Volition of God.}
\end{quote}
\end{center}
\vspace{12pt}
 from Eureka - A Prose Poem (1848), written 81 years
before Edwin Hubble's observations of the expansion of the universe.

\begin{abstract}

In this thesis we investigate certain cosmological brane world
models of the Randall-Sundrum type. The models are motivated by
string theory but we focus on the phenomenology of the cosmology.

Two models of specific interest are the Dvali-Gabadadze-Porrati
(DGP, induced-gravity) model, where the brane action is modified,
and the Gauss-Bonnet model where the bulk action is modified. Both
of these modifications maybe motivated by string theory.

We provide a brief review of Randall-Sundrum models and then
consider the Kaluza-Klein modes on Minkowski and de Sitter branes,
in both the two and one brane cases. The spectrum obtained for the
de Sitter branes is a new result. We then consider a
Friedmann-Robertson-Walker brane in order to investigate the
cosmological dynamics on the brane.

We present a brief discussion of the DGP and Gauss-Bonnet brane
worlds. We then investigate the Gauss-Bonnet-Induced-Gravity (GBIG)
model where the Gauss-Bonnet (GB) bulk term is combined with the
induced-gravity (IG) brane term of the DGP model. We present a
thorough investigation of cosmological dynamics, in particular
focusing on GBIG models that behave like self-accelerating DGP
models at late times but at early times show the remarkable feature
of a finite-temperature Big Bang. We also discuss the constraints
from observations, including ages and Big Bang nucleosynthesis.

\end{abstract}
\begin{preface}
The work of this thesis was carried out at the Institute of
Cosmology \& Gravitation, University of Portsmouth, United Kingdom.

The following chapters are based on published work:
\begin{itemize}
\item Chapter~\ref{GBIGB} - R. A. Brown,
Roy Maartens, Eleftherios Papantonopoulos and  Vassilis Zamarias.
``A late-accelerating universe with no dark energy-and a
finite-temperature big bang'', JCAP 0511 (2005) 008, gr-qc/0508116.
\item Chapter~\ref{GGBIG} - R. A. Brown. ``Brane Universes with Gauss-Bonnet-Induced-Gravity'',
accepted for publication in ``General Relativity and Gravitation'',
gr-qc/0602050.
\end{itemize}

\end{preface}
\begin{acknowledgements}
First and foremost I would like to thank Roy Maartens without whose
support, guidance and infinite patience this thesis would never have
been completed. Other people who directly influenced this work and
require my thanks include Mariam Bouhmadi-L\'{o}pez, Chris Clarkson,
Kazuya Koyama and Sanjeev Seahra the Maple Guru. I must also include
David Wands and Andrew Mennim for useful discussions. Other people
who deserve thanks for their support and making my stay in the ICG
more pleasurable are Kishore Ananda, Frederico Arroja, Iain Brown,
Chris Byrnes, Dan Carson, Robert Crittenden, Mathew Smith and David
Wake. Kishore Ananda deserves special thanks for putting up with me
as a flat mate for the three years of my PhD study.

I would also like to thank all the teaching staff in the Physics and
Astronomy department at the University of Hertfordshire. Without the
foundation degree course provided there I would never have got into
cosmology in any practical way.

I must also thank dear friends such as Dan Mee, my connection to the
real world outside academia in Portsmouth. Also all the Hatfield
guys who are able to make me forget all about work within a few
minutes of meeting up. I want to thank all the Haslemere crew as
well, whose friendship I know I can rely upon however long it is
between meeting up.

Last but not least I must thank my family whose support is one of
the few constants in life that can be relied upon however bad things
get.

Without all these people, starting, let alone finishing, this
project would never have happened.

\end{acknowledgements}
\end{preliminary}

\startthesis

\include{Intros}
\include{CHRSBranes}
\include{CHGBIGBranes}

\include{CHGeneralGBIGBranes}
\include{Conclusions}

\appendix
\include{ApCon}

\addcontentsline{toc}{chapter}{References}

\bibliographystyle{h-elsevier2}
\bibliography{Biblio}

\end{document}

%% file: Intros.tex
\chapter{Introduction}\label{intro}

Quantum mechanics and General Relativity are two very successful and
well validated theories within their own domains. The problem is
that we have no way of unifying them into a single consistent
theory. One of the most promising models of unification is String
Theory (for an introduction to some of the concepts of String Theory
see~\cite{Sevrin:2004bu}). String Theories remove the infinities
that are present in a classical unification by describing particles
as extended 1-dimensional strings rather than point particles. These
1D strings live in a 10 dimensional space, or 11D for supergravity.
The idea of extra dimensions was first put forward by
Kaluza~\cite{Kaluza:1921up} and Klein~\cite{Klein:1926tv} in order
to unify general relativity and electromagnetism. A perturbative
approach is usually applied when working in String Theory due to the
complexity of the equations. It has been shown that there are five
different formulations of perturbative String Theory and that these
are dual to each other, under a certain set of transformations. It
has been suggested that these five theories are aspects of one
underlying theory, M Theory.

Two classes of strings are the closed and open strings. Gravity is
described by closed strings and matter is described by open strings.
In non-pertubative string theory there exist extended objects known
as D (Dirichlet) branes. These are surfaces where the open strings
must start and finish. This provides an alternative to the
Kaluza-Klein approach, where matter penetrates the extra dimensions,
leading to strong constraints from collider physics. If matter is
confined to a 3-dimensional brane, the extra dimensions can be
larger, since the constraints on gravity are weaker.

In this thesis we consider brane world models. These models are
inspired by String Theory, in particular by the Ho\v{r}ava and
Witten model~\cite{Horava:1995qa}. Brane world models are
characterized by the feature that standard model matter is confined
to a 1+3 dimensional brane while gravity propagates in the higher
dimensional bulk. This means that gravity is fundamentally a higher
dimensional interaction and we only see the effective 4D theory on
the brane. We can not see or measure (with existing technology)
Planck scale compact dimensions as they require Planck energy scales
in order to probe them. But if an extra dimension is large (relative
to the Planck scale) then there may be signatures of this in
colliders and table-top experiments, as well as in cosmological
observations.

In the model of Arkani-Hamed, Dimopoulos and Dvali
(ADD)~\cite{Arkani-Hamed:1998rs}, the weakness of gravity on scales
$\gtrsim 1{\rm mm}$ (Newton's law has only been verified down to a
scale of $\sim0.1\textrm{mm}$~\cite{Chung:2000rg} ) is due to the
existence of $n\geqslant2$ extra compact dimensions. This model can
be embedded in String Theory~\cite{Antoniadis:1998ig} and has the
advantage of putting forward a mechanism for explaining the
hierarchy problem. The hierarchy problem is the discrepancy between
the gravity scale ($M_{\rm Planck}\sim 10^{19}{\rm GeV}$) and the
electro-weak scale ($M_{\rm EW}\sim 10^{3}{\rm GeV}$). The $10^{16}$
orders between these two scales is too vast to be considered
natural. In the ADD model, length scales smaller then the scale of
the extra dimensions ($r<L$) have a $4+n$ dimensional gravitational
potential:

\begin{equation}\label{ADD1}
V\sim r^{-(1+n)}.
\end{equation}
On scales larger than $L$, $r\sim L$ in the $n$ extra dimensions.
Therefore the potential is given by:

\begin{equation}\label{ADD2}
V\sim L^{-n}r^{-1},
\end{equation}
on large scales. The observed Planck scale is then the product of
the fundamental Planck scale and the volume of the extra
dimensions~\cite{Maartens:2003tw}:

\begin{equation}\label{ADDP}
M^2_{Planck}\sim M^{2+n}_{4+n}L^{n}.
\end{equation}
Therefore in the ADD model the fundamental Planck scale could be
much smaller than the Planck scale and the observed value is due to
the size of the extra dimensions.

\begin{figure}
\begin{center}
\includegraphics[scale=0.75]{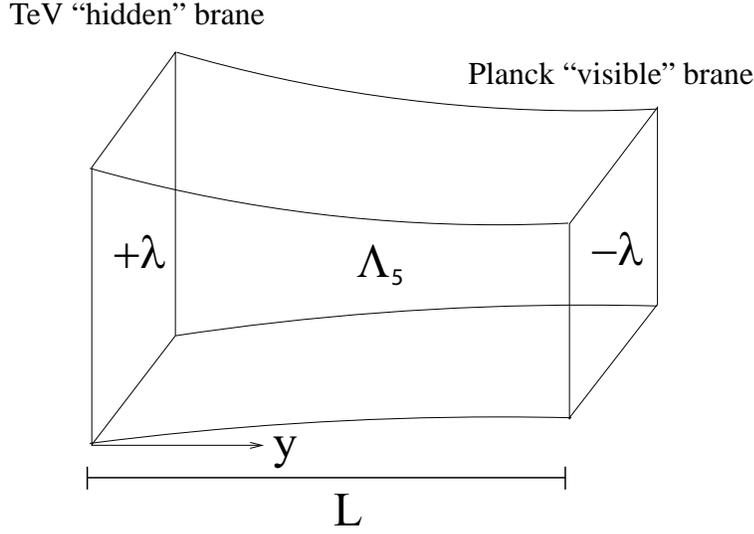}
\rput(-9.8,7){TeV ``hidden''
brane}\rput(-3.5,6.2){Planck ``visible'' brane} \caption{The RS1
configuration} \label{RS1fig}
\end{center}
\end{figure}

The Randall-Sundrum model of two, $Z_2$ symmetric, 3-branes living
in an anti-de Sitter (AdS) spacetime provides an alternative to the
ADD model for $n=1$~\cite{Randall:1999ee} (see Fig.~\ref{RS1fig}).
The second Randall-Sundrum model has only one
brane~\cite{Randall:1999vf} .

\subsubsection{The Randall-Sundrum 2 Brane (RS1) Model}

The RS1 model has two Minkowski branes living in a 5D AdS bulk,
Fig.~\ref{RS1fig}. The bulk metric is given, in a Gaussian normal
coordinate system, by:

\begin{equation}\label{LE}
ds^2=a^2(y)\eta_{\mu\nu}dx^\mu dx^\nu+dy^2,
\end{equation}
where $\eta_{\mu\nu}$ is the Minkowski metric. We use Latin indices
for 5D spacetime and Greek indices for 4D spacetime throughout this
work. $a(y)$ is the warp factor, given by:

\begin{equation}\label{a}
a(y)=e^{-|y|/\ell},
\end{equation}
where $\ell$ is the AdS curvature scale. The extra dimension is
compact. In order for the bulk to be compact we invoke $Z_2$
symmetry so that:

\begin{equation}\label{Z2}
y\leftrightarrow -y,~~~L+y\leftrightarrow L-y,
\end{equation}
where $L$ is the inter-brane separation.

As we shall see in the next chapter the branes have equal and
opposite tensions. At $y=0$ is the positive tension, ``TeV'' or
``hidden'' brane. At $y=L$  is the negative tension, ``Planck'' or
``visible'' brane. The brane tensions $\lambda$ are given by:

\begin{equation}\label{inbt}
\lambda_{y=0}=-\lambda_{y=L}=\frac{6}{\kappa^2_5\ell}\equiv\lambda,
\end{equation}
where $\kappa^2_5=8\pi G_5=M^{-3}_5$. We reside on the negative
tension brane in order to solve the hierarchy problem. The
``hidden'' brane has fundamental energy scale $M_5$. Then due to the
warping of the bulk the effective fundamental energy scale on the
``visible'' brane is $M_{Planck}$, where:

\begin{equation}\label{MpdS}
M_{Planck}=M^3_5\ell\left[1-e^{-2L/\ell}\right].
\end{equation}

The low energy effective theory on the branes is of Brans-Dicke
type~\cite{Gen:2000nu}. The Brans-Dicke parameter ($\omega$) on each
brane has the same sign as the brane tension~\cite{Garriga:1999yh}.
Nucleosynthesis constrains
$\omega\gtrsim40000$~\cite{Bertotti:2003rm,DeFelice:2005bx}. This is
a problem due to the fact that in this model we reside on the
negative tension brane and as such we should measure a negative
value of $\omega$. A radion is required to stabilise the inter-brane
separation in order for 4D General Relativity to be achieved at low
energies~\cite{Goldberger:1999uk}.

\begin{figure}
\begin{center}
\includegraphics[scale=0.75]{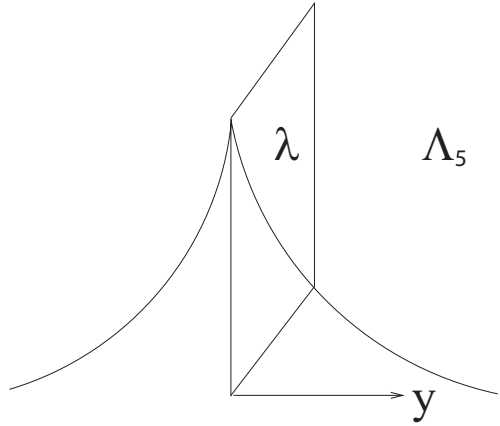}
\caption{The RS2 configuration} \label{RS2fig}
\end{center}
\end{figure}

\subsubsection{The Randall-Sundrum 1 Brane (RS2) Model}

The one brane (RS2)~\cite{Randall:1999vf} model has a single
positive tension brane at $y=0$ as seen in Fig.~\ref{RS2fig}. The
bulk is infinite in extent. As there is only one brane there is no
mechanism for solving the hierarchy problem but the low energy
theory on the brane is general relativity~\cite{Garriga:1999yh}.
Gravity is localised to the brane, in the low energy limit, via the
warping of the bulk. At high energies the warping is insufficient to
localise the gravity to the brane and observers on the brane
perceive the 5D nature of gravity. The fundamental energy and the
Planck scales are related by:

\begin{equation}\label{HP2}
M^2_{Planck}=M^3_5\ell,
\end{equation}
i.e. we have effectively sent the second brane off to infinity
($L=\infty$).

The Randall-Sundrum model exhibits its 5D nature, to a brane
observer, at high energies, when the warp factor is insufficient to
localise gravity to the brane. The deviation between the 4D and 5D
nature can be seen in the gravitational potential. The modified
gravitational potential is~\cite{Garriga:1999yh}:

\begin{equation}\label{MGPl}
V(r)\approx\frac{GM}{r}\left(1+\frac{2\ell^2}{3r^2}\right).
\end{equation}
In Ref.~\cite{Randall:1999vf} this calculation was carried out with
the result having a factor of $1$ instead of $2/3$ in front of the
$r^{-2}$ term. The result in~\cite{Garriga:1999yh} was obtained by
including the coupling of the graviton to matter on the brane.

At high energies we can not get away from the 5D nature of the
universe but at low energies ($r\gg\ell$) we are given a constraint
on the AdS curvature scale. As we have seen no deviation from
Newton's laws down to $\approx 0.1{\rm mm}$ in the lab,
Eq.~(\ref{MGPl}) gives us the constraint that $\ell\lesssim 0.1{\rm
mm}$. This constraint means that the brane tension,
Eq.~(\ref{inbt}), and fundamental energy scale, Eq.~(\ref{HP2}), are
also constrained. These constraints are:

\begin{equation}\label{FES2}
\lambda>(1~~{\rm TeV})^4,~~~~~M_5>10^5~~{\rm TeV}.
\end{equation}

\vspace{10mm}

In the RS models the 4-D cosmological constant on the brane is given
by:

\begin{equation}\label{Lb}
\Lambda_4=\frac{1}{2}\left ( \Lambda_5 + \kappa^2_4 \lambda \right
).
\end{equation}
Since the branes are Minkowski, we require the fine tuning:

\begin{equation}\label{L5R}
\Lambda_5+\kappa^2_4\lambda=0,
\end{equation}
where $\kappa^2_4=8\pi G_4=M_{Planck}^{-2}$, so that $\Lambda_4=0$.
Breaking this fine tuning leads to different cosmologies on the
branes~\cite{Maartens:2003tw}:

\begin{eqnarray}\label{Lbs}
\lambda <-\frac{\Lambda_5}{\kappa^2_4} &\Rightarrow &
\textrm{Anti-de
 Sitter,}\\\nonumber \\
 \lambda =-\frac{\Lambda_5}{\kappa^2_4}  &\Rightarrow &
 \textrm{Minkowski,}\\\nonumber \\
 \lambda >-\frac{\Lambda_5}{\kappa^2_4} &\Rightarrow & \textrm{de
 Sitter.}
\end{eqnarray}
The bulk cosmological constant is given by:

\begin{equation}\label{inBCC}
\Lambda_5=-\frac{6}{\ell^2},
\end{equation}
where $\ell$ is the AdS curvature scale.

The Randall-Sundrum models provides a simple way of investigating
aspects of the phenomenology of the very complicated String
Theories. In this thesis, we focus on brane world cosmological
models based on generalisations of the RS scenario (see the
reviews~\cite{Maartens:2003tw,Perez-Lorenzana:2004na,Csaki:2004ay,
Langlois:2003yy,Brax:2003fv,Kim:2003pc,Langlois:2002ux}). These
include models that can explain the late-time acceleration of the
universe without the need for dark energy.

%% file: CHRSBranes.tex
\chapter{RS Brane Worlds}\label{RSB}

\section{Introduction}
In this chapter we will look in some detail at the Randall-Sundrum
(RS) model. We will start with the simplest model, the Minkowski
brane. We shall then consider a de Sitter brane before looking at a
cosmological solution (the Friedmann-Robertson-Walker brane). In the
first two cases we will look at the equations in the bulk and on the
brane. We will also look at the spectrum of Kaluza-Klein modes
generated due to perturbations on the brane. For the cosmological
brane we will only consider the equations on the brane and how they
differ from the standard general relativity result. It is these
brane Friedmann equations that will be of interest in the following
chapters.


\section{Minkowski Branes}

Minkowski branes are the most simple branes possible. They are flat,
empty and static. We start by obtaining the background form of the
bulk Einstein field equations. Using the bulk field equations we can
work out the induced equations on the brane. Later we shall consider
bulk gravitational waves. These give rise to the Kaluza-Klein modes
on the brane.

\subsection{The background}
The line element for Minkowski branes in an anti de Sitter bulk was
considered in the previous chapter, Eq.~(\ref{LE}).  For simplicity
we take $y\geq0$ and implicity assume $Z_2$ symmetry. We can write
the metric for this model as:

\begin{equation}\label{metric}
g_{ab}=a^2(y)\eta_{\mu\nu}\delta^\mu_a\delta^\nu_b+\delta^y_a\delta^y_b.
\end{equation}
This is a solution of the 5-dimensional Einstein equation:

\begin{equation}\label{5DEE}
G^{(5)}_{ab}=-\Lambda_5 g^{(5)}_{ab}.
\end{equation}
Using Eq.~(\ref{a}) we can write out the following relationships for
the warp factor:

\begin{equation}\label{diffa}
a'=-\frac{a}{\ell},~~\Rightarrow~~a''=\frac{a}{\ell^2},
\end{equation}
where a prime represents differentiation with respect to $y$. This
implies:

\begin{equation}\label{diffMet}
g_{ab,y}=g_{ab}'=-\frac{2}{\ell}\left (
g_{ab}-\delta^y_a\delta^y_b\right ),
\end{equation}
and similarly:

\begin{equation}\label{diffMetU}
g^{ab}_{\;\;\;,y}={g^{ab}}' = \frac{2}{\ell}\left (
g^{ab}-\delta_y^a\delta_y^b\right ).
\end{equation}
By Eq.~(\ref{metric}):

\begin{equation}\label{gdrel1}
g_{ay}=\delta^{y}_{a},~~\Rightarrow~~g^{ay}=\delta_y^a.
\end{equation}
These will be useful later.

In order to construct the Einstein field equations we require both
the Ricci curvature tensor and the Ricci scalar. We start by
obtaining the RS Riemann tensor. The only non-zero derivatives of
$g_{ab}$ are those with respect to $y$. Using Eq.~(\ref{diffMet}) we
can write the Christoffel symbols as:

\begin{equation}\label{CHmS}
\Gamma^a_{bc}= -\frac{1}{\ell}\left(\delta^y_c \delta^a_b
+\delta^y_b \delta^a_c-\delta^a_y
g_{bc}-\delta^a_y\delta^y_b\delta^y_c\right),
\end{equation}
so that:

\begin{equation}\label{CHmdiff}
\Gamma^a_{bc,d}= -\frac{2}{\ell^2}\delta^a_y \delta^y_d\left(
g_{bc}-\delta^y_b\delta^y_c\right).
\end{equation}
The Riemann tensor is defined as:

\begin{equation}
R^a_{\;bcd}=\Gamma^a_{bd,c}-\Gamma^a_{bc,d}+
\Gamma^e_{bd}\Gamma^a_{ec}-\Gamma^e_{bc}\Gamma^a_{ed}.
\end{equation}
So for the Randall-Sundrum brane model the Riemann tensor has the
anti de Sitter form:

\begin{equation}\label{RF1}
R^a_{\;bcd}=-\frac{1}{\ell^2}\left(\delta^a_c g_{bd}-\delta^a_d
g_{bc}\right).
\end{equation}
The RS Ricci tensor and scalar are given by:

\begin{equation}\label{Ri2}
R_{ab}=-\frac{4}{\ell^2}g_{ab},
\end{equation}
and:
\begin{equation}\label{RiS2}
R=-\frac{20}{\ell^2}.
\end{equation}
The Einstein tensor is given by:

\begin{equation}\label{Ein2}
G_{ab}=-\frac{4}{\ell^2}g_{ab}-\frac{1}{2}\left(-\frac{20}{\ell^2}\right)
g_{ab}=\frac{6}{\ell^2}g_{ab},
\end{equation}
which satisfies the field equation:

\begin{equation}\label{EinL}
G_{ab}=-\Lambda_{5}g_{ab},
\end{equation}
where the 5D bulk cosmological constant is given by:

\begin{equation}\label{L5}
\Lambda_{5}=-\frac{6}{\ell^2}.
\end{equation}
This also means that we can write the Ricci tensor as:

\begin{equation}\label{RiL}
R_{ab}=\frac{2}{3}\Lambda_{5}g_{ab}.
\end{equation}

Using the solution for the RS bulk cosmological constant in
Eqs.~(\ref{Lbs}) we find that in order for the brane(s) to have a
Minkowski geometry we require the brane tension(s) to be fined tuned
to:

\begin{equation}\label{Tm4}
\lambda_\pm=\pm\frac{6}{\ell^2\kappa^2_4},
\end{equation}
where the $\pm$  refer to the two different branes. The brane
tensions can also be found from the extrinsic curvature of the
brane. The extrinsic curvature depends on how the brane is embedded
in the higher dimensional space.  As we are working in Gaussian
normal coordinates we can write the extrinsic curvature
as~\cite{Wald}:

\begin{equation}\label{EKd}
K_{\mu\nu}=\frac{1}{2}\partial_y g_{\mu\nu}.
\end{equation}
For Minkowski branes we find that the extrinsic curvature is given
by:

\begin{equation}\label{EKM}
K_{\mu\nu}=\frac{g_{\mu\nu}}{\ell}.
\end{equation}
The extrinsic curvature on the brane can also be obtained by
considering the Israel-Darmois junction
conditions~\cite{Israel:1966rt}:

\begin{eqnarray}
g^+_{\mu\nu}-g^-_{\mu\nu}&=&0 ,\\
K^+_{\mu\nu}-K^-_{\mu\nu}&=&-\kappa^2_5\left[T^{brane}_{\mu\nu}-\frac{1}{3}T^{brane}g_{\mu\nu}\right],
\end{eqnarray}
where:

\begin{equation}\label{ted}
T^{brane}_{\mu\nu}=T_{\mu\nu}-\lambda g_{\mu\nu}.
\end{equation}
$T^{brane}_{\mu\nu}$ is the total energy-momentum tensor of the
brane. This is the combination of the standard matter $T_{\mu\nu}$
on the brane and the brane tension $\lambda$ which acts as a brane
cosmological constant.

Using $Z_2$ symmetry, $K^-_{\mu\nu}=-K^+_{\mu\nu}$, the extrinsic
curvature can be written as~\cite{Maartens:2003tw} (where we drop
the $\pm$ superscript):

\begin{equation}\label{EKR}
K_{\mu\nu}=-\frac{1}{2}\kappa^2_5\left[T_{\mu\nu}+\frac{1}{3}(\lambda-T)g_{\mu\nu}\right],
\end{equation}
where $T=T^\mu_\mu$. Using Eq.~(\ref{EKR}) for a Minkowski brane i.e
$T_{\mu\nu}=0$ and Eq.~(\ref{EKd}) we find that the brane tension is
given by:

\begin{equation}\label{Tm5}
\lambda_\pm=\pm\frac{6}{\kappa^2_5\ell}.
\end{equation}
Using this form of the brane tension with that in Eq.~(\ref{Tm4})
gives us the following relationship for Minkowski branes:

\begin{equation}\label{kappaRel}
\kappa^2_5=\ell\kappa^2_4.
\end{equation}

\subsection{Field equations on the brane}

For a general cosmological brane, in a general bulk, the
5-dimensional field equations are:

\begin{equation}\label{En}
G^{(5)}_{ab}=\kappa^2_5T_{ab}^{(5)},
\end{equation}
where the energy-momentum tensor is given by:

\begin{equation}\label{EMT}
T^{a\,(5)}_b=T^a_b|_{bulk}+T^a_b|_{brane}.
\end{equation}
The bulk part of $T^{(5)}_{ab}$ is given by:
\begin{equation}\label{EMT1}
T^a_b|_{bulk}=-\frac{\Lambda_5}{\kappa^2_5}\delta^a_b,
\end{equation}
which is just the 5D bulk cosmological constant.

The induced field equations on the brane are given
by~\cite{Maartens:2003tw}:

\begin{equation}\label{FEB}
G_{\mu\nu}=-\Lambda_4
g_{\mu\nu}+\kappa_4^2T_{\mu\nu}+\frac{6\kappa_4^2}{\lambda}S_{\mu\nu}
-{\cal E}_{\mu\nu}+4\frac{\kappa^2_4}{\lambda}F_{\mu\nu}.
\end{equation}
The first two terms on the right hand side are the standard GR
contributions of the matter on the brane and the cosmological
constant. $\Lambda_4$ and $\kappa_4$ are as defined in Chapter
\ref{intro}. The other terms are unique to brane models. The tensor
$S_{\mu\nu}$ is the high-energy correction term and is quadratic in
the energy-momentum tensor:

\begin{equation}\label{SQEMT}
S_{\mu\nu}=\frac{1}{12}TT_{\mu\nu}-\frac{1}{4}T_{\mu\alpha}T^\alpha_{\nu}
+\frac{1}{24}g_{\mu\nu}\left[3T_{\alpha\beta}T^{\alpha\beta}-T^2\right].
\end{equation}
The fourth term in the induced field equations, ${\cal E}_{\mu\nu}$,
is the projection of the bulk Weyl tensor orthogonal to $n^a$:

\begin{equation}\label{WeylProj}
{\cal E}_{\mu\nu}=C^{(5)}_{acbd}~n^cn^d~g^{(5)\,a}_{~~~~\,\mu}~
g^{(5)\,b}_{~~~~\,\nu}.
\end{equation}
The Weyl tensor describes the free gravitational field. The
$F_{\mu\nu}$ tensor is defined as:

\begin{equation}\label{FEF}
F_{\mu\nu}=T^{(5)}_{ab}g_{\mu}^ag_{\nu}^b+\left[T^{(5)}_{ab}n^an^b-
\frac{1}{4}T^{(5)}\right]g_{\mu\nu},
\end{equation}
and includes contributions on the brane of the 5D energy-momentum
tensor (if there is a bulk scalar field this is where it is felt).
In the special case of Minkowski branes and an empty bulk, the Weyl
tensor is zero, so that ${\cal E}_{\mu\nu}=0$ and due to the empty
bulk $F_{\mu\nu}=0$. Also in Minkowski spacetime we have
$T_{\mu\nu}=0$ and therefore $S_{\mu\nu}=0$, and $\Lambda_4=0$. The
induced field equations on the brane are simply given by:

\begin{equation}\label{FEBSD}
G_{\mu\nu}=0.
\end{equation}

\subsection{Kaluza-Klein Modes}\label{CKKMB}

In this section we shall look at the Kaluza-Klein (KK) modes on the
Minkowski brane. Before we look at the solutions we should consider
the polarisations of the 5D graviton as this generalises the
equivalent 4D polarisations.

\subsubsection{Polarisations of the 5D graviton}
The number of polarisations of the graviton is given by the number
of degrees of freedom in the perturbation ($h_{ab}$) of the bulk
metric tensor:

\begin{equation}\label{pmet}
g^{(5)}_{ab}\rightarrow g^{(5)}_{ab}+h_{ab}.
\end{equation}
The perturbation initially has 15 components since the metric is
symmetric. We then apply the Randall-Sundrum gauge, which is the
Gaussian normal and the transverse-traceless gauges combined. The
Gaussian normal gauge ($h_{ay}=0$) removes 5 degrees of freedom. The
transverse-traceless gauge (after the application of the Gaussian
normal gauge) is:

\begin{equation}\label{TTg}
\nabla^\nu h_{\mu\nu}=0=g^{(5)\,\mu\nu}h_{\mu\nu}.
\end{equation}
This removes another 5 degrees of freedom. This means that the 5D
graviton has 5 degrees of freedom and therefore 5 polarisation
states. These polarisation states (as felt on the brane) are
decomposed into~\cite{Maartens:2003tw}:

\begin{itemize}
  \item 2 polarisations in 4D spin-2 graviton modes.
  \item 2 polarisations in 4D spin-1 gravi-vector modes.
  \item 1 polarisation in 4D spin-0 gravi-scalar modes.
\end{itemize}
The normal 4D graviton is the massless spin-2 mode. The massive
modes of the graviton are felt as massive modes in all of these
fields on the brane, i.e massive graviton, gravi-vector and
gravi-scalar.

\subsubsection{Obtaining the gravitational wave equation}

The perturbed line element is:

\begin{equation}\label{LEP}
ds^2=\left(a^2(y)\eta_{\mu\nu}+\varepsilon h_{\mu\nu} \right )dx^\mu
dx^\nu+dy^2,
\end{equation}
where $\left|\varepsilon\right|\ll 1$. The modified metric can be
written as:

\begin{eqnarray}\label{Pmet}
g_{ab}^{(5)}&=&\overline{g}_{ab}+\delta g_{ab}, \nonumber \\
&=&\overline{g}_{ab}+\varepsilon h_{\mu\nu}\delta^\mu_a\delta^\nu_b,
\end{eqnarray}
where $\overline{g}_{ab}$ is the background metric (all subsequent
barred terms are background quantities). Therefore for the inverse
metric we have:

\begin{equation}\label{PmetIn}
g^{(5)\,ab}=\overline{g}^{ab}-\varepsilon
h^{\mu\nu}\delta_\mu^a\delta_\nu^b.
\end{equation}
The perturbed Christoffel symbol is:

\begin{equation}\label{CHPFh}
\delta\Gamma^{(5)\,a}_{~~~~\,bc}=\frac{\varepsilon}{2}\left\{\overline{g}^{ad}
\left(h_{bd,c}+h_{dc,b}-h_{bc,d}\right)-
h^{ad}\left(\overline{g}_{bd,c}+\overline{g}_{dc,b}-\overline{g}_{bc,d}\right)\right\}.
\end{equation}
The background Christoffel symbol is given by Eq.~(\ref{CHmS}),
$h_{ay}=0$ and $\overline{g}_{ay,b}=0$. This becomes:

\begin{equation}\label{CHPFhv}
\delta\Gamma^{(5)\,a}_{~~~~\,bc}=\frac{\varepsilon}{2}\left\{
\overline{g}^{a\nu}\left(h_{b\nu,c}+h_{\nu
c,b}-h_{bc,\nu}\right)-g^{ay}h_{bc,y} h^{a\nu}
\left(\overline{g}_{b\nu,c}+\overline{g}_{\nu
c,b}-\overline{g}_{bc,\nu}\right)\right\}.
\end{equation}
Then by using the relationships in Eqs.~(\ref{diffMet}),~
(\ref{diffMetU}) and defining $\tilde{h}_{\mu\nu}=a^{-2}h_{\mu\nu}$
this can be written as:

\begin{eqnarray}\label{DLF}
\delta\Gamma^{(5)\,a}_{~~~~\,bc}&=&\frac{\varepsilon}{2}\left\{
\delta^a_\gamma
\left[\tilde{h}^{\;\;\gamma}_{b\;\;,c}+\tilde{h}^\gamma_{\;\;c,b}-\tilde{h}^{\;\;\;\;\gamma}_{bc,}
-\frac{2}{\ell}\left(\delta^y_c\tilde{h}_b^{\;\;\gamma}+\delta^y_b\tilde{h}_c^{\;\;\gamma}\right)\right]\right.\nonumber\\
&&~~~~+\left.a^2\delta^a_y\left(\frac{2}{\ell}\tilde{h}_{bc}-\tilde{h}_{bc,y}\right)+\frac{2}{\ell}\left(\delta^y_c\delta^\mu_b\tilde{h}^a_{\;\;\mu}+
\delta^y_b\delta^\alpha_c\tilde{h}^a_{\;\;\alpha}\right)\right\}.
\end{eqnarray}
In order to write out the perturbed Riemann tensor we need the
explicit form of the derivative of Eq.~(\ref{DLF}). We find this to
be:

\begin{eqnarray}\label{DLFD}
\delta\Gamma^{(5)\,a}_{~~~~\,bc,d}&=&\frac{\varepsilon}{2}\left\{
\delta^a_\gamma
\left[\tilde{h}^{\;\;\gamma}_{b\;\;,cd}+\tilde{h}^\gamma_{\;\;c,bd}-\tilde{h}^{\;\;\;\;\gamma}_{bc,\;\;d}
-\frac{2}{\ell}\left(\delta^y_c\tilde{h}_{b\;\;,d}^{\;\;\gamma}+\delta^y_b\tilde{h}_{c\;\;d}^{\;\;\gamma}\right)\right]
\right.\nonumber\\
&&~~~~+\left.a^2\delta^a_y\left(\frac{2}{\ell}\tilde{h}_{bc,d}-\tilde{h}_{bc,yd}\right)-
\frac{2a^2}{\ell}\delta^a_y\delta^y_d\left(\frac{2}{\ell}\tilde{h}_{bc}-\tilde{h}_{bc,y}\right)\right.\nonumber\\
&&~~~~+\left.\frac{2}{\ell}\left(\delta^y_c\delta^\mu_b\tilde{h}^a_{\;\;\mu,d}+
\delta^y_b\delta^\alpha_c\tilde{h}^a_{\;\;\alpha,d}\right)\right\}.
\end{eqnarray}

We now have all the required quantities for the perturbed Riemann
tensor which is given by:

\begin{equation}\label{VRfa}
\delta R^{(5)\,a}_{~~~~\,bcd}=\delta
\Gamma^{(5)\,a}_{~~~~\,bd,c}-\delta
\Gamma^{(5)\,a}_{~~~~\,bc,d}+\overline{\Gamma}^e_{bd}\delta
\Gamma^{(5)\,a}_{~~~~\,ec}+\overline{\Gamma}^a_{ec}\delta
\Gamma^{(5)\,e}_{~~~~\,bd}-\overline{\Gamma}^e_{bc}\delta
\Gamma^{(5)\,a}_{~~~~\,ed}-\overline{\Gamma}^a_{ed}\delta
\Gamma^{(5)\,e}_{~~~~\,bc}.
\end{equation}
In order for us to obtain the perturbed Einstein equations we do not
need to explicitly write out the full perturbed Riemann tensor. We
require the perturbed Ricci tensor so we can just consider each term
and perform the Ricci contraction on each in turn.

We must also consider the Randall-Sundrum gauge requirements. We
have already used $h_{ay}=0$ but we also have
$h^\nu_{\;\;\mu,\nu}=0$ and $h^\mu_\mu=0$. Using these and the Ricci
contraction over the 1st and 3rd indices we find that
Eq.~(\ref{DLFD}) (the 2nd term in the perturbed Riemann tensor) has
zero contribution to the Ricci tensor. Writing out Eq.~(\ref{DLFD})
for the 1st term in the perturbed Riemann tensor and carrying out
the same procedure we find the perturbed Ricci terms to be:

\begin{equation}\label{RiC1}
\frac{\varepsilon}{2}\left\{-\tilde{h}_{bd,\;\;\gamma}^{\;\;\;\;\gamma}-\frac{4a^2}{\ell^2}\tilde{h}_{bd}
+\frac{4a^2}{\ell}\tilde{h}_{bd,y}-a^2\tilde{h}_{bd,yy}\right\}.
\end{equation}
The 3rd term in the perturbed Riemann tensor
($\overline{\Gamma}^e_{bd}\delta \Gamma^{(5)\,a}_{~~~~\,ec}$) also
gives us zero when we perform the Ricci contraction.  The 4th term
in the perturbed Riemann tensor ($\overline{\Gamma}^a_{ec}\delta
\Gamma^{(5)\,e}_{~~~~\,bd}$) gives us the following components of
the Ricci tensor:

\begin{equation}\label{RiC3}
-\frac{\varepsilon}{2}\left\{\frac{8a^2}{\ell^2}\tilde{h}_{bd}-\frac{4a^2}{\ell}\tilde{h}_{bd,y}\right\}.
\end{equation}
The 5th term in the Riemann tensor ($\overline{\Gamma}^a_{ed}\delta
\Gamma^{(5)\,e}_{~~~~\,bc}$) gives us:

\begin{equation}\label{RiC5}
-\frac{\varepsilon}{2}\left\{\frac{2a^2}{\ell^2}\tilde{h}_{bd}-\frac{2a^2}{\ell}\tilde{h}_{bd,y}\right\}.
\end{equation}
The final term in the Riemann tensor
($\overline{\Gamma}^e_{bc}\delta \Gamma^{(5)\,a}_{~~~~\,ed}$) also
gives us the following terms in the Ricci tensor:

\begin{equation}\label{RiC6}
-\frac{\varepsilon}{2}\left\{\frac{2a^2}{\ell^2}\tilde{h}_{bd}-\frac{2a^2}{\ell}\tilde{h}_{bd,y}\right\}.
\end{equation}
Then by combining
Eqs.~(\ref{RiC1}),~(\ref{RiC3}),~(\ref{RiC5}),~(\ref{RiC6}) and
making use of the fact that we can change the indices as long as the
summations stay correct we find the perturbed Ricci tensor to be:

\begin{equation}\label{PRi}
\delta R^{(5)}_{bd}=\frac{\varepsilon}{2}\left \{
-\tilde{h}_{bd,\;\;\gamma}^{\;\;\;\;\gamma}-\frac{8a^2}{\ell^2}\tilde{h}_{bd}+
\frac{4a^2}{\ell}\tilde{h}_{bd,y}-a^2\tilde{h}_{bd,yy}\right\}.
\end{equation}
Using Eq.~(\ref{Ri2}) we can write the following:

\begin{equation}\label{PRi2}
\delta R^{(5)}_{bd}=-\frac{4}{\ell^2}\varepsilon h_{bd}.
\end{equation}
By combing these two forms of the Ricci tensor we find that the
following wave equation must be satisfied:

\begin{equation}\label{PRi3}
-\tilde{h}_{bd,\;\;\gamma}^{\;\;\;\;\gamma}+
\frac{4a^2}{\ell}\tilde{h}_{bd,y}-a^2\tilde{h}_{bd,yy}=0.
\end{equation}
We can write the wave equation (\ref{PRi3}) as:

\begin{equation}\label{PRiGf}
\left [ a^{-2}\Box+\partial^2_y-\frac{4}{\ell}\partial_y \right ]
\tilde{h}_{\mu\nu}=0,
\end{equation}
where $\Box$ is the 4D Minkowski d'Alembertian operator. This can be
equivalently written with $\tilde{h}_{bd}$ instead of
$\tilde{h}_{\mu\nu}$. The wave equation (\ref{PRiGf}) is also:

\begin{equation}\label{PRiG5B}
\Box^{(5)}\tilde{h}_{\mu\nu}=0.
\end{equation}
Using $h_{\mu\nu}$ the wave equation (\ref{PRiGf}) becomes:

\begin{equation}\label{PRiGfh}
\left [ a^{-2}\Box+\partial^2_y-\frac{4}{\ell^2}\right ]
h_{\mu\nu}=0,
\end{equation}
which agrees with the form given in~\cite{Langlois:2002ux}. The
junction conditions can be enforced via a delta term which takes
into account the presence of the brane at $y=0$:
\begin{equation}\label{PRiGfhf}
\left [
a^{-2}\Box+\partial^2_y-\frac{4}{\ell^2}+\frac{4}{\ell}\delta(y)\right
] h_{\mu\nu}=0.
\end{equation}
The two forms of the line element seen so far are written in
Gaussian normal coordinates (the extra dimension is normal to a
given hypersurface, the Minkowski spacetime brane in this case).
Another coordinate system that can be used is the Poincar\'{e}
system. This gives us a line element of the form:

\begin{equation}\label{PCLE}
ds^2=\frac{\ell^2}{z^2} \left [ dz^2+\eta_{\mu\nu}dx^\mu dx^\nu
\right ],
\end{equation}
where $z$ is defined as:

\begin{equation}\label{z}
z=\ell e^{y/\ell}.
\end{equation}
The perturbed line element in these coordinates is given by:

\begin{equation}\label{PCLEP}
ds^2=\frac{\ell^2}{z^2} \left [
dz^2+\left(\eta_{\mu\nu}+\varepsilon\tilde{h}_{\mu\nu}\right)dx^\mu
dx^\nu \right ].
\end{equation}

\subsubsection{Solving the wave equation}

The solutions to the wave equation (\ref{PRiGfh}) can be written as
the superposition of Fourier modes of the form:

\begin{equation}\label{sol1}
h_{\mu\nu}(x^\lambda,y)=U_m(y)e^{\imath k_\lambda
x^\lambda}\epsilon_{\mu\nu},
\end{equation}
where $\epsilon_{\mu\nu}$ is a (constant) polarisation tensor. We
are interested in $U_m(y)$, the bulk gravitational wave part of the
decomposition. By using Eq.~(\ref{sol1}) in Eq.~(\ref{PRiGf}) we
find the $y$ dependent wave equation to be:

\begin{equation}\label{WEm}
\left[e^{2y/\ell}m^2+\partial_y^2-\frac{4}{\ell^2}
\right]e^{-2y/\ell}U_m=0,
\end{equation}
where $m$ can be interpreted as the 4D mass of the 5D graviton. The
mass is due to the fact that an observer on the brane would measure
the timelike 4D projection on the brane of the gravitons null
5-momentum:

\begin{equation}\label{m}
-m^2=k_\lambda k^\lambda.
\end{equation}
The zero mode satisfies the boundary conditions given by $U_0=C$
where $C$ is a constant. For $m>0$, we now define the function:

\begin{equation}\label{WEmf}
F_m(y)=e^{-2y/\ell}U_m(y),
\end{equation}
and change variable to:

\begin{equation}\label{x}
X=e^{y/\ell}m\ell,
\end{equation}
to rewrite Eq.~(\ref{WEm}) as:

\begin{equation}\label{WEmx}
X^2\frac{\partial^2 F_m}{\partial X^2}+X\frac{\partial F_m}{\partial
X}+\left(X^2-4\right)F_m=0.
\end{equation}
This is a Bessel equation, with solution:

\begin{equation}\label{solFY}
F_m=Z_2 \left(e^{y/\ell}m\ell\right),
\end{equation}
where:

\begin{equation}\label{BesZ}
Z_p(S)=AJ_p(S)+BY_p(S),
\end{equation}
with $A$ and $B$ constant and $p$ is the order of the Bessel
function. By Eq.~(\ref{sol1}) the full solution for $m>0$ is:

\begin{equation}\label{hsol}
h_{\mu\nu}(x^\lambda,y)=e^{2y/\ell}Z_2
\left(e^{y/\ell}m\ell\right)e^{\imath k_\lambda
x^\lambda}\epsilon_{\mu\nu}.
\end{equation}

\subsubsection{Schr\"odinger form of the wave equation}

\begin{figure}
\begin{center}
\includegraphics[height=3in,width=4.0in,angle=0]{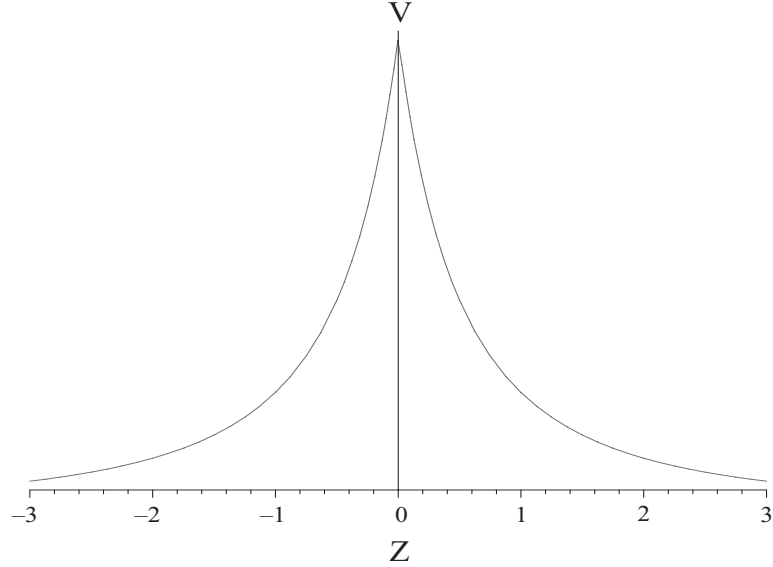}
\caption{A plot of the potential in equation (\ref{potZ}) with
$\ell=1$} \label{potZplot}
\end{center}
\end{figure}

The wave equation (\ref{WEm}) can be rewritten in a form similar to
the Schr\"odinger wave equation. We define two new variables as:

\begin{equation}\label{psim}
\psi _m = a^{3/2} U_m,
\end{equation}
\begin{equation}\label{Z}
Z= \ell (e^{y/\ell} -1).
\end{equation}
This gives:

\begin{equation}\label{SE}
-\frac{\partial ^2 \psi_m }{ \partial Z^2}+V(Z)\psi_m=m^2\psi_m.
\end{equation}
This is a Schr\"odinger equation with a potential $V(Z)$ given by:

\begin{equation}\label{potZ}
V(Z)=\frac{15}{4(Z+\ell)^2}.
\end{equation}
This is plotted in Fig.~\ref{potZplot}.

\subsubsection{Specific Solutions}

In order to obtain specific solutions we must apply boundary
conditions to the general solution. The boundary conditions for the
perturbations are:

\begin{equation}\label{BC}
\frac{\partial h_{\mu\nu}(y)}{\partial y}\Big{|}_{brane}=0.
\end{equation}
This means RS1 has two boundary conditions whilst RS2 only has one.
\begin{figure}
\begin{center}
\includegraphics[height=4in,width=3in,angle=270]{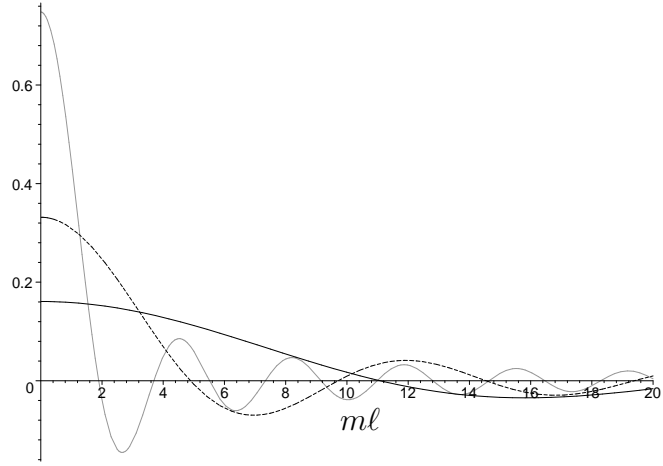}
\rput(-5,-6.3){$m\ell$} \caption{A plot of equation
(\ref{BesDetEQF}) with $L/\ell=1/4,~1/2$ and $L/\ell=1$ for the
solid, dashed and pale solid lines respectively.}
\label{bessSolplot}
\end{center}
\end{figure}
For RS1:

\begin{equation}\label{RS1BCEQ0}
A \frac{\partial}{\partial
y}\left(e^{2y/\ell}J_2(e^{y/\ell}m\ell)\right)|_{(y=0)}+ B
\frac{\partial}{\partial
y}\left(e^{2y/\ell}Y_2(e^{y/\ell}m\ell)\right)|_{(y=0)} =0,
\end{equation}
\begin{equation}\label{RS1BCEQL}
A \frac{\partial}{\partial
y}\left(e^{2y/\ell}J_2(e^{y/\ell}m\ell)\right)|_{(y=L)}+ B
\frac{\partial}{\partial
y}\left(e^{2y/\ell}Y_2(e^{y/\ell}m\ell)\right)|_{(y=L)} =0.
\end{equation}
These can then be combined in the matrix equation:

\begin{equation}\label{Besmatrix}
\left[\begin{array}{cc} J_2'(0)& Y_2'(0) \\J_2'(L)& Y_2'(L)
\end{array}\right] \left[
\begin{array}{c} A \\ B \end{array}\right]=0,
\end{equation}
where $J_2'(0)=\frac{\partial}{\partial
y}J_2(e^{y/\ell}m\ell)|_{(y=0)}$ and so on.

In order for us to obtain a unique and non-trivial solution we
require that:

\begin{equation}\label{BesDetEQ}
  J_2'(0)Y_2'(L)-J_2'(L) Y_2'(0)=0,
\end{equation}
i.e.:

\begin{equation}\label{BesDetEQF}
J_1\left(m\ell\right)Y_1\left(e^{L/\ell
}m\ell\right)-J_1\left(e^{L/\ell}m\ell\right)Y_1\left(m\ell\right)=0.
\end{equation}
In Fig.~\ref{bessSolplot} we see that for a particular value of
$L/\ell$ we have a series of values of $m\ell$ that satisfy the
above equation. We see that as we increase the brane separation the
mode spacings decrease. We find that we have a zero mode solution
with an infinite tower of discrete massive Kaluza-Klein modes. In
the RS2 scenario we only have a single brane and therefore only one
boundary condition. This means we can not solve for $m$
independently of $A$ and $B$ and therefore obtain a specific
solution. Thus RS2 has an infinite continuous tower of Kaluza-Klein
modes. This is confirmed by taking the RS1 solution and sending the
second brane off to infinity.


\section{de Sitter Brane Worlds}

The de Sitter brane is the simplest expanding solution. A de Sitter
universe expands at a uniform rate given by the Hubble constant $H$.

\subsection{The background}
In Poincar\'{e} coordinates the spacetime is described
by~\cite{Gen:2000nu}:

\begin{equation}\label{GSPdS}
ds^2=A(z)^2\left(dz^2+\gamma_{\mu\nu}dx^\mu dx^\nu\right),
\end{equation}
where:

\begin{equation}\label{dSit}
\gamma_{\mu\nu}dx^\mu dx^\nu=-dt^2+e^{2Ht}d\vec{x}^2.
\end{equation}
and the warp factor is:

\begin{equation}\label{a(z)}
A(z)=\frac{\ell H}{\textrm{sinh}\left(Hz\right)}.
\end{equation}
On the visible (positive tension) brane, $z=z_+$, we need to have
the standard de Sitter solution ($A(z_+)=1$), so that:

\begin{equation}\label{HB21}
z_+=H^{-1}\textrm{sinh}^{-1}\left(H\ell\right).
\end{equation}

Let us consider the two brane case where the second brane is at
$z=z_->z_+$. The extrinsic curvature in this coordinate system is
given by:

\begin{equation}\label{Ka}
K_{\mu\nu}=-\frac{1}{2}A^{-1}\partial_z g_{\mu\nu},
\end{equation}
leading to:

\begin{equation}\label{PEKF}
K_{\mu\nu}=\left(\frac{\textrm{cosh}(Hz)}{\ell}\right) g_{\mu\nu}.
\end{equation}
The extrinsic curvature as found from the junction conditions,
Eq.~(\ref{EKR}), is independent of the brane geometry. For de Sitter
branes $T_{\mu\nu}^{\pm}=-\rho_{\pm} g_{\mu\nu}^{\pm}$ with
$\rho_{\pm}$ constant, unlike the Minkowski case, where
$T_{\mu\nu}^{\pm}=0$. The extrinsic curvature for a de Sitter brane
can then be written as:

\begin{equation}\label{EKdS2}
 K_{\mu\nu}=-\frac{\kappa_5^2}{6}\sigma g_{\mu\nu}.
\end{equation}
where $\sigma=\lambda+\rho$ is the total energy density on the
brane. By Eq.~(\ref{PEKF}) we obtain the following relationship:

\begin{equation}\label{dStension}
\sigma_{\pm}=\pm\frac{6}{\kappa_5^2\ell}\textrm{cosh}(Hz_{\pm}).
\end{equation}
There is some degeneracy in the brane tension and the energy density
on the brane. We can rewrite the above equation in terms of the
brane tensions as:

\begin{equation}\label{dStensionEx}
\lambda_{\pm}=\pm\frac{6}{\kappa_5^2\ell}\textrm{cosh}(Hz_{\pm})-\rho_{\pm}.
\end{equation}
Now if we evaluate the metric on each of the two branes:

\begin{itemize}
\item Positive tension brane:
\begin{equation}\label{+BM}
ds^2|_{z_+}=-dt^2+e^{2Ht}d\vec{x}^2.
\end{equation}
This is just the standard de Sitter line element as we have set the
observer on this brane.
\item Negative tension brane:
\begin{eqnarray}\label{-BM}
ds^2|_{z_-}&=&A^2(z_-)\left(-dt^2+e^{2Ht}d\vec{x}^2\right),\nonumber
\\&=&-dt_-^2+e^{2H_-t_-}d\vec{x}^2,
\end{eqnarray}
\end{itemize}
where we have defined the proper time on the negative tension brane
as:

\begin{equation}\label{NTPT2}
t_-=A(z_-)t+C,
\end{equation}
with $C$ constant. The Hubble rate on the negative brane is also
modified by the warp factor:

\begin{equation}\label{NH4}
H_-=\frac{H}{A(z_-)}.
\end{equation}
This means that the warp factor on the negative tension brane is
less than that on the positive tension brane~\cite{Felder:2001da}.

The form of the Hubble rates given above means that the two are
dependent only on the brane position ($z$). We can view the
expansion of the brane as a consequence of the brane moving through
the warped bulk. We can write the Hubble constant on each brane as:

\begin{equation}\label{HBrel}
H_\pm=\frac{\textrm{sinh}H_+z_\pm}{\ell},
\end{equation}
where $H_+=H$ (the Hubble rate on the positive tension brane). Using
this result in the brane tension, Eq.~(\ref{dStension}), we find
that the tensions are given, as functions of the Hubble rate, by:

\begin{equation}\label{dStensionF}
\lambda_{\pm}=\pm\frac{6}{\kappa_5^2\ell}\sqrt{1+(\ell
H_\pm)^2}-\rho_{\pm}.
\end{equation}

High energy branes are characterised by $H\ell\gg1$, so that:

\begin{equation}\label{dStensionFHE}
\lambda_{\pm}\approx\pm\frac{6}{\kappa_5^2}H_\pm-\rho_{\pm}.
\end{equation}
On low energy ($H\ell\ll 1$) branes:

\begin{equation}\label{dStensionFLE}
\lambda_{\pm}\approx\pm\frac{6}{\kappa_5^2\ell}-\rho_{\pm}.
\end{equation}
This reduces to the Minkowski relation when $\rho_\pm=0$.

In the de Sitter case:

\begin{equation}\label{SQEMT2}
S_{\mu\nu}=-\frac{1}{12}\rho^2g_{\mu\nu}.
\end{equation}
The bulk is still empty in this case so the other two terms (${\cal
E}_{\mu\nu}$ and $F_{\mu\nu}$) are still zero. The induced field
equations on the brane are then given by:

\begin{equation}\label{FEBS}
G_{\mu\nu}=-\Lambda_4 g_{\mu\nu}-\kappa_4^2\rho
g_{\mu\nu}-\frac{\kappa^2_4}{2\lambda}\rho^2g_{\mu\nu},
\end{equation}
where

\begin{equation}\label{kappalambdaRel+}
\kappa^2_{4}=\frac{\lambda}{6}\kappa^4_5.
\end{equation}
Using Eqs.~(\ref{Lb}) and (\ref{L5}) we can write the field
equations on the brane as:

\begin{equation}\label{FEB2}
G_{\mu\nu}^\pm=-\left\{\frac{\lambda_\pm\kappa_5^4}{6}\rho_\pm+
\frac{\kappa_5^4}{12}\rho^2_\pm+\frac{1}{2}\left[\frac{\lambda_\pm^2\kappa_5^4}{6}-\frac{6}{\ell^2}
\right]\right\}g_{\mu\nu}^\pm.
\end{equation}

\subsection{Kaluza-Klein modes}\label{CKKMdSB}

In Gaussian normal coordinates, the metric for a de Sitter brane is:

\begin{equation}\label{KLE}
ds^2=dy^2+w^2(y)\gamma_{\mu\nu}dx^\mu dx^\nu,
\end{equation}
where warp factor is given by:

\begin{equation}\label{W}
w(y)=H\ell\textrm{sinh}\left(y/\ell\right).
\end{equation}
If $y_1$ is the position of the visible brane then:

\begin{equation}\label{HB22}
H=\frac{1}{\ell\textrm{sinh}\left(y_1/\ell\right)}.
\end{equation}

The perturbed metric is:

\begin{equation}\label{PM}
ds^2=dy^2+w^2(y)\left(\gamma_{\mu\nu}+f_{\mu\nu}\right)dx^\mu
dx^\nu,
\end{equation}
where:

\begin{equation}\label{Pm2}
\gamma^{\mu\nu}f_{\mu\nu}=0=\nabla^{\nu}f_{\mu\nu}.
\end{equation}
The perturbation can be split into Fourier modes:

\begin{equation}\label{PS}
f_{\mu\nu}\left(t,\vec{x},y\right)\rightarrow
f_m(y)\psi_m\left(t,\vec{k}\right)e_{\mu\nu}\left(\vec{x}\right).
\end{equation}
Using these in the perturbed field equations we obtain the following
two wave equations~\cite{Langlois:2000ns}:

\begin{equation}\label{PWEsy}
f''_m+4\frac{w'}{w}f'_m+\left(\frac{m}{w}\right)^2f_m=0,
\end{equation}
and
\begin{equation}\label{PWEst}
\ddot{\psi}_m+3H\dot{\psi}_m+
\left[\left(\frac{k}{e^{Ht}}\right)^2+m^2\right]\psi_m=0,
\end{equation}
where $m$ is the separation constant and again can be considered as
the effective mass of the 4D graviton. Eq.~(\ref{PWEsy}) becomes:

\begin{equation}\label{PWEsy2}
f''_m(y)+4\frac{\textrm{cosh}(y/\ell)}{\ell\textrm{sinh}(y/\ell)}f'_m(y)+
\left(\frac{m}{H\ell\textrm{sinh}(y/\ell)}\right)^2f_m(y)=0.
\end{equation}
The zero-mode satisfying the boundary condition is given by:

\begin{equation}\label{zm}
f_0={\rm constant}.
\end{equation}
For the massive modes, $m>0$, we rewrite Eq.~(\ref{PWEsy2}) as:

\begin{equation}\label{PWEsy3}
\left[\frac{1}{\textrm{sinh}^4(y/\ell)}\partial_y(\textrm{sinh}^4(y/\ell)\partial_y)+
\left(\frac{m}{H\ell\textrm{sinh}(y/\ell)}\right)^2\right]f_m(y)=0.
\end{equation}

The wave equation for (\ref{PWEst}) can be written as:

\begin{equation}\label{WEX}
\left[\Box-m^2\right]\psi_m=0.
\end{equation}

If we make the following substitutions:

\begin{eqnarray}\label{subs}
U=\textrm{sinh}^{3/2}(y/\ell)\psi(y),~~Z=\textrm{cosh}(y/\ell),
\end{eqnarray}
we can write Eq.~(\ref{PWEsy3}) as:

\begin{equation}\label{AL1}
(1-Z^2)\frac{d^2U}{dZ^2}-2Z\frac{dU}{dZ}+\left[\nu(\mu+1)-\frac{\mu^2}{1-Z^2}\right]U=0,
\end{equation}
where $\nu=\frac{3}{2}$ and $\mu=i\sqrt{((m/H)^2-9/4)}=ip$. The
solutions to this equation are associated  Legendre functions. The
general solution is a linear combination of the two associated
Legendre functions $P^{\mu}_{\nu}$ and $Q^{\mu}_{\nu}$. We find the
profile for the KK modes are given by:

\begin{eqnarray}\label{WE}
f_m(y)=\textrm{sinh}^{-3/2}\left(y/\ell\right)
\left\{A_pP^{i\sqrt{(m/H)^2-9/4}}_{3/2}\left(\textrm{cosh}\left(y/\ell\right)\right)
\right.\nonumber\\\left.+B_pQ^{i\sqrt{(m/H)^2-9/4}}_{3/2}\left(\textrm{cosh}\left(y/\ell\right)\right)\right\}.
\end{eqnarray}
The two branes, at $y_1$ and $y_2$, are related by:

\begin{equation}\label{L}
y_2-y_1=L.
\end{equation}
We assume $y_1>0$, $y_2>y_1$. The boundary conditions are the same
as in Eq.~(\ref{BC}). To obtain forms for the derivatives we use the
following relationship~\cite{ZieleckandGreg}:

\begin{equation}\label{ALFR}
\frac{dP^\nu_\mu(Z)}{dZ}=\left(Z^2-1\right)^{-1}\left\{\mu
ZP^\nu_\mu(Z)-(\mu+\nu)P^\nu_{\mu-1}(Z)\right\},
\end{equation}
which is valid for $Re(Z)>1$. $Q^{\mu}_{\nu}(Z)$ obeys the same
relations. Then:

\begin{figure}
\begin{center}
\includegraphics[height=4in,width=3in,angle=270]{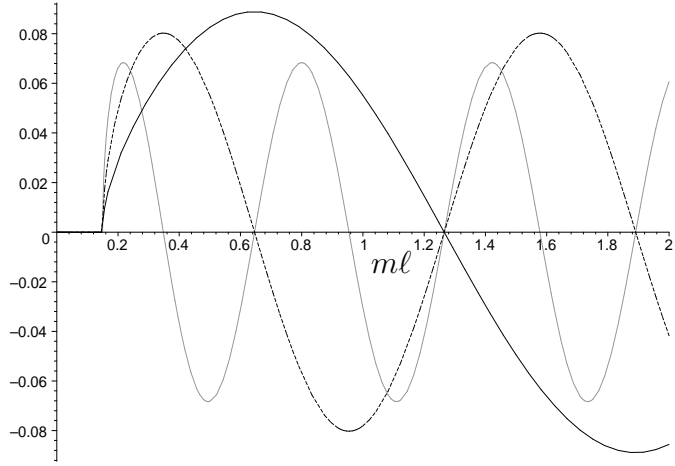}
\rput(-4.8,-4.2){$m\ell$} \caption{Determining the KK mass
eigenvalues for two de Sitter branes with $L/\ell=1/4,~1/2$ and
$L/\ell=1$ for the solid, dashed and light solid lines respectively.
The y-axis is the left-hand side of Eq.~(\ref{ASS}). $H\ell=0.1$ in
this plot.} \label{ModesLl}
\end{center}
\end{figure}

\begin{eqnarray}\label{TdP}
\frac{df_m(y)}{dy}=-\left\{\frac{3/2+i\sqrt{\left(m/H\right)^2-9/4}}{\ell
\textrm{sinh}^{5/2}\left(y/\ell\right)
}\right\}\left[A_mP^{i\sqrt{(m/H)^2-9/4}}_{1/2}\left(\textrm{cosh}\left(y/\ell\right)\right)\right.\nonumber\\
+\left.B_mQ^{i\sqrt{(m/H)^2-9/4}}_{1/2}\left(\textrm{cosh}\left(y/\ell\right)\right)\right].\nonumber\\
\end{eqnarray}
We then obtain the specific solution to the boundary conditions:

\begin{equation}\label{ASS}
P^{ip}_{1/2}\left(\textrm{cosh}\left(y_1/\ell\right)\right)Q^{ip}_{1/2}\left(\textrm{cosh}\left(y_2/\ell\right)\right)-
P^{ip}_{1/2}\left(\textrm{cosh}\left(y_2/\ell\right)\right)Q^{ip}_{1/2}\left(\textrm{cosh}\left(y_1/\ell\right)\right)
=0,
\end{equation}
where:

\begin{equation}\label{Y1}
\frac{y_1}{\ell}=\textrm{arcsinh}\left(\frac{1}{H\ell}\right),
~~\frac{y_2}{\ell}=\frac{y_1}{\ell}+\frac{L}{\ell}.
\end{equation}
The KK mass eigenvalues defined by this equation are illustrated in
Fig.~\ref{ModesLl}. As far as we are aware, these eigenvalues have
not previously been found.

By rewriting Eq.~\ref{PWEsy3} in Schr\"odinger  form, it is possible
to show that for normalisable modes \cite{Langlois:2000ns}:

\begin{equation}\label{MassGap}
m^2>\frac{9}{4}H^2.
\end{equation}
In the general solution, Eq.~(\ref{WE})  , the associated Legendre
function $P^{i\sqrt{(m/H)^2-9/4}}_{3/2}$ diverges for large $y$ if
$i\sqrt{(m/H)^2-9/4}$ is real i.e. for $m^2<\frac{9}{4}H^2$.  There
is a mass gap between the zero mode and the normalisable KK tower in
both the one and two brane set-up.

\begin{figure}
\begin{center}
\includegraphics[height=4in,width=3in,angle=270]{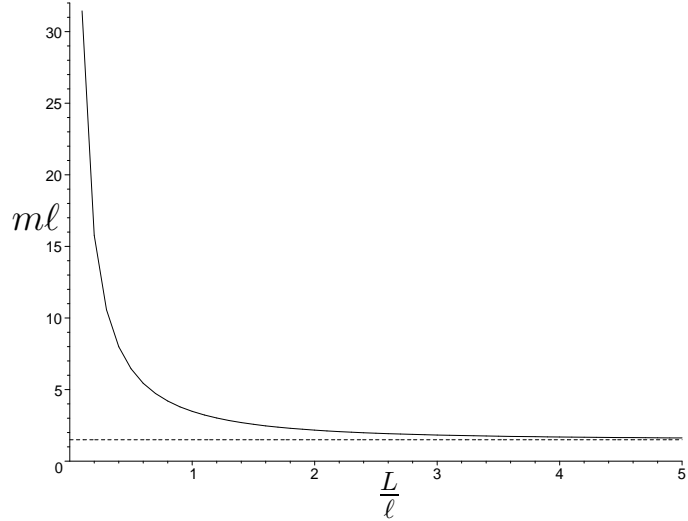}
\rput(-5,-7.3){\large $\frac{L}{\ell}$}\rput(-9.7,-3.6){\large
$m\ell$}\caption{A plot of the first massive mode as a function of
$L/\ell$ (solid line), and the mass-gap (dashed line). We see that
as we increase $L/\ell$ the first massive mode approaches $3H/2$.
$H\ell=1$ in this plot.} \label{MG}
\end{center}
\end{figure}

As with the Minkowski solution if we increase the separation of the
branes the mode separation decreases (see Fig.~\ref{ModesLl}). In
Fig.~\ref{MG} we can see the spacing between the zero mode and the
first massive mode as a function of $L/\ell$.
If we consider the RS2 model and send the second brane off to
infinity we find that there is again a continuum of massive modes
starting at $m^2>\frac{9}{4}H^2$ ($m^2=\frac{9}{4}H^2$ is not a
solution).

\begin{figure}
\begin{center}
\includegraphics[height=4in,width=3in,angle=270]{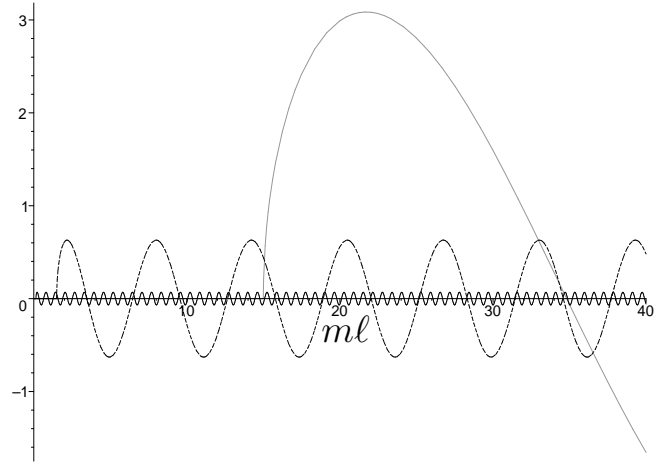}
\rput(-5.1,-5.1){\large $m\ell$} \caption{The KK mass eigenvalues
for two de Sitter branes with different values of $H\ell$, i.e.
different energy regimes. $H\ell=0.1,~1$ and $H\ell=10$ for the
solid, dashed and light solid lines respectively. The y-axis is the
left-hand side of Eq.~(\ref{ASS}). $L/\ell=1$ in this plot.}
\label{SitHigh}
\end{center}
\end{figure}

\begin{figure}
\begin{center}
\includegraphics[height=4in,width=3in,angle=270]{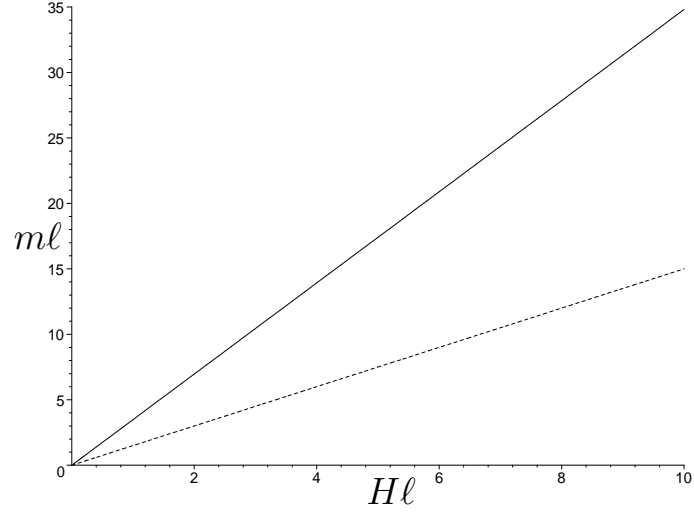}
\rput(-9.7,-3.8){\large $m\ell$} \rput(-5,-7.2){\large $H\ell$}
\caption{A plot of the first massive mode as a function of $H\ell$
(solid line), and the mass-gap (dashed line). We see that as we
increase $H\ell$ the first massive mode diverges from $3H/2$.
$L/\ell=1$  in this plot.} \label{MGHl}
\end{center}
\end{figure}

In Fig.~\ref{SitHigh} we see the effect of going from the low to the
high energy regime. At high energy the first KK modes are much more
massive than in the low energy regime. The spacings are also much
larger. Fig.~\ref{MGHl} shows the spacing between the first massive
mode and the mass gap as a function of $H\ell$. As we increase the
energy ($H\ell$ increasing) the mass gap increases linearly. The
mass of the first KK mode also increases linearly but at a faster
rate than the mass gap. Therefore at higher energies we must put in
even more energy, relative to the low energy regime, in order to
excite the KK modes.

\section{Friedmann Equations}

Here we shall consider Friedmann-Robertson-Walker (FRW) branes. This
is a general cosmological solution where the Hubble rate decreases
as the energy density goes down. The FRW solution describes the
uniform expansion of a homogenous and isotropic perfect fluid.

We shall derive the Friedmann equations, that govern the evolution
of the model, for the standard GR case before we consider the brane
results.

\subsection{The GR result}
The FRW metric is:

\begin{equation}\label{FRWLE}
ds^2=-dt^2+a^2(t)\left\{\frac{dr^2}{1-Kr^2}+r^2d\theta^2+r^2\sin^2\theta\right\},
\end{equation}
where $t={\rm constant}$ are maximally symmetric 3-spaces. The scale
factor is $a(t)$ and $K$ is the spatial curvature. We then use the
Einstein Field equations:

\begin{equation}\label{EFE}
G_{\mu\nu}=-\Lambda g_{\mu\nu}+\kappa^2 T_{\mu\nu}.
\end{equation}
For $\Lambda=0$, $\kappa^2=8\pi G$:

\begin{equation}\label{EFE2}
 R_{\mu\nu}=8\pi G\left(T_{\mu\nu}-\frac{1}{2}g_{\mu\nu}T\right).
\end{equation}
We use the energy-momentum tensor for a perfect fluid:

\begin{equation}\label{EMTPF}
T^{\mu\nu}=\left(\rho+P\right)U^\mu U^\nu+Pg^{\mu\nu},
\end{equation}
where $U^{\mu}$ is the 4-velocity:

\begin{equation}\label{4V}
U^{\mu}=\frac{dx^{\mu}}{ds},~~U^{\mu}U_{\mu}=-1.
\end{equation}
Therefore:

\begin{equation}\label{T2}
T=-\rho+3P.
\end{equation}
The spacetime is expanding but the fluid is at rest in a comoving
frame i.e:

\begin{equation}\label{4VR}
U^\mu=(1,0,0,0).
\end{equation}
The other ingredient that we need is the conservation of energy
equation:

\begin{equation}\label{CE}
\nabla_{\mu}T^{\mu}_{\nu}=0.
\end{equation}
The time component gives:

\begin{equation}\label{CE2}
\dot{\rho}=-3\rho H(1+w),
\end{equation}
where $w=P/\rho$ is the equation of state ($|w|\leq 1$). If $w$ is
constant then:

\begin{equation}\label{ra}
\rho\propto a^{-3(1+w)}.
\end{equation}
Throughout this work we assume that $w$ is constant. Physically this
is not true as the effective equation of state will decrease from
$w=1/3$ to $w=0$ as the universe evolves from radiation domination
to matter domination. But is reasonable to assume $w$ is constant as
we either consider early time, radiation domination, or late time,
matter domination.

By Eq.~(\ref{EFE2}) we find the pure temporal component of the Ricci
tensor to be:

\begin{equation}\label{R00}
R_{00}=4\pi G (\rho+3P).
\end{equation}
From the metric we can show that:

\begin{equation}\label{R002}
R_{00}=-3\frac{\ddot{a}}{a}.
\end{equation}
So we have:

\begin{equation}\label{R003}
\frac{\ddot{a}}{a}=-\frac{4}{3}\pi G (\rho+3P),
\end{equation}
which is one of the Friedmann equations. We now consider the spatial
directions. In fact we only need to look at one as the expansion is
isotropic. If we look at the $\theta\theta$ component, from the
metric we have:

\begin{equation}\label{Rtheta}
R_{\theta\theta}=r^2(a\ddot{a}+2\dot{a}^2+2K).
\end{equation}
Eq.~(\ref{EFE2}) gives us:

\begin{equation}\label{Rij}
R_{ij}=4\pi G(\rho-P)g_{ij},
\end{equation}
so the $\theta\theta$ component is:

\begin{equation}\label{Rtheta2}
R_{\theta\theta}=4\pi G(\rho-P)a^2r^2.
\end{equation}
By Eqs.~(\ref{Rtheta}) and~(\ref{Rtheta2}) we have:

\begin{equation}\label{thetatheta}
\frac{\ddot{a}}{a}+2\frac{\dot{a}^2}{a^2}+2\frac{K}{a^2}=4\pi
G(\rho-P).
\end{equation}
Using this with Eq.~(\ref{R003}) we get the main Friedmann equation:

\begin{equation}\label{Fried}
\left(\frac{\dot{a}}{a}\right)^2=\frac{8\pi G}{3}\rho-\frac{K}{a^2}.
\end{equation}
This is usually written in terms of the Hubble parameter which is
defined as:

\begin{equation}\label{H}
 H\equiv\frac{\dot{a}}{a}.
\end{equation}
If we include a cosmological constant in the field equations the
Friedmann equations become:

\begin{equation}\label{Fried211}
\frac{\ddot{a}}{a}=-\frac{4}{3}\pi G (\rho+3P)+\frac{\Lambda}{3},
\end{equation}
and:
\begin{equation}\label{Fried21}
H^2=\frac{8\pi G}{3}\rho-\frac{K}{a^2}+\frac{\Lambda}{3}.
\end{equation}
These are the standard GR Friedmann equations. When we consider the
braneworld we will require the Israel junction conditions. This is
because we now live on a hypersurface so the Friedmann equations
only apply to a restricted part of the spacetime.

\subsection{RS result}

We consider a single Friedmann brane model (i.e RS2), so that
gravity is described by General Relativity in the low energy limit
(and not a Brans-Dicke theory). The line element
is~\cite{Binetruy:1999hy}:

\begin{equation}\label{RSLE}
ds^2_{(5)}=-N(t,y)dt^2+A^2(t,y)\gamma_{ij}dx^idx^j+B^2(t,y)dy^2,
\end{equation}
where $\gamma_{ij}$ is the maximally symmetric space given by:

\begin{equation}\label{gamma}
\gamma_{ij}dx^idx^j=\frac{dr^2}{1-Kr^2}+r^2d\theta^2+r^2\sin^2\theta
d\phi^2.
\end{equation}
The 5D field equations are:

\begin{equation}\label{En0}
G^{(5)}_{ab}=\kappa^2_5T_{ab}^{(5)},
\end{equation}
where the energy-momentum tensor is given by:

\begin{equation}\label{EMT0}
T^{(5)\,a}_b=T^a_b|_{bulk}+T^a_b|_{brane}.
\end{equation}
The bulk part of $T^{(5)}_{ab}$ is given by:
\begin{equation}\label{EMT10}
T^a_b|_{bulk}=-\frac{\Lambda_5}{\kappa^2_5}\delta^a_b,
\end{equation}
which is just the 5D bulk cosmological constant. The brane term is
given by:

\begin{equation}\label{EMT2}
T^a_b|_{brane}=\frac{\delta(y)}{b}{\rm
diag}(-\rho-\lambda,P-\lambda,P-\lambda,P-\lambda,0),
\end{equation}
where $\lambda$ is the brane tension. The standard conservation of
energy equation applies on the brane (only gravity can leak off the
brane into the bulk). If we follow \cite{Binetruy:1999hy} and look
at the non-zero components of the Einstein tensor we get for the
$tt$ component ($'=\partial/\partial y$ and
$\dot{}=\partial/\partial t$):

\begin{equation}\label{Gtt}
G^{(5)}_{tt}=3\left\{\daa\left(\daa+\dbb\right)-\frac{n^2}{b^2}
\left(\ppaa+\paa\left(\paa-\pbb\right)\right)+K\frac{N^2}{A^2}\right\}.
\end{equation}
For the three brane dimensions we get:

\begin{eqnarray}\label{Gij}
G^{(5)}_{ij}&=&\frac{A^2}{B^2}\gamma_{ij}\left\{\paa\left(\paa+2\pnn\right)
-\pbb\left(\pnn+2\paa\right)+2\ppaa+\ppnn\right\}
+\nonumber\\&&\frac{A^2}{N^2}\gamma_{ij}\left\{\daa\left(2\dnn-\daa\right)
-2\ddaa+\dbb\left(\dnn-2\daa\right)-\ddbb\right\}-\gamma_{ij}K.\nonumber\\
\end{eqnarray}
For the bulk dimension we get two non-zero components:

\begin{equation}\label{Gty}
G^{(5)}_{ty}=3\left\{\daa\pnn+\paa\dbb-\frac{\dot{A}'}{A}\right\},
\end{equation}
and:

\begin{equation}\label{Gyy}
G^{(5)}_{yy}=3\left\{\paa\left(\paa+\pnn\right)-
\frac{B^2}{N^2}\left(\daa\left(\daa-\pnn\right)+\ddaa\right)
-K\frac{B^2}{A^2}\right\}.
\end{equation}
Since $T^{(5)}_{ty}=0$ (and therefore $G^{(5)}_{ty}=0$), this means
that there is no flow of matter along the extra dimension. Therefore
we get the following results from Eq.~(\ref{Gty}):

\begin{equation}\label{bd}
\dbb=\frac{\dot{A}'}{A'}-\pnn\frac{\dot{A}}{A'}.
\end{equation}
In Ref.~\cite{Binetruy:1999hy} they have shown that if we define the
function:

\begin{equation}\label{bF}
F(t,y)\equiv\frac{(A'A)^2}{B^2}-\frac{(\dot{A}A)^2}{N^2}-KA^2,
\end{equation}
we can rewrite the Einstein equations in a simpler form. Using
Eqs.~(\ref{Gtt}),~(\ref{bd}) and (\ref{bF}) we can show that:

\begin{equation}\label{F2}
F'=\frac{2A'A^3}{3}\kappa^2T^t_t|_{bulk}.
\end{equation}
Using Eqs.~(\ref{Gyy}),~(\ref{bd}) and (\ref{bF}) we can show that:

\begin{equation}\label{F22}
\dot{F}=\frac{2\dot{A}A^3}{3}\kappa^2T^y_y|_{bulk}.
\end{equation}
Thus:

\begin{equation}\label{Fint}
F=\frac{2\kappa^2_5T^t_t|_{bulk}}{3}\int
A'A^3dy=-\frac{A^4\Lambda_5}{6}+{\mathcal C}.
\end{equation}
So if we now use the form of $F$ from Eq.~(\ref{bF}) we get:

\begin{equation}\label{fr}
\left(\frac{\da}{AN}\right)^2=\left(\frac{A'}{AB}\right)^2
-\frac{K}{A^2}+\frac{\Lambda_5}{6}+\frac{\mathcal{C}}{A^4}
\end{equation}
We have the freedom in the coordinate system to redefine the extra
dimension coordinate, so we set $B=1$. We can also define $t$ to be
the proper time on the brane so we can set $N(t,0)=1$, which implies
that:

\begin{equation}\label{fr2}
\left(\frac{\da}{A}\right)^2=\left(\frac{A'}{A}\right)^2-\frac{K}{A^2}
+\frac{\Lambda_5}{6}+\frac{\mathcal{C}}{A^4}
\end{equation}
We must now consider the $A'/A$ term, and we use the Israel junction
conditions:

\begin{equation}\label{K}
K_{\mu\nu}=-\frac{\kappa^2_5}{2}\left(T_{\mu\nu}-
\frac{1}{3}Tg_{\mu\nu}\right),
\end{equation}
where:

\begin{equation}\label{K2}
K_{\mu\nu}=-\frac{1}{2}\partial_yg_{\mu\nu}.
\end{equation}
Using this we can show that we can write the extrinsic curvature of
the brane as:

\begin{equation}\label{Kb}
K^{\mu}_{\nu}={\rm
diag}\left(\frac{N'}{N},\frac{A'}{A},\frac{A'}{A},\frac{A'}{A}\right).
\end{equation}
By Eq.~(\ref{K}):

\begin{equation}\label{Kr}
\left.\frac{A'}{A}\right|_{brane}=-\frac{\kappa^2_5}{6}(\rho+\lambda).
\end{equation}
By Eqs.~(\ref{Lb}) and (\ref{Tm4}) we can write the Friedmann
equation as:

\begin{equation}\label{RSF}
H^2=\frac{\kappa_4^2}{3}\rho\left(1+\frac{\rho}{2\lambda}\right)
-\frac{K}{a^2}+\frac{\Lambda_4}{3}+\frac{\mathcal{C}}{a^4},
\end{equation}
where $\mathcal{C}$ is the constant of integration from
Eq.~(\ref{Fint}) (it can also be obtained from the electric
(Coulomb) part of the bulk Weyl tensor~\cite{Ida:1999ui}). This term
is called the dark radiation term from the bulk due to its $a^{-4}$
behaviour. It can be shown that $\mathcal{C}$ is the mass of the
black hole in the Schwarzschild anti de Sitter bulk
\cite{Collins:2000yb,Barcelo:2000re}. The main differences between
this Friedmann equation and the GR result is that we have a $\rho^2$
term driving the expansion. When $\rho\gg\lambda$, i.e. at early
times (high energies), we see that $H\sim\rho$, unlike the GR case,
$H\sim\sqrt{\rho}$. Therefore the scale factor evolves like (taking
$\mathcal{C}=0$):

\begin{equation}\label{ERS}
a\propto t^{1/3(1+w)},
\end{equation}
rather than:

\begin{equation}\label{EGR}
a\propto t^{2/3(1+w)},
\end{equation}
as it does in the standard GR case. In the low energy limit
$\rho\ll\lambda$ (with $\mathcal{C}=0$ again) the Friedmann equation
gives us the same result as in the GR case.

The Raychaudhuri equation for the FRW brane follows from the
Friedmann equation (\ref{RSF}) and the conservation equation
(\ref{CE2}) as:

\begin{equation}\label{RSR}
\dot{H}=-\frac{\kappa^2_4}{2}(\rho+p)\left(1+\frac{\rho}{2\lambda}\right)
+\frac{K}{a^2}- 2\frac{\mathcal{C}}{a^4}.
\end{equation}
The RS form of the Friedmann and Raychaudhuri equations are going to
modify the predictions of primordial nucleosynthesis due to the
presence of the $\rho^2$ and dark radiation terms. In order for the
RS model to mimic the GR result at nucleosynthesis we require:

\begin{equation}\label{btnuc}
 \lambda>>\rho_{nuc}>>\frac{\mathcal{C}}{a^4}.
\end{equation}
We shall consider the nucleosynthesis constraint further in
Chapter~\ref{GBIGB}.


\section{Conclusions}

In this chapter we have looked at the field equations in the bulk
and how they are projected onto the brane for Minkowski and de
Sitter brane worlds. These equations could be much more complicated
if we had considered a scalar field to be present in the bulk. In
general a scalar field should be present in order to stabilise the
brane positions when there are two branes.

We have seen how the brane tensions are related on the two branes in
both the Minkowski and de Sitter cases. The brane tensions are
required to be fine tuned in order to obtain Minkowski or de Sitter
branes. We have also seen how the Hubble rate on the two de Sitter
branes are related.

We have looked at the Kaluza-Klein modes on Minkowski and de Sitter
branes in both the RS1 and RS2 models. The Kaluza-Klein modes are
decomposed into five different polarisations in three different
fields (gravi-scalar, gravi-vector and the graviton). We have found
the wave equation for bulk gravitational waves. We worked out the
spectrum of the KK modes in both the Minkowski and de Sitter brane
cases, the de Sitter case being a new result. In the RS1 case there
is a discrete spectrum, described by associated  Legendre functions,
beginning above a mass gap. The spectrum is altered due to the
curvature scale and the inter-brane separation. In the RS2 model,
with an infinite brane separation, we obtain a continuum of massive
KK modes, starting above the mass gap. The mass-gap is a function of
the Hubble rate on the brane.

We then looked at a more general and physically relevant solution,
that of the Friedmann-Robertson-Walker brane. We saw how the
Friedmann equations were obtained in the standard GR setup and how
we extend this to the brane world case. We saw that the brane setup
modifies the high energy Hubble rate but at low energies we obtain
the same result as in the GR case. In the next chapter we shall
consider more complicated brane models which can give us more
drastic modifications to GR.

%% file: CHGBIGBranes.tex
\chapter{The DGP, GB and GBIG Brane World Cosmologies}\label{GBIGB}

\section{Introduction}

So far we have been considering the Randall-Sundrum brane models in
which gravity appears 4D on the brane via the warping of the bulk
dimension. This warping means that to a brane observer gravity
appears 5D at high energy and 4D in the low energy regime. It is
also possible to have 5D effects at low
energy~\cite{Gregory:2000jc,Dvali:2000hr}. The DGP (Dvali, Gabadadze
and Porrati) model achieves this via a Einstein-Hilbert brane
action; this brane Ricci scalar can be interpreted as arising from a
quantum effect due to the interaction between the bulk gravitons and
the matter on the brane~\cite{Dvali:2000hr} (see
also~\cite{Collins:2000yb,Shtanov:2000vr}). The induced gravity term
on the brane dominates at higher energies, below a certain length
scale $r$, so that gravity becomes 4D at high energy. As the bulk is
no longer required to be warped, the DGP model lives in a Minkowski
bulk. When a Friedmann brane in used within the DGP model it is
possible to show that one of the solutions gives rise to late time
acceleration without the presence of a dark energy
field~\cite{Deffayet:2000uy}.

We can also construct brane models using the Gauss-Bonnet (GB)
higher order curvature terms in the bulk action (see,
e.g.,~\cite{Kim:2000pz,Cho:2001nf,Charmousis:2002rc,Nojiri:2002hz,Davis:2002gn,Gravanis:2002wy,
Lidsey:2003sj,Maeda:2003vq,Sami:2004xk,Tsujikawa:2004dm,Nojiri:2004bx,Rizzo:2004rq,Brax:2004np}).
These models modify gravity at high energy, unlike the DGP model.
The Gauss-Bonnet term may be interpreted as a low-energy String
Theory correction to the Einstein-Hilbert action. These models are
akin to the Randall-Sundrum ones, in the sense that it is the nature
of the bulk that gives rise to the 4D and 5D regimes on the brane,
unlike the DGP model.

We would like to have a model that has 5D effects at both high
energy and low energy but 4D gravity in between. This is because we
would like a model that can modify the early universe and explain
the acceleration of the universe that we are currently experiencing.
We require 4D gravity between these two limits in order for
nucleosynthesis and other constraints to be obeyed.

In this chapter we shall start by taking a brief look at the DGP and
GB models and see how they modify the cosmology on the brane. We
shall then consider the effect of combining these two.

\section{DGP Branes}

In 2000 Dvali, Gabadadze and Porrati (DGP)~\cite{Dvali:2000hr} put
forward a brane model that has low energy (infrared or IR)
modifications through an induced gravity term. The induced gravity
term (4D Ricci scalar term in the brane action) can be motivated by
possible quantum effects of the interaction between matter on the
brane and the bulk gravity. (String theories with a ghost free GB
term in the bulk give rise to induced gravity terms on the
brane~\cite{Mavromatos:2005yh}.)

At early times, the 4D term dominates and General Relativity is
recovered (in the background -- note that the perturbations are not
General
Relativistic~\cite{Tanaka:2003zb,Deffayet:2004ru,Koyama:2005br}).
Gravity on the brane is 4D until the ``cross-over'' scale $r$ is
reached and we enter a 5D regime. At scales greater than $r$ gravity
``leaks'' off the brane (the 5D Ricci term in Eq.~(\ref{AcDGP})
begins to dominate over the 4D Ricci term), and appears to observers
on the brane to be 5D. As we have 4D gravity on the brane from the
induced gravity we do not require the warping of the bulk space in
order to recover 4D gravity. So the DGP model lives in a infinite
Minkowski bulk. The DGP model has two branches, one of which is very
interesting for cosmology as it ends with a phase of
``self-acceleration''~\cite{Deffayet:2000uy}. This is useful as we
observe the universe to be in a period of acceleration which is
usually explained via a dark energy field. The DGP model explains
this with modified gravity. The DGP models are in some sense
``unbalanced", since they do not include ultra-violet modifications
to cosmological dynamics. We would like a brane world model that
modifies gravity at early times as this is where string and quantum
effects must eventually dominate.

With induced gravity on the brane we have a gravitational action of
the form~\cite{Deffayet:2000uy}:

\begin{eqnarray}\label{AcDGP}
S_{grav}&=&\frac{1}{2\kappa_5^2}\int d^5 x\sqrt{- g^{(5)}}\left[
R^{(5)}-2\Lambda_5\right]\nonumber\\&+&\frac{r}{\kappa^2_5}\int_{y=0}d^4x\sqrt{-
g}\left[R-2\kappa^2_4\lambda\right],
\end{eqnarray}
where $r$ is the cross-over scale and is defined as:

\begin{equation}\label{r}
r=\frac{\kappa^2_5}{2\kappa^2_4}=\frac{M^2_4}{2M^3_5}.
\end{equation}
The sign of the induced gravity on the brane must be the same as the
5D part. We have an effective 4D gravitational constant which can be
written as $\kappa^2_4=\kappa^2_5/2r$.

The bulk field equations are of the same form as in the RS model.
The 4D energy-momentum tensor is modified, in comparison to the RS
model, as it includes the contribution from the induced gravity term
in the action. By using the same assumptions and conditions as in
the RS case (a perfect fluid undergoing isotropic and homogenous
expansion) we obtain the induced gravity Friedmann equation to
be~\cite{Deffayet:2000uy}:

\begin{equation}\label{DGPF}
\epsilon\sqrt{H^2-\frac{\mathcal{C}}{a^4}-\frac{\Lambda_5}{6}+\frac{K}{a^2}}=r
\left(H^2+\frac{K}{a^2}\right)-\frac{\kappa^2_5}{6}(\rho+\lambda),
\end{equation}
where $\epsilon=\pm1$ and $\mathcal{C}$ is the bulk black hole mass
(the DGP limit has a Minkowski bulk $\Lambda_5=0$ with
$\mathcal{C}=0$). If we consider a Minkowski bulk ($\Lambda_5=0$,
$\mathcal{C}=0$) with $\lambda=0$ on a flat brane ($K=0$) then we
have:

\begin{equation}\label{DGPFL2}
H^2=\pm\frac{H}{r}+\frac{\kappa^2_4}{3}\rho.
\end{equation}

The two different solutions corresponding to the two different
values of $\epsilon$ correspond to two different embeddings of the
brane within the bulk. Both branches have a 4D limit at high
energies:
\begin{equation}\label{bbdgp}
\text{DGP($\pm$):}~~H\gg r^{-1}~\Rightarrow ~
H^2=\frac{\kappa^2_4}{3} \rho\,,
\end{equation}
while at low energies both have modified 5D limits:
\begin{eqnarray}
\text{DGP(+):} && \rho \to 0 ~\Rightarrow ~ H\to {1 \over r}\,,\\
\text{DGP(--):} && \rho \to 0 ~\Rightarrow ~ H=0\,.
\end{eqnarray}
DGP(--) has a non-standard (and non-accelerating) late universe. The
self-accelerating DGP(+) branch is of most interest for cosmology.

\begin{figure}
\begin{center}
\includegraphics[height=3in,width=2.75in,angle=270]{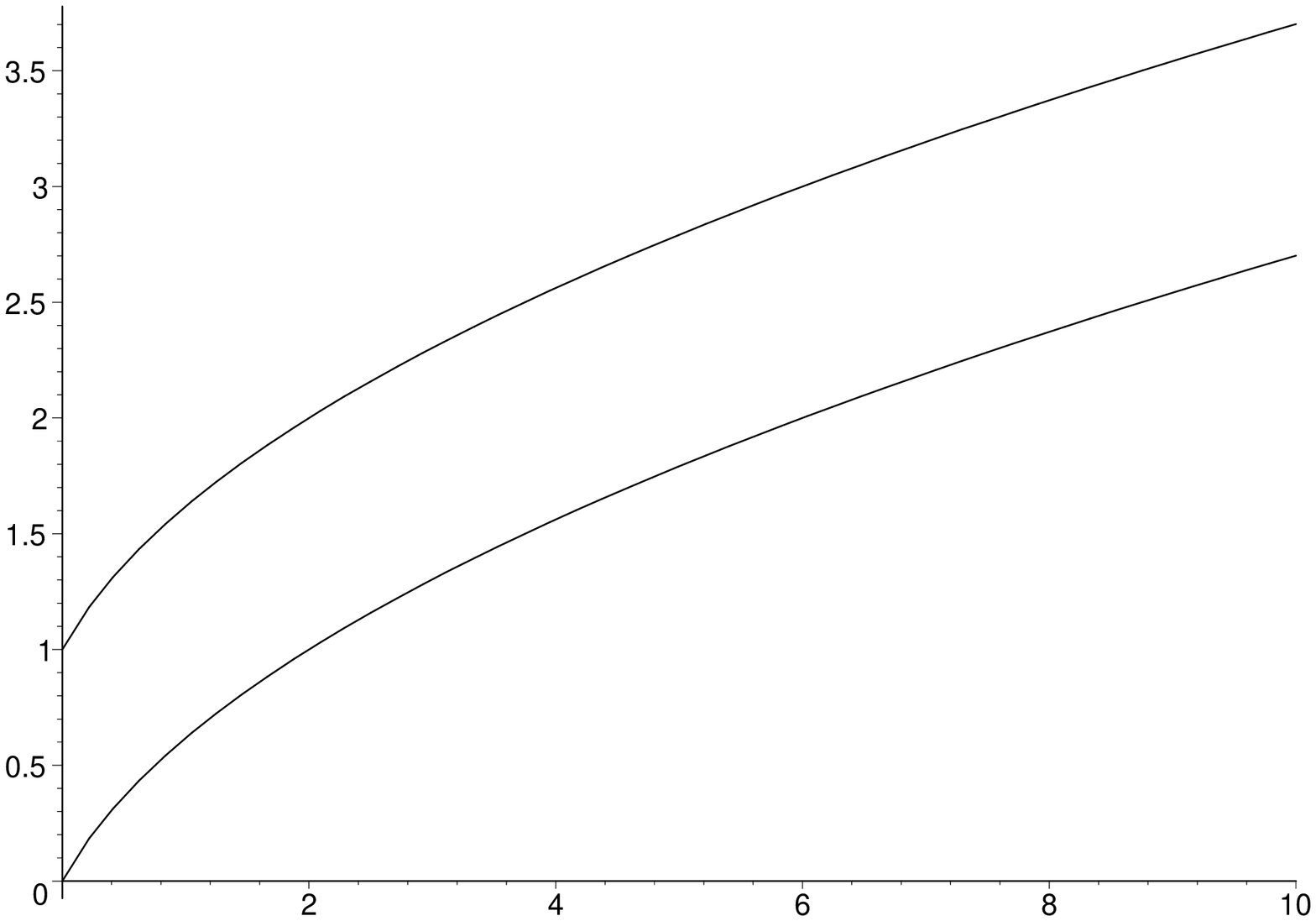}
\rput(-3,-1.1){DGP (+)} \rput(-3,-3.7){DGP
(-)}\rput(-3.9,-6.4){\large $\mu$}\rput(-7.4,-3.6){\large
$h$}\caption{Solutions of the DGP Friedmann equation ($h$ vs $\mu$)
with $\mathcal{C}=\Lambda_5=\lambda=0$. (Brane proper time $\tau$
flows from right to left, with $\tau=\infty$ at $\mu=0$.)}
\label{DGPSP}
\end{center}
\end{figure}

In dimensionless variables:

\begin{equation}\label{DGPDV}
h=Hr,\,\,\,\mu=\frac{r\kappa^2_5}{6}\rho,\,\,\,\tau=\frac{t}{r},
\end{equation}
we can write the Friedmann equation (in the case
$\mathcal{C}=\Lambda_5=\lambda=0$) as:

\begin{equation}\label{DGPFDV}
h^2=\pm h+\mu.
\end{equation}
The solutions for $h$ can be written in terms of $\mu$ as:

\begin{equation}\label{DGPFDV2}
h=\frac{\sqrt{1+4\mu}}{2}\pm \frac{1}{2}.
\end{equation}
These solutions are shown in Fig.~\ref{DGPSP}

In the $+$ branch we have late time self-acceleration ($H=1/r$) i.e
we end in a vacuum de Sitter state. In the $-$ branch we have $H=0$
so we end up with a Minkowski state. The DGP(+) model gives us late
time acceleration without dark energy with early universe dynamics
that agrees with GR so we do not need to modify baryogenesis and
other early universe events. The DGP late time acceleration has been
tested against the supernova luminosity and baryon acoustic
oscillations~\cite{Alcaniz:2004kq,Fairbairn:2005ue,Maartens:2006yt}.
But the DGP is ``unbalanced'' as we would expect high energy
(ultraviolet or UV) modifications as well. If we generalise the DGP
model by introducing a negative bulk cosmological constant and a
brane tension, so that the bulk is anti de Sitter, we do not get the
desired results. It has been
shown~\cite{Kiritsis:2002ca,Tanaka:2003zb,Papantonopoulos:2004bm}
that if we have an induced gravity brane in a warped bulk we either
have 4D at early and late times with a period of 5D gravity in the
middle or we have 4D gravity at all times. So we require another
mechanism to modify early times in a induced gravity brane model.

\section{Gauss-Bonnet Brane Worlds}

The Gauss-Bonnet term added to the Einstein-Hilbert term, gives the
most general action in 5D with $2^{nd}$ order field equations, as
shown by Lovelock~\cite{Lovelock:1971yv}. In four dimensions the
Gauss-Bonnet term is a topological invariant for compact manifolds
without a boundary. Therefore Lovelock type gravity in 4D reduces to
standard GR. The Gauss-Bonnet terms have also been shown to be a low
order string correction to the action (see,
e.g.,~\cite{Kim:2000pz,Cho:2001nf,Charmousis:2002rc,Nojiri:2002hz,Davis:2002gn,Gravanis:2002wy,
Lidsey:2003sj,Maeda:2003vq,Sami:2004xk,Tsujikawa:2004dm,Nojiri:2004bx,Rizzo:2004rq}).
In a 5D theory the GB terms are only present in the bulk. The bulk
action is~\cite{Charmousis:2002rc}:

\begin{eqnarray}\label{AcGB}
S_{grav}&=&\frac{1}{2\kappa_5^2}\int d^5 x\sqrt{-
g^{(5)}}\left[ R^{(5)}-2\Lambda_5\right.\nonumber\\
&+&\left.\alpha\left( R^{(5)\,2}-4R^{(5)}_{ab}R^{(5)\,ab}+
R^{(5)}_{abcd}R^{(5)\,abcd}\right)\right],
\end{eqnarray}
where $\alpha$ is the Gauss-Bonnet coupling. In a classical
Gauss-Bonnet theory $\alpha$ can be of either sign. It has been
shown in Ref.~\cite{Davis:2004yf} that for Gauss-Bonnet brane worlds
a negative value of $\alpha$ gives rise to antigravity or tachyon
modes on the brane. However if a bulk scalar field is present a
negative $\alpha$ can be allowed without these effects occurring.

The field equations are much more complicated with the Gauss-Bonnet
term in the action. The Friedmann equation that we obtain is of the
form~\cite{Davis:2002gn}:

\begin{equation}\label{FredGB}
H^2=\frac{C_++C_--2}{8\alpha}-\frac{K}{a^2},
\end{equation}
where:

\begin{equation}\label{GBC}
C_{\pm}=\left(\sqrt{\left(1+\frac{4}{3}\alpha\Lambda_5+
8\alpha\frac{\mathcal{C}}{a^4}\right)^{3/2}+\frac{\alpha\kappa^4_5(\rho+\lambda)^2}{2}}
\pm\kappa^2_5(\rho+\lambda)\sqrt{\frac{\alpha}{2}}\right)^{2/3}.
\end{equation}
The bulk cosmological constant $\Lambda_5$ has a modified relation
to the curvature scale due to the GB gravity~\cite{Dufaux:2004qs}:

\begin{equation}\label{L5GB}
\Lambda_5=-\frac{6}{\ell^2}+\frac{12\alpha}{\ell^4}.
\end{equation}
The first term is the standard RS result while the second is the
Gauss-Bonnet contribution. This also gives us a constraint on
$\alpha$ as Eq.~(\ref{L5GB}) implies:

\begin{equation}\label{L5GBq}
\frac{1}{\ell^2}=\frac{1}{4\alpha}\left[1-\sqrt{1+
\frac{4}{3}\alpha\Lambda_5}\right].
\end{equation}
We have only one root here as this is the only one that gives the
correct RS limit. The plus root could be written as:

\begin{equation}\label{L5GBq2}
\frac{4\alpha}{\ell^2}=1+\sqrt{1+\frac{4}{3}\alpha\Lambda_5}.
\end{equation}
The RS limit ($\alpha\rightarrow0$) is not consistent with this
solution.

From Eq.~(\ref{L5GBq}) we see that:

\begin{equation}\label{alCon}
\alpha\leq\frac{\ell^2}{4}.
\end{equation}
As $\sqrt{1+ \frac{4}{3}\alpha\Lambda_5}>0$, this also ensures that
$\Lambda_5<0$.

If we consider the case $\mathcal{C}=0=\Lambda_5$ and
$\rho\gg\lambda$ for a flat brane ($K=0$) we see, from
Eq.~(\ref{FredGB}), that $H\sim\rho^{1/3}$. The scale factor behaves
as:

\begin{equation}\label{EGB}
a\propto t^{1/(1+w)}.
\end{equation}
So the Gauss-Bonnet bulk gravity effects the early universe. At late
times $\mathcal{C}/a^4,~\rho\rightarrow 0$ with
$\lambda=0=\Lambda_5$ we can show that we end with $H=0$ so again we
end with a Minkowski brane.

This GB Friedmann equation has no 4D limit:
\begin{eqnarray}
\text{GB high energy:} && H\gg \alpha^{-1/2} ~\Rightarrow ~
H^2\propto \rho^{2/3} \,,\label{bbgb}\\
\text{GB low energy:} && H\ll \alpha^{-1/2} ~\Rightarrow ~
H^2\propto \rho^2\,.
\end{eqnarray}

\begin{figure}
\begin{center}
\includegraphics[height=3in,width=2.75in,angle=270]{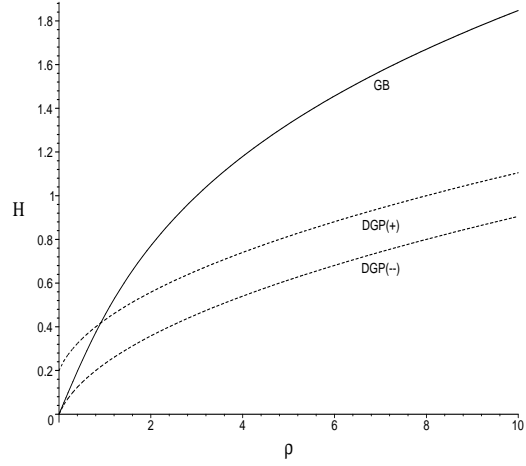}
\caption{DGP and GB solutions of the Friedmann equation ($H$ vs
$\rho$) for a Minkowski bulk. (Brane proper time $t$ flows from
right to left, with $t=\infty$ at $\rho=0$.)} \label{GBDGP}
\end{center}
\end{figure}

The Friedmann equations for pure DGP and pure GB models with a
Minkowski bulk are compared in Fig.~\ref{GBDGP}. The GB Friedmann
equation is a cubic, so has three roots. We can show that for
$\rho>0$ there is only one real root. This is shown in
Fig.~\ref{GBDGP}. When $\rho=0$ there is the repeated root $H=0$.

\section{Gauss-Bonnet-Induced Gravity (GBIG) Branes}

In this section and the next chapter we shall consider what happens
when we combine the induced gravity terms of the DGP model with a
Gauss-Bonnet bulk. We will start by presenting the model's governing
equations in full generalisation. We shall then consider the
modifications in relation to the pure DGP model. This means we shall
only consider models with zero brane tension in a Minkowski bulk. In
the next chapter we shall extend this to the non-zero brane tension
in Anti-de Sitter bulk cases. We will see that even in the simplest
case we get some striking new effects such as a finite density big
bang. Some of the work in this section was first presented
in~\cite{Brown:2005ug}.

\subsection{Field Equations}
The generalised gravitational action contains the Gauss-Bonnet (GB)
term in the bulk and the induced gravity (IG) term on the brane:
\begin{eqnarray}\label{AcGBIG}
S_{\rm grav}&=&\frac{1}{2\kappa_5^2}\int d^5 x\sqrt{-
g^{(5)}}\left\{ R^{(5)} -2\Lambda_5 +\alpha\left[ R^{(5)2}-4
R^{(5)}_{ab}R^{(5)ab}+ R^{(5)}_{abcd}R^{(5)abcd}\right]\right\}
\nonumber\\&+&\frac{r}{\kappa^2_5} \int_{\rm brane}d^4x\sqrt{-
g}\,\left[R-\frac{\kappa^2_5}{r}\lambda\right]\,,
\end{eqnarray}
where $\alpha\, (\geq0)$ is the GB coupling constant and $r\, (\geq
0)$\footnote{Note: the cross-over scale used is the one in
Ref.~\cite{Dvali:2000hr} and is half that used in
Refs.~\cite{Brown:2005ug,Brown:2006mh}.} is the IG cross-over scale
and $\lambda$ is the brane tension. As in the previous models we
assume $Z_2$ symmetry about the brane.

The standard energy conservation equation holds on the brane,
\begin{equation}\label{ec}
\dot \rho+3H(1+w)\rho=0\,,~w=p/\rho\,.
\end{equation}
The modified Friedmann equation was found in the most general case
(where the bulk contains a black hole and a cosmological constant,
and the brane has tension) in Ref.~\cite{Kofinas:2003rz}:

\begin{eqnarray}\label{Ff}
\left[1+\frac{8}{3} \alpha\left(H^2+\frac{\Phi}{2}+\frac{K}{a^2}
\right)\right]^2 \left(H^2-\Phi+\frac{K}{a^2} \right)=\left[rH^2
+r\frac{K}{a^2} -\frac{\kappa^2_5}{6}(\rho+\lambda)
\right]^2\!,\nonumber\\
\end{eqnarray}
where $K$ is the brane curvature and $\Phi$ is determined by:
\begin{equation}\label{Phif}
\Phi+2\alpha\Phi^2=\frac{\Lambda_5}{6}+\frac{\mathcal{C}}{a^4}\,,
\end{equation}
where $\mathcal{C}$ is the bulk black hole mass. In the correct
limits this Friedmann equation reduces to forms equivalent to those
previously seen.

In the rest of this chapter we shall only consider a Minkowski bulk
($\Lambda_5=0$) without a bulk black hole ($\mathcal{C}=0$), so that
$\Phi$ is a solution of:
\begin{equation}\label{Phi}
\Phi+2\alpha\Phi^2=0\,.
\end{equation}
Equation~(\ref{Phi}) has solutions $\Phi=0, -1/2\alpha$, but here we
only consider $\Phi=0$, since the second solution has no IG limit
and thus does not include the DGP model. We shall consider
$\Phi=-1/2\alpha$ in Chapter~\ref{GGBIG}. In this bulk we shall
consider a spatially flat brane ($K=0$) without tension
($\lambda=0$). Therefore the modified Friedmann equation is given
by:

\begin{eqnarray}\label{F}
\left(1+\frac{8}{3} \alpha H^2\right)^2 H^2 =\left(rH^2
-\frac{\kappa^2_5}{6}\rho \right)^2\!.
\end{eqnarray}

\subsection{DGP brane with GB bulk gravity: combining UV and IR
modifications of GR}\label{sec}

By defining the dimensionless variables:
\begin{eqnarray}\label{DV}
\gamma&=&\frac{8\alpha}{3r^2}\,,~ h=Hr\,,~ \mu
=\frac{r\kappa^2_5}{6}\rho\,,~\tau={t \over r}\,,
\end{eqnarray}
the GBIG Friedmann equation becomes:
\begin{equation}\label{DGPF2}
\left(\gamma h^2+1\right)^2h^2=\left(h^2-\mu\right)^2,
\end{equation}
while the conservation equation becomes:
\begin{equation}\label{ec2}
\mu'+3h(1+w)\mu=0\,,
\end{equation}
where a dash denotes $d/d\tau$, and $h=a'/a$. In defining the
variable $\gamma$ as above, we have both the IG and GB contributions
within one parameter. This reduces the dimensionality of the
solution state space.

Combining Eqs.~(\ref{DGPF2}) and (\ref{ec2}), we find the modified
Raychaudhuri equation:
\begin{equation}\label{Ray}
{h}'=\frac{3\mu(1+w)(h^2-\mu)}{(\gamma h^2+1)(3\gamma
h^2+1)-2(h^2-\mu)}\,.
\end{equation}
The acceleration $a''/a=h'+h^2$ is then given by:
\begin{equation}\label{acc}
{a'' \over a}={h^2(\gamma h^2+1)(3\gamma h^2+1)-(h^2-\mu)
[2h^2-3(1+w)\mu] \over (\gamma h^2+1)(3\gamma h^2+1)-2(h^2-\mu)}.
\end{equation}
These shall be of use later.

\begin{figure}
\includegraphics[height=3in,width=2.75in,angle=270]{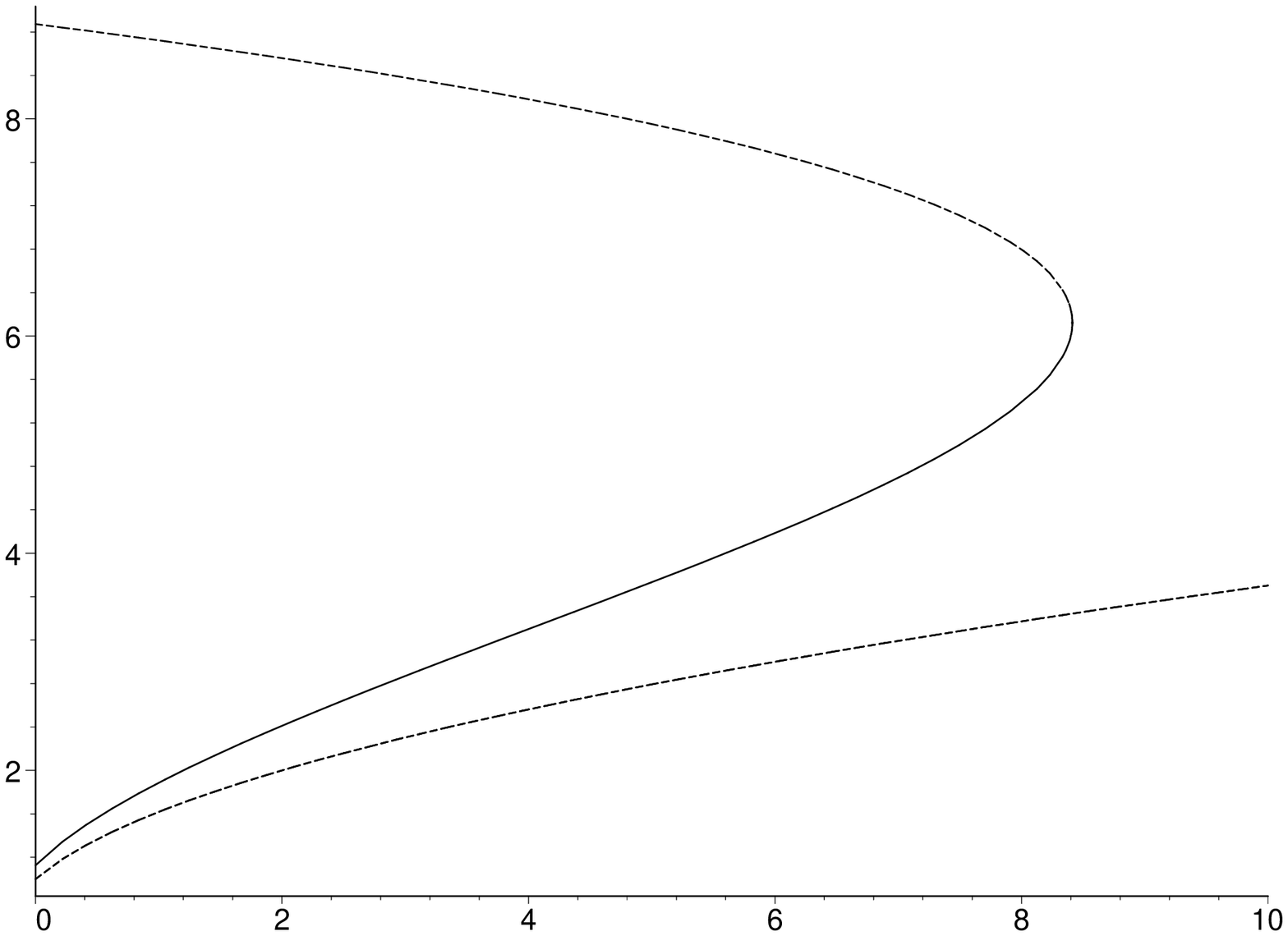}
\includegraphics[height=3in,width=2.75in,angle=270]{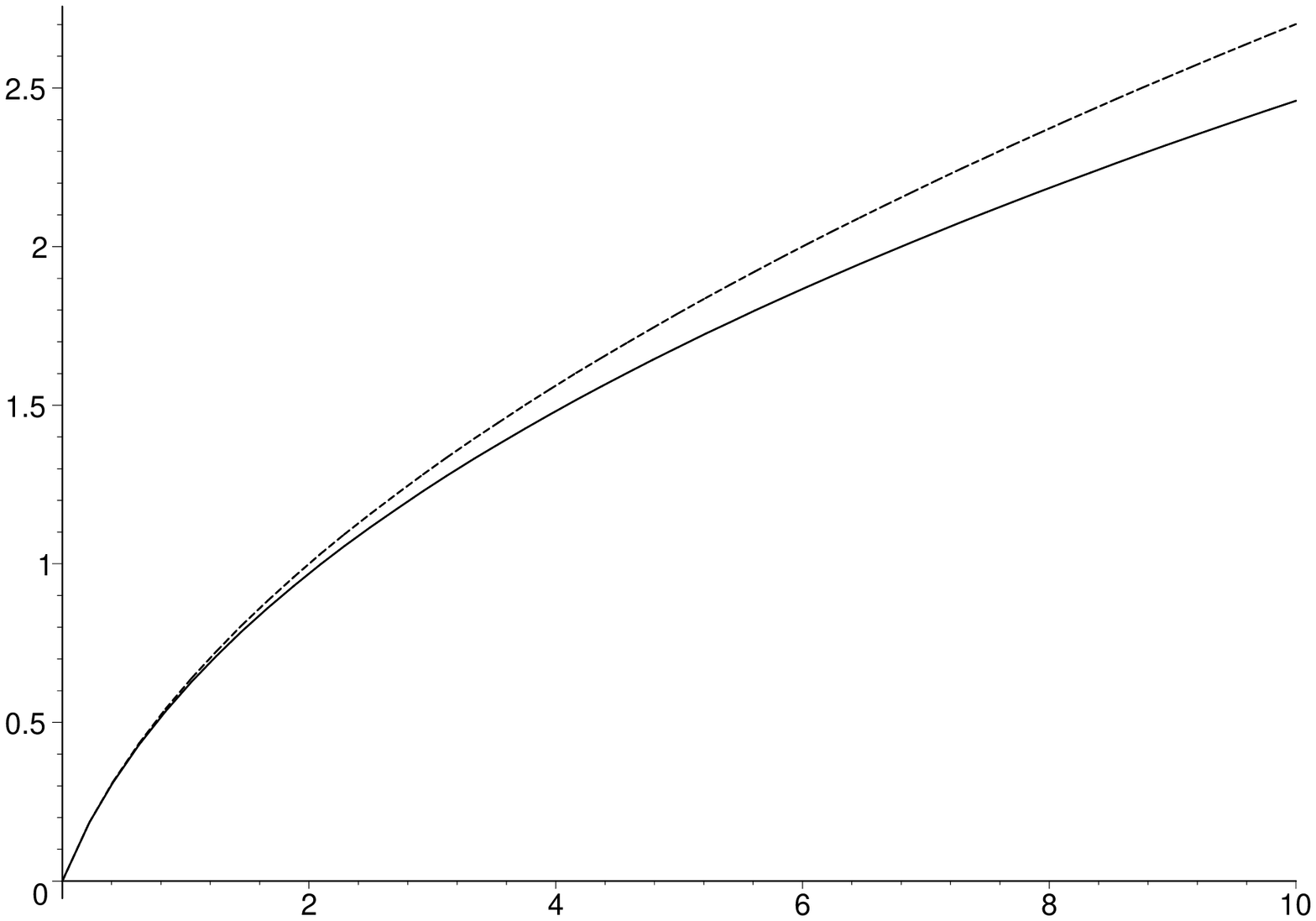}
\rput(0.6,3.4){\large $h$}\rput(3.8,0.4){\large
$\mu$}\rput(5,5.6){\small GBIG 2}\rput(5,2.8){\small GBIG
1}\rput(5,2){\small DGP (+)} \rput(8.2,3.4){\large
$h$}\rput(11.5,0.4){\large $\mu$}\rput(12,5.4){\small DGP
(--)}\rput(12,3.8){\small GBIG 3}
 \caption{Solutions of the GBIG Friedmann
equation ($h(\mu)$) with positive $\gamma$ ($\gamma=0.1$). On the
left is the DGP (+) model and its Gauss-Bonnet corrections, GBIG 1
and GBIG 2. On the right is the DGP (--) model and its GB
correction, GBIG 3. The curves are independent of the equation of
state $w$. Brane proper time $\tau$ flows from right to left, with
$\tau=\infty$ at $\mu=0$.} \label{GBIGDGP}
\end{figure}

The GB correction, via a non-zero value of $\gamma$, introduces
significant complexity to the Friedmann equation, which becomes
cubic in $h^2$, as opposed to the quadratic DGP~($\pm$) case,
$\gamma= 0$, for which:
\begin{equation}\label{dd}
h^2=\frac{1}{2}\left\{1+2\mu\pm\sqrt{1+4\mu}\right\}\,,
\end{equation}
as seen in Eq.~(\ref{DGPFDV2}). This additional complexity has a
dramatic effect on the dynamics of the DGP (+) model, as shown in
Fig.~\ref{GBIGDGP}. The contribution of GB gravity at early times
removes the infinite density big bang, and the universe starts at
finite maximum density and finite pressure (but, as we show below,
with infinite curvature). Furthermore, there are two such solutions,
each with late-time self-acceleration, marked GBIG 1 and 2 on the
plots. Since GBIG 2 is accelerating throughout its evolution
(actually super-inflating, $h'>0$), the physically relevant
self-accelerating solution is GBIG 1.

\begin{figure}
\includegraphics[height=3in,width=2.75in,angle=270]{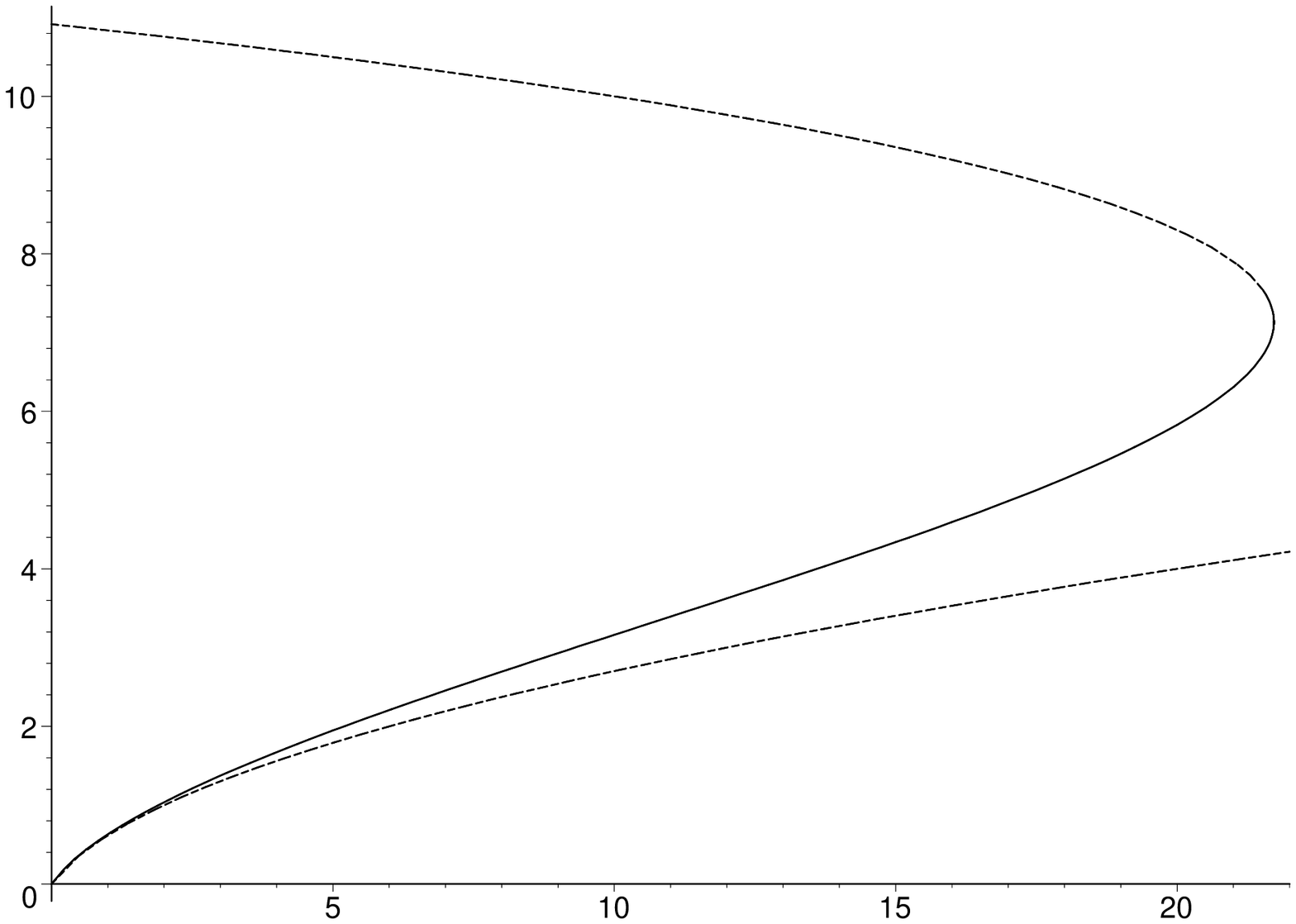}
\includegraphics[height=3in,width=2.75in,angle=270]{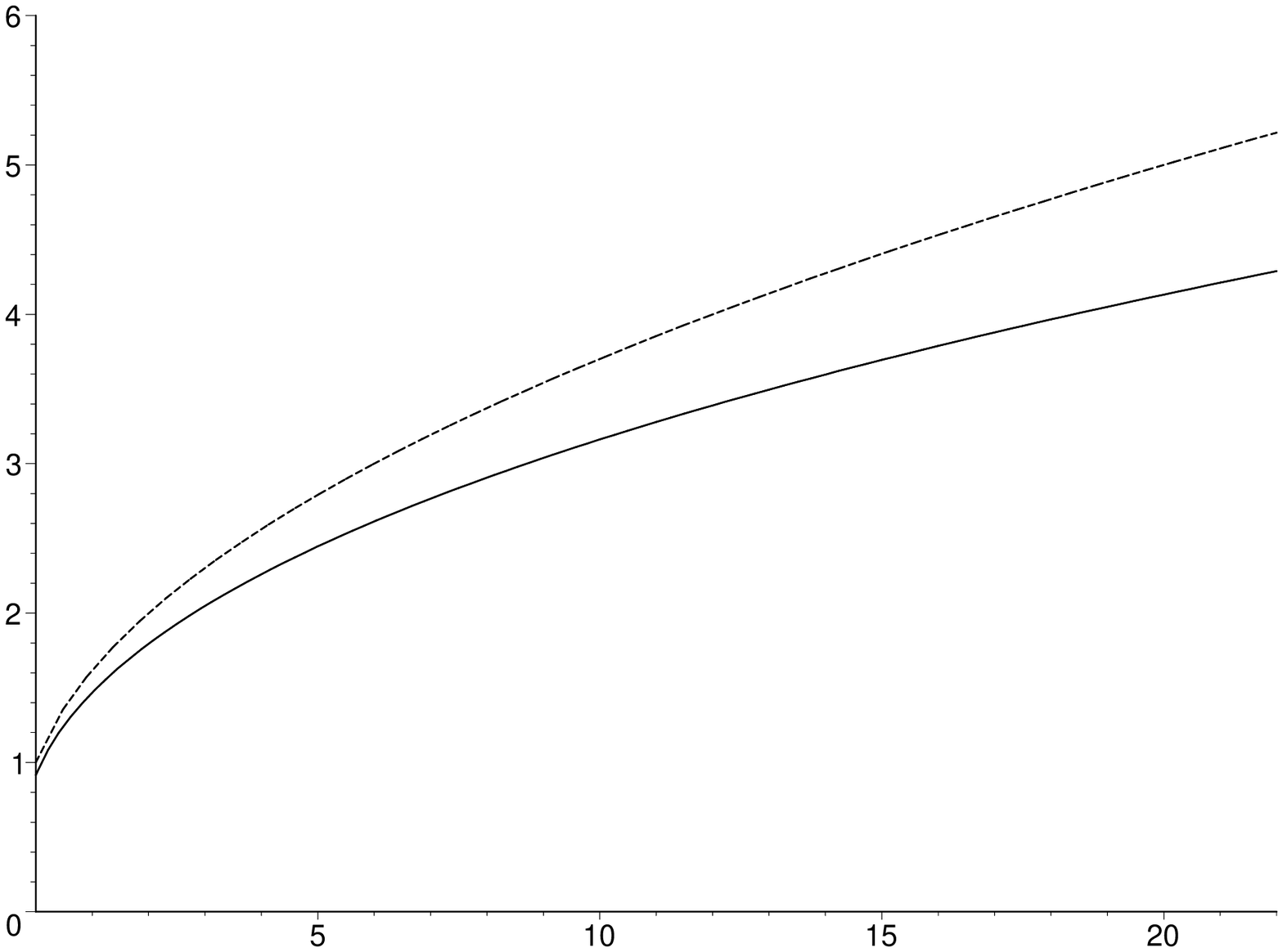}
\rput(0.5,3.2){\large $h$}\rput(3.8,0.4){\large
$\mu$}\rput(5,2){\small DGP (--)} \rput(8.2,3.2){\large
$h$}\rput(11.5,0.4){\large $\mu$}\rput(12,5.4){\small DGP (+)}
\caption{Solutions of the GBIG Friedmann equation ($h(\mu)$) with
negative $\gamma$ ($\gamma=-0.1$). On the left is the DGP (--) model
and its Gauss-Bonnet corrections, GBIG 1 and GBIG 2. On the right is
the DGP (+) model and its GB correction, GBIG 3. The curves are
independent of the equation of state $w$. Brane proper time $\tau$
flows from right to left, with $\tau=\infty$ at $\mu=0$.}
\label{GBIGDGP2}
\end{figure}

In Ref.~\cite{Davis:2004yf} it is shown that a negative value of
$\alpha$ leads to antigravity or tachyon modes on the brane. A
negative value of $\alpha$ can give regular solutions if a bulk
scalar field is present. If we allow $\gamma<0$ the effect of the GB
terms is to produce a finite density big bang on the DGP (--)
branch, Fig.~\ref{GBIGDGP2}. The DGP (--) branch is wrapped back on
itself instead of the DGP (+) branch as in the $\gamma>0$ case. The
DGP (+) branch is now modified in a similar manner as the DGP (--)
was in the previous case. The consequence of this is that the finite
density big bang model no longer self-accelerates. We could get
acceleration at late times by including positive brane tension as we
shall see later. In the rest of this chapter we shall confine
ourselves to $\gamma>0$.

The cubic in $h^2$, Eq.~(\ref{DGPF2}), has three real roots when
$0<\gamma<1/4$ (see below). Two of these roots correspond to GBIG
1 and GBIG 2, which are modifications of the DGP (+) model. The
third root GBIG 3 is a modification of the DGP (--) model, as
illustrated in Fig.~\ref{GBIGDGP}. Note that the curves in these
figures are independent of the equation of state $w$ of the matter
content of the universe -- $w$ will determine the time evolution
of the universe along the curves, via the conservation
equation~(\ref{ec2}).

The plots show that GBIG 3 starts with a hot big bang,
$\rho=\infty$, in common with the DGP ($\pm$) and GB models in
Fig.~\ref{GBDGP}. By contrast, GBIG 1 and GBIG 2 have no big bang,
since the density is bounded above:
\begin{equation}
\mu \leq \mu_{\rm i}\,,
\end{equation}
where $\mu_{\rm i}$ (which is positive only for $\gamma<1/4$), is
found below, in Eq.~(\ref{rhomax}).

The finite-density beginning was pointed out in
Ref.~\cite{Kofinas:2003rz}, where the cubic for the general case
(i.e., with brane tension, bulk cosmological constant and bulk black
hole) was qualitatively analysed. The analysis shows that one
solution, GBIG 3, is not bounded, which was not noticed in
Ref.~\cite{Kofinas:2003rz}. The numerical plots of the Friedmann
equation in Fig.~\ref{GBIGDGP} are crucial to a proper understanding
of the algebraic analysis of the cubic roots.

A detailed analysis~\cite{Kofinas:2003rz,Marium} of the cubic
equation~(\ref{DGPF2}) confirms the numerical results, and shows
that (for $\mu>0$):
\begin{eqnarray}
0<\gamma<{1\over 4} &:& \mbox{3 real roots, GBIG 1--3,}
\label{class}
\\ \gamma\geq {1\over 4} &:& \mbox{1 real root,~ GBIG 3.}
\end{eqnarray}
The real roots are given as follows:\\

\noindent$\bullet$ For $0<\gamma<1/4$: the roots GBIG 1--2 are
\begin{equation}\label{roots1}
\gamma^2 h^2= {1-2\gamma \over 3}+ 2\sqrt{-Q}
\cos\left(\theta+{n\pi \over 3}\right) ~\mbox{for}~\mu \leq
\mu_{\rm i}\,,
\end{equation}
where $n=4$ for GBIG 1, $n=2$ for GBIG 2, and the root GBIG 3 is
\begin{equation}\label{roots2}
\gamma^2 h^2= {1-2\gamma \over 3}+\left\{ \begin{array}{ll}
2\sqrt{-Q} \cos\theta & \mbox{for}~\mu \leq \mu_{\rm i},
\\ \\ S_++S_- & \mbox{for}~\mu\geq \mu_{\rm i}\,.
\end{array}\right.
\end{equation}
$\bullet$ For $\gamma\geq 1/4$: the only real root GBIG 3 is
\begin{equation}\label{roots3}
\gamma^2 h^2= {1-2\gamma \over 3}+ S_++S_- \,.
\end{equation}\\

In the above, $S_\pm, Q,R,\theta$ are defined by
\begin{eqnarray}
&&S_\pm = \left[R\pm \sqrt{R^2+Q^3}\right]^{1/3},\\\nonumber\\
&& Q = {\gamma^2 \over 3}(1+2\mu)-{1\over 9}(2\gamma-1)^2\,,\\\nonumber\\
&&R = \frac{\gamma^4\mu^2}{2}+ {\gamma^2 \over
6}(2\gamma-1)(1+2\mu) -{(2\gamma-1)^3\over 27}\!, \label{R}
\\\nonumber\\ && \cos 3\theta = R/ \sqrt{-Q^{3}}\,. \label{theta}
\end{eqnarray}

The GBIG 1 and GBIG 2 solutions agree with those in
Ref.~\cite{Kofinas:2003rz} (where the roots are given in the fully
general case, with brane tension and a bulk black hole and
cosmological constant).

The explicit form of the solutions can be used to confirm the
features in Figs.~\ref{GBIGDGP} and \ref{GBIGDGP2}.
Equations~(\ref{roots2})--(\ref{theta}) show that GBIG 3 starts with
a big bang, $h,\mu\to\infty$, with $h^2 \sim \mu^{2/3}$ near the big
bang. This is the same as the high-energy behaviour of the pure GB
model, as shown by Eq.~(\ref{bbgb}) -- the GB effect dominates at
high energies in GBIG 3. This is not the case for GBIG 1--2, where
the high energy behaviour is completely different from the pure GB
model (and from the DGP(+) model).

The maximum density feature of GBIG 1--2 is more easily confirmed by
analysing the turning points of $\mu$ as a function of $h^2$. The
GBIG Friedmann equation~(\ref{DGPF2}) gives
\begin{equation}\label{in}
{d\mu \over d(h^2)}={2(h^2-\mu)-(\gamma h^2+1)(3\gamma h^2+1)\over
2(h^2-\mu)}\,.
\end{equation}
Substituting $d\mu/d(h^2)=0$ into Eq.~(\ref{DGPF2}), we find that
\begin{eqnarray}
h_{\rm i}&=&\frac{1\pm\sqrt{1-3\gamma}}{3\gamma}\,,
\\
\mu_{\rm i} &=&
\frac{2-9\gamma\pm2\left(1-3\gamma\right)^{3/2}}{27\gamma^2}\,.
\end{eqnarray}
The second equation shows that positive maximum density only
arises for the upper sign and with $\gamma<1/4$, in agreement with
the cubic analysis.
\begin{equation}\label{b}
\infty>\mu_{\rm i}>0~~\Rightarrow~~0<\gamma<\frac{1}{4}.
\end{equation}
Thus the initial Hubble rate and density for GBIG 1--2 are
\begin{eqnarray}
h_{\rm i}&=&\frac{1+\sqrt{1-3\gamma}}{3\gamma}\,, \label{Hrhomax}
\\
\mu_{\rm i} &=&
\frac{2-9\gamma+2\left(1-3\gamma\right)^{3/2}}{27\gamma^2}\,.
\label{rhomax}
\end{eqnarray}
If $\gamma=0$, then GBIG 1--2 reduce to DGP(+), and $h_{\rm
i}=\mu_{\rm i}=\infty$. Note that
\begin{equation}
h_{\rm i}\geq2\,\Leftrightarrow H_{\rm i}\geq\frac{2}{r}.
\end{equation}
The case $\gamma=1/4, \mu_{\rm_i}=0$ corresponds to a vacuum brane
with de Sitter expansion, with $h=h_{\rm i}=2$, generalizing the
DGP(+) vacuum de Sitter solution~\cite{Deffayet:2000uy}.

The late-time asymptotic value of the expansion rate, as $\mu\to
0$, is
\begin{equation}\label{GBIGhinfin}
h_\infty=\frac{1\pm\sqrt{1-4\gamma}}{2\gamma},
\end{equation}
where the minus sign corresponds to GBIG 1 and the plus sign to
GBIG 2. In the limit $\gamma\to 0$, GBIG 1 recovers the DGP(+)
case, $h_\infty=1$, while for GBIG 2, $h_\infty\to\infty$; the
parabolic GBIG 1--2 curve in Fig.~\ref{GBIGDGP} ``unwraps" and
transforms into the DGP(+) curve. Equations~(\ref{class}) and
(\ref{GBIGhinfin}) show that
\begin{equation}\label{cross}
1\leq h_\infty <2 ~~\mbox{for GBIG 1,}
\end{equation}
while:
\begin{equation}\label{cross2}
2<h_\infty<\infty ~~\mbox{for GBIG 2.}
\end{equation}

\begin{figure}
\begin{center}
\includegraphics[height=3in,width=2.75in,angle=270]{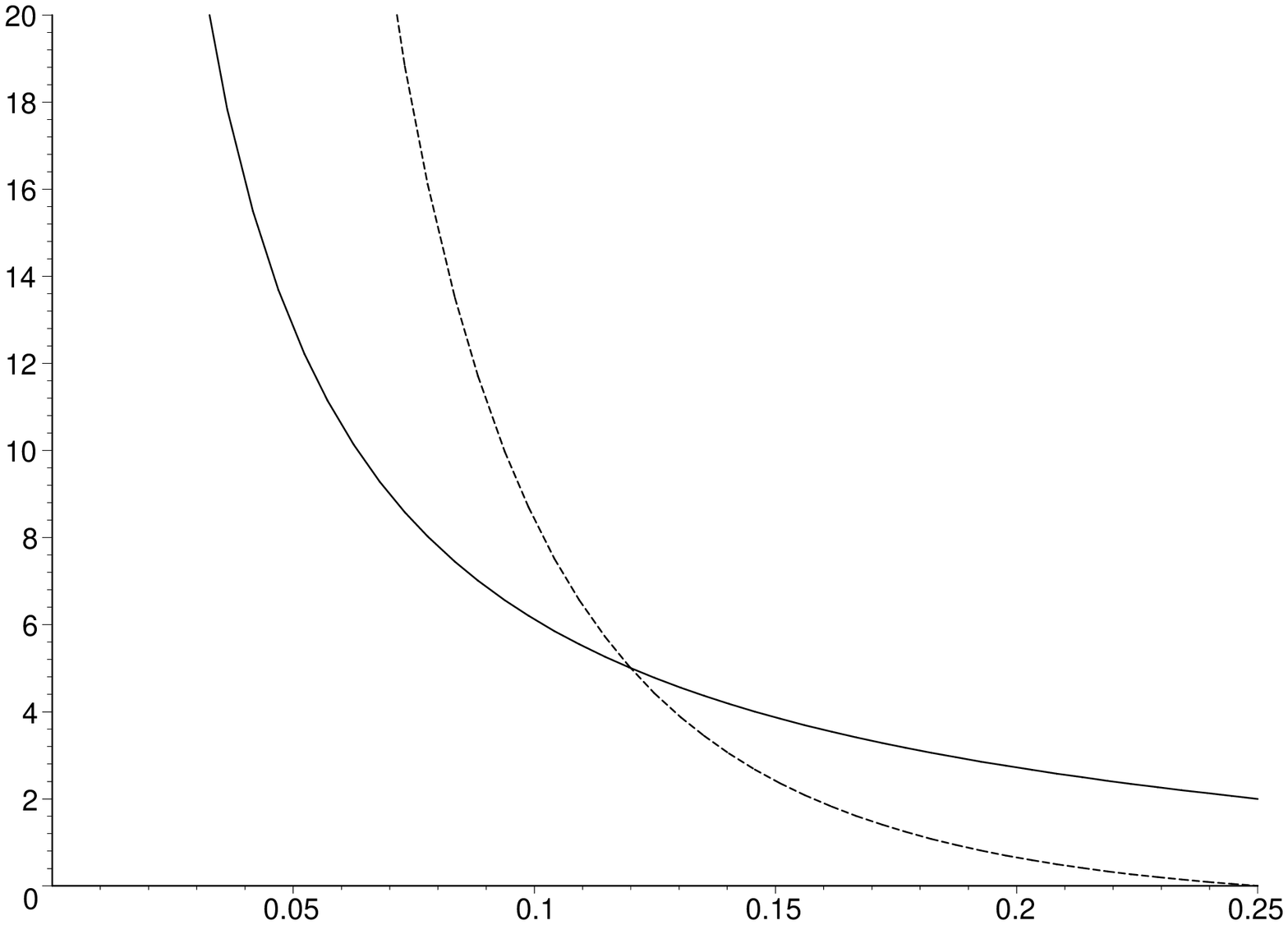}
\rput(-4,-6.7){\large $\gamma$}\rput(-6.4,-2){ $h_i$}
\rput(-4.6,-2){ $\mu_i$} \caption{The dependence in GBIG 1--2 of
the initial expansion rate and density on $\gamma$. }
\label{himui2}
\end{center}
\end{figure}

\begin{figure}
\begin{center}
\includegraphics[height=3in,width=2.75in,angle=270]{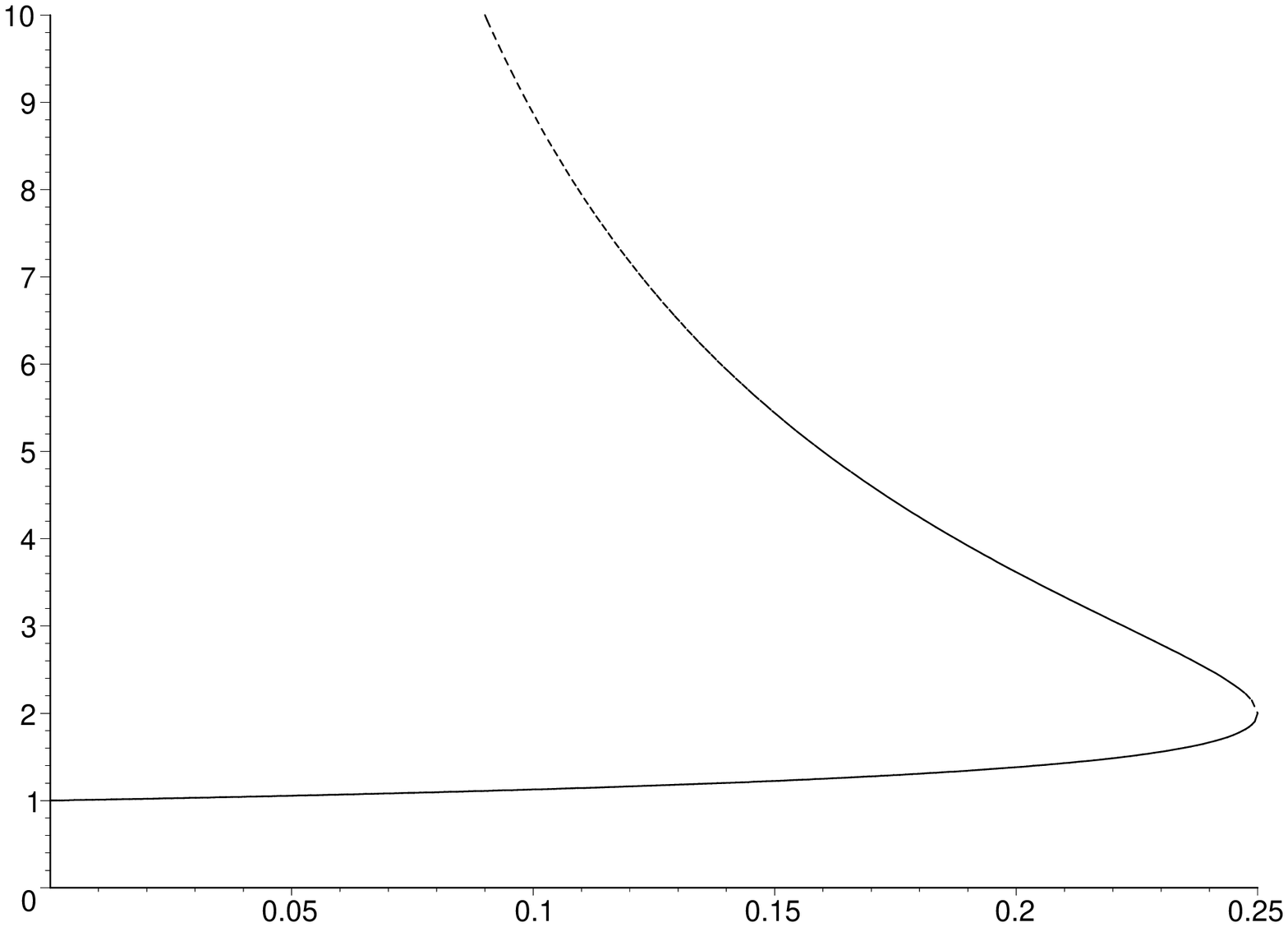}
\rput(-7.4,-3.2){\large $h_{\infty}$}\rput(-3.9,-6.7){\large
$\gamma$}\rput(-3,-5.3){\small GBIG 1} \rput(-3,-2.4){\small GBIG
2} \caption{The GBIG 1--2 late-time asymptotic expansion rate as a
function of $\gamma$. } \label{hinfin}
\end{center}
\end{figure}

The behaviour of the key GBIG 1--2 parameters is illustrated in
Figs.~\ref{himui2} and \ref{hinfin}.

\subsection{Cosmological Dynamics}

The GBIG 1 model, which is the physically relevant generalisation of
the DGP(+) model, exists if Eq.~(\ref{class}) holds. By
Eq.~(\ref{DV}), this means that the GB length scale $L_{\rm
gb}=\sqrt{\alpha}$ must be below a maximum threshold determined by
the IG cross-over scale:
\begin{equation}\label{gam2}
\gamma<{1\over 4} \,\Leftrightarrow\, L_{\rm gb}\equiv \sqrt
\alpha < {1\over 4}\sqrt{3 \over 2}\,r\,.
\end{equation}
If the GB term is taken as the correction term in certain string
theories, then $L_{\rm gb}\sim L_{\rm string}$, while $r\sim
H_0^{-1}$, so that this bound is easily satisfied.

When Eq.~(\ref{gam2}) holds, the universe starts with a maximum
density $\rho_{\rm i}$ and maximum Hubble rate $H_{\rm i}$, and
evolves to an asymptotic vacuum de Sitter state:
\begin{eqnarray}
0<\rho &<& \rho_{\rm i}= \frac{r^3}{16\kappa^2_5\alpha^2}\left\{1-
\frac{12\alpha}{r^2}+\left(1-\frac{8\alpha}{r^2}\right)^{3/2}\right\},\\
H_\infty< H &<& H_{\rm i}= {r \over 8\alpha}\left(
1+\sqrt{1-{8\alpha \over r^2}}\right).
\end{eqnarray}
At any epoch $t_0$, the proper time back to the beginning is:
\begin{equation}
t_0-t_{\rm i}=r\int_{a_{\rm i}}^{a_0}\,{da \over ah}\,.
\end{equation}
Since $a$ and $h$ are nonzero on the interval of integration, the
time back to the beginning is finite.

The current Hubble rate can be approximated by the final de Sitter
Hubble rate, $H_0\sim H_\infty$, so that by Eq.~(\ref{cross}), the
cross-over scale obeys:
\begin{equation}\label{hub}
H_0^{-1} \lesssim r \lesssim 2H_0^{-1}\,.
\end{equation}
In the DGP(+) limit, $r\sim 1.2H_0^{-1}$~\cite{Deffayet:2002sp}. The
effect of GB gravity is to allow for increased $r$ but not beyond
$r\sim 2H_0^{-1}$.

However, there is a UV-IR ``bootstrap" operating to severely limit
the GB effect at late times. The key point is that appreciable
late-time GB effects require an increase in $\gamma$ (see
Fig.~\ref{hinfin}), whereas the primordial Hubble rate $H_{\rm i}$
is suppressed by an increase in $\gamma$ -- as shown in
Fig.~\ref{himui2}. Equations~(\ref{Hrhomax}) and (\ref{GBIGhinfin})
imply that:
\begin{equation}\label{boot}
H_{\rm i}\gg H_0~ \Rightarrow~ \gamma \ll {1\over 4}\,.
\end{equation}
Thus the GBIG 1 model does not alleviate the DGP(+) fine-tuning
problem of a very large cross-over scale, $r\sim H^{-1}_0 \sim
(10^{-33}\,\mbox{eV})^{-1}$.

The GBIG 1 Friedman equation~(\ref{roots1}) gives:
\begin{equation}
H^2=\frac{3r^2}{64\alpha^2}\left\{1-\frac{16\alpha}{3r^2}
+2\sqrt{\left(\frac{16\alpha}{3r^2}-1\right)^2-
\frac{64\alpha^2}{9r^4}\left(3+2r^2\kappa^2_4\rho\right)}\cos\left[\theta(\rho)
+{4\pi \over 3} \right]\right\}, \label{root}
\end{equation}
where:
\begin{equation}
\cos3 \theta(\rho) = \frac{2048 \alpha^4 \mu^2
+96\alpha^2r^4(1+2\mu)\left(\frac{16\alpha}{3r^2}-1\right)
-2\left(\frac{16\alpha}{3r^2}-1\right)^3}{
3r^8\left[\left(\frac{16\alpha}{3r^2}-1\right)^2
-\frac{64\alpha^2}{3r^4}(1-2\mu)\right]^{3/2} }.
\end{equation}
A more convenient form of the Friedmann equation follows from
solving Eq.~(\ref{DGPF2}) for $\mu$:
\begin{equation}\label{fnew}
\mu=h^2-h(\gamma h^2+1)\,,~~ h_\infty \leq h < h_{\rm i}\,.
\end{equation}
By expanding to first order in $h^2-h_\infty^2$, we find that at
late times:
\begin{equation}\label{late}
h^2 = h_\infty^2+2\left({h_\infty \over 2 -h_\infty}\right) \mu
+O(\mu^2)\,.
\end{equation}
Taking the DGP(+) limit $h_\infty\to 1$, and comparing with
Eqs.~(\ref{DGPFL2}) and (\ref{dd}), we find that the effective
Newton constant in GBIG 1 is:
\begin{equation}\label{newt}
G=\left({h_\infty \over 2 -h_\infty}\right){G_5 \over r} \,,
\end{equation}
where $G_5=\kappa_5^2/8\pi$ is the fundamental, 5D gravitational
constant. In the DGP (+) case, $G=G_5/r$.

Equation~(\ref{newt}) gives a relation for the fundamental Planck
scale $M_5$:
\begin{equation}\label{newt2}
M_5^3\sim \left({rH_0 \over 2 -rH_0}\right){M_{\rm p}^2 \over r}
\,,
\end{equation}
where $M_{\rm p}$ is the effective 4D Planck scale, and we used
$H_\infty \sim H_0$. As $r\to 2H_0^{-1}$ (its upper limit), so $M_5$
increases. This is very different from the DGP(+) case, where:
\begin{equation}\label{M5}
M_5^3= M_{\rm p}^2/r,
\end{equation}
so that $M_5$ is constrained to be very low, $M_5 \lesssim
100\,$MeV. In principle, GB gravity allows us to solve the problem
of a very low fundamental Planck scale in DGP(+)-- but in practice,
the UV-IR bootstrap, Eq.~(\ref{boot}), means that $\gamma\sim 0$ so
that $M_5$ is effectively the same as in the DGP(+) case. Thus the
GB modification of the DGP(+) does not change the fine tuning of the
cross-over scale $r$, nor the consequent low value of $M_5$. This is
because $\gamma$ is forced to be close to $0$ if we want
self-acceleration to replace dark energy, as in the DGP(+) case.
However, no matter how small $\gamma$ is, a nonzero $\gamma$
dramatically alters the early universe, by removing the
infinite-density big bang.

What is the nature of the beginning of the universe in GBIG 1? We
can use Eq.~(\ref{in}) in Eq.~(\ref{Ray}), for matter with $w>-1$,
to analyze the initial state, $d\mu/d(h^2)\to 0+$. We find that
$h'_{\rm i}=-\infty$, i.e., infinite deceleration:
\begin{equation}
a''_{\rm i}=-\infty\,.
\end{equation}
(For GBIG 2, with $d\mu/d(h^2)\to 0-$, we have $h'_{\rm
i}=+\infty$.) The initial state has no big bang, but it has infinite
deceleration, and thus infinite Ricci curvature. The brane universe
is born in a ``quiescent" singularity. Although similar
singularities may be found in induced gravity
models~\cite{Shtanov:2002ek}, they arise from the special extra
effect of a bulk black hole or a negative brane tension which is
very different from the gravitational GB effect that is operating in
the GBIG singularity. A key point is that neither the DGP(+) model
nor the GB model avoid the big bang, as shown in Eqs.~(\ref{bbdgp})
and (\ref{bbgb}). But together, the IG and GB effects combine in a
``nonlinear" way to produce entirely new behaviour. If we switch off
either of these effects, the big bang reappears.

This singularity is reminiscent of the ``sudden" (future)
singularities in General Relativity~\cite{Barrow:2004xh} -- but
unlike those singularities, the GBIG 1--2 singularity has finite
pressure. The initial curvature singularity signals a breakdown of
the brane spacetime. The (Minkowski) bulk remains regular, but the
imbedding of the brane becomes singular. Higher-order
quantum-gravity effects will be needed to cure this singularity.

By performing an expansion near the initial state, using
Eq.~(\ref{fnew}), we find that the primordial Hubble rate in GBIG 1,
after the infinite deceleration at the birth of the universe, is
given by:
\begin{equation}
H^2\approx H_{\rm i}^2-\frac{r\kappa_4}{\sqrt{3}} \left[\frac
{3H^2_{\rm i}-\kappa_4^2\rho_{\rm i}}{ 8\alpha\left(1+4\alpha
H^2_{\rm i}\right)-1 } \right]^{1/2}(\rho_{\rm i}-\rho)^{1/2}.
\end{equation}
This is independent of the equation of state $w$ so that the
universe decelerates for a finite time after its
infinite-deceleration birth, regardless of the matter content. If
there is primordial inflation in the GBIG 1 universe, then the
acceleration $a''$ will become positive. For a realistic model
(satisfying nucleosynthesis and other constraints), $a''$ must
subsequently become negative, so that the universe decelerates
during radiation- and early matter-domination. Finally, $a''$ will
become positive again as the late universe self-accelerates due to
the IG effects.

\begin{figure}
\begin{center}
\includegraphics[height=3in,width=2.75in,angle=270]{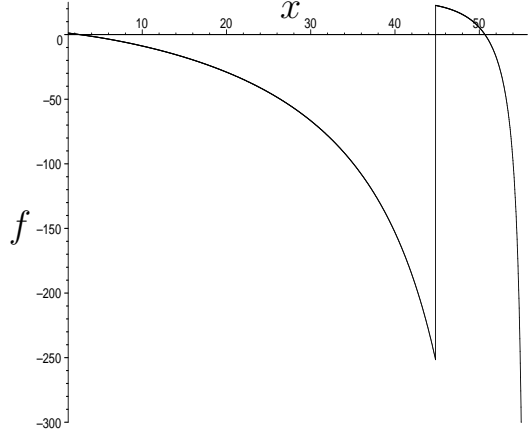}
\rput(-4,-0.8){\large $x$}\rput(-7.6,-3.7){\large $f$} \caption{The
acceleration $f=a''/a$ vs $x=h^2$, for a GBIG 1 cosmology with
inflation, followed by radiation domination, followed by late-time
self-acceleration. Brane proper time flows from right to left. Here
$\gamma=1/12$, and $n=0.8$ in Eq.~(\ref{eos}).  } \label{cos}
\end{center}
\end{figure}

We can simplify the expression~(\ref{acc}) for the acceleration in
GBIG 1 via Eq.~(\ref{fnew}):
\begin{equation}
f=\frac{x(3\gamma x+1-2\sqrt{x}) +3(1+w)[x\sqrt{x}-{x}(\gamma
x+1)]}{3\gamma x+1-2\sqrt{x}}\!,
\end{equation}
where $f\equiv a''/a, x\equiv h^2$. For a given $w(x)$, we can plot
$f(x)$. We show an example in Fig.~\ref{cos} of a simple model, with
primordial inflation followed by radiation domination, followed by
late-time self-acceleration. We have used the effective equation of
state:
\begin{equation}\label{eos}
w= \left\{ \begin{array}{lll} -0.9 && n(h_{\rm i}^2 -
h_{\infty}^2) + h_{\infty}^2< x<h_{\rm i}^2\,,\\1/3 &&
h_{\infty}^2 < x < n(h_{\rm
i}^2 - h_{\infty}^2) + h_{\infty}^2\,.\\
\end{array} \right.
\end{equation}
Here $0<n<1$ is a parameter determining the time of reheating (with
$n=0$ corresponding to no inflation and $n=1$ to no reheating/
radiation). The values used for the equation of state are chosen in
order to mimic slow-roll inflation in the first epoch ($w=-0.9$) and
radiation in the second ($w=1/3$).

\subsection{Nucleosynthesis and the age of the universe}\label{ages}

We have noted the UV-IR bootstrap, which enforces a very small value
on $\gamma$ if the universe is to be old enough. In order to make
this more quantitative we impose constraints from nucleosynthesis
and the age of the universe.

As already stated the DGP cross over-scale must be of the order of
the Hubble scale or larger, therefore:

\begin{equation}\label{Nuc1}
 r\geqslant H_0^{-1}\Rightarrow Hr\geqslant\frac{H}{H_0},
\end{equation}
where $H_0$ is the current Hubble rate. Taking nucleosynthesis to
have occurred at an energy scale of $1$MeV and the observed universe
to be at a scale of $10^{-33}$eV we have the constraint that:

\begin{equation}\label{Nuc2}
Hr\gtrsim\frac{1{\rm MeV}}{10^{-33}{\rm eV}}=10^{39}.
\end{equation}
We can also put a constraint on $\gamma$ from requiring $H_{\rm i}$
to have occurred at an energy scale greater than $1$TeV. This is in
order to put it outside the range of collider experiments. Using the
solution for $h_{\rm i}$ in Eq.~(\ref{Hrhomax}) we can write
$\gamma$ as:

\begin{equation}\label{Nuc3}
\gamma=\frac{2h_{\rm i}-1}{3h^2_{\rm i}},
\end{equation}
where:
\begin{equation}\label{Nuc4}
h_{\rm i}=H_{\rm i}r>10^{45}.
\end{equation}
Therefore $\gamma$ is constrained by:

\begin{equation}\label{Nuc5}
\gamma\lesssim 10^{-45}.
\end{equation}
Taking $r\sim H^{-1}_0 \sim (10^{-33}\,\mbox{eV})^{-1}$ the above
constraint gives:

\begin{equation}\label{Nuc6}
\alpha\lesssim 10^{21}\mbox{eV}^{-2}.
\end{equation}
The string energy scale ($\alpha^{-1/2}$) is then constrained by:

\begin{equation}\label{sc}
\alpha^{-1/2}\gtrsim 10^{-10}\mbox{eV}.
\end{equation}
This is compatible with the constraint from proton decay which
gives~\cite{Burikham:2005wj}:

\begin{equation}\label{sesc2}
\alpha^{-1/2}> 10^{17}\mbox{eV}.
\end{equation}

The GBIG Friedmann equation, in the case $\lambda=0=\Phi$, can be
written as:

\begin{equation}\label{FE}
\left(H^2-\frac{\kappa^2_4}{3}\rho\right)^2=
\frac{H^2}{r^2}\left(1+\frac{8\alpha}{3}H^2\right)^2,
\end{equation}
where the left hand side is exactly zero in the GR limit. To see if
the GBIG model approximates GR in the era of nucleosynthesis we
write the above Friedmann equation in a dimensionless form and apply
the constraints above. The dimensionless Friedmann equation,
Eq.~(\ref{FE}), can be written as:

\begin{equation}\label{DFE}
\left(1-\frac{\kappa^2_4\rho}{3H^2}\right)^2=
\frac{1}{H^2r^2}\left(1+\gamma H^2r^2\right)^2.
\end{equation}
Expanding out the right hand side and using the constraints above we
find:

\begin{equation}\label{DFE2}
\left(1-\frac{\kappa^2_4\rho}{3H^2}\right)^2\lesssim10^{-12},
\end{equation}
during nucleosynthesis. Thus the model will safely meet
nucleosynthesis constraints if Eq.~(\ref{Nuc4}) holds.

In order to calculate the age of the universe we define a new set of
dimensionless density parameters:

\begin{equation}\label{O}
\Omega_{\alpha}=\frac{16}{3}\alpha H^2_0,~~\Omega_{\rm r_{\rm
c}}=\frac{1}{4r^2H^2_0},~~\Omega_{\phi}=\frac{\Phi}{2H^2_0},~~\Omega_{\rm
m}=\frac{\kappa^2_4}{3H^2_0}\rho,~~\Omega_{\lambda}=\frac{\kappa^2_4}{3H^2_0}\lambda.
\end{equation}
In terms of the $\Omega$ parameters the (general) GBIG Friedmann
equation, Eq.~(\ref{Ff}), can be written as:

\begin{equation}\label{Fr2}
\Omega_{\rm
r_c}\left\{2+\Omega_{\alpha}(E^2+\Omega_{\phi})\right\}^2\left(E^2-2\Omega_{\phi}\right)
=\left[E^2-\Omega_{\rm m}(1+z)^3-\Omega_{\lambda}\right]^2,
\end{equation}
where $E=H/H_{0}$.

In order to reduce the number of free parameters in the model we
consider the constraint on $\Omega_{\alpha}$ at the present time
($z=0$, $H=H_0$). This gives:

\begin{equation}\label{ac}
\Omega_{\alpha_{\pm}}=\frac{1}{(1+\Omega_{\phi})}\left\{\pm\frac{(1-\Omega_{\rm
m}-\Omega_{\lambda})}{\sqrt{\Omega_{\rm
r_c}(1-2\Omega_{\phi})}}-2\right\}.
\end{equation}

\begin{figure}
\includegraphics[height=3in,width=2.75in,angle=270]{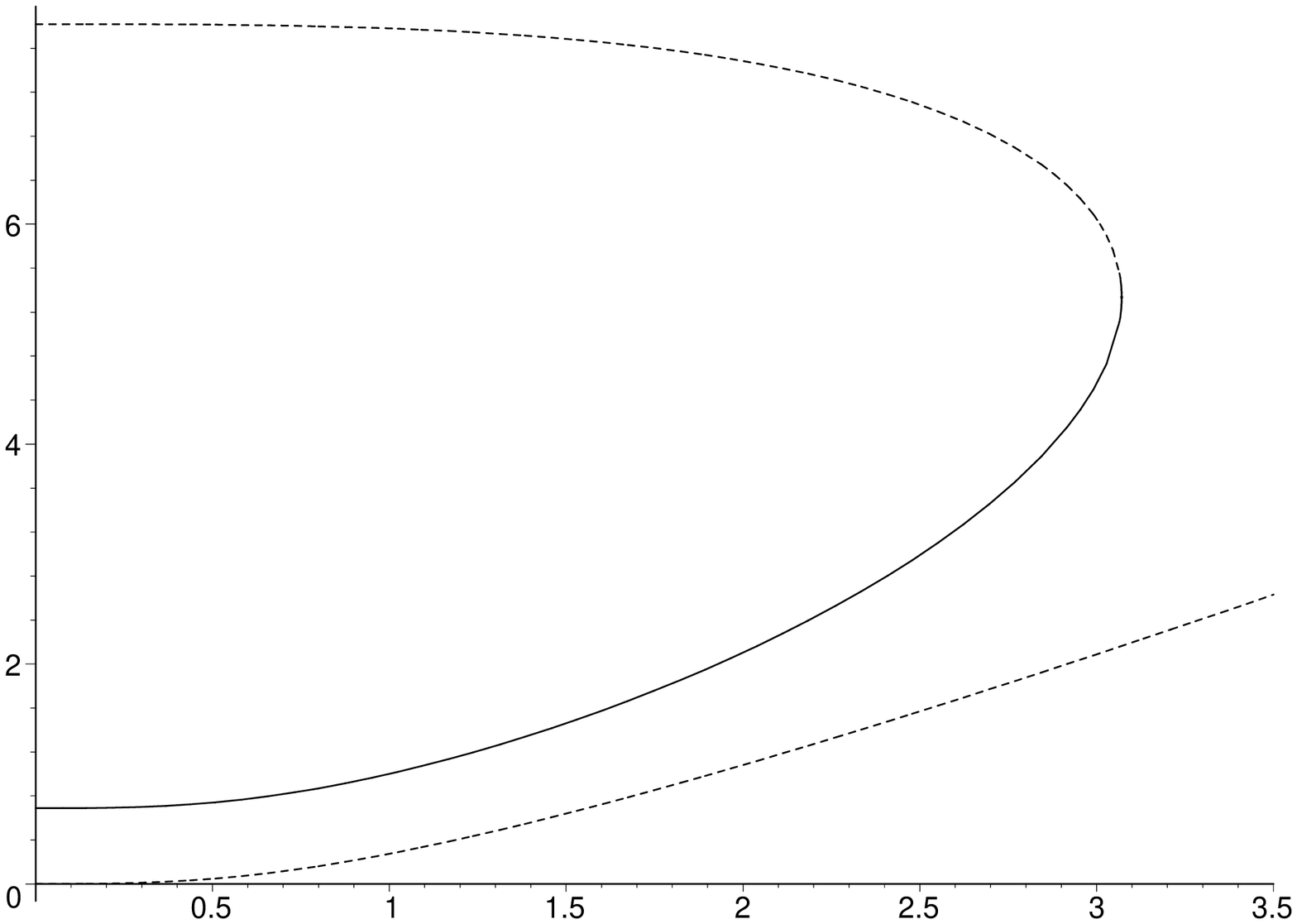}
\includegraphics[height=3in,width=2.75in,angle=270]{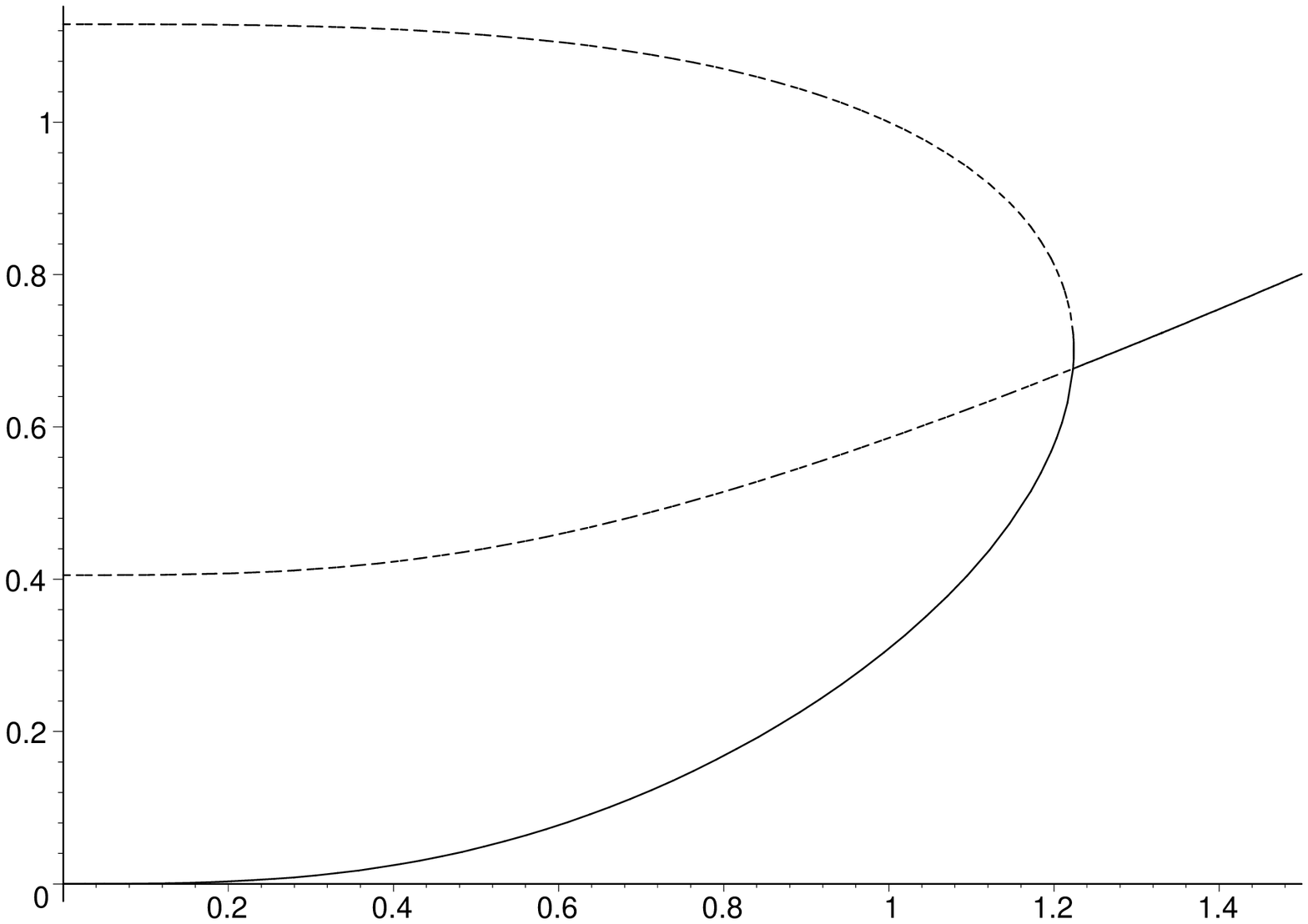}
\includegraphics[height=3in,width=2.75in,angle=270]{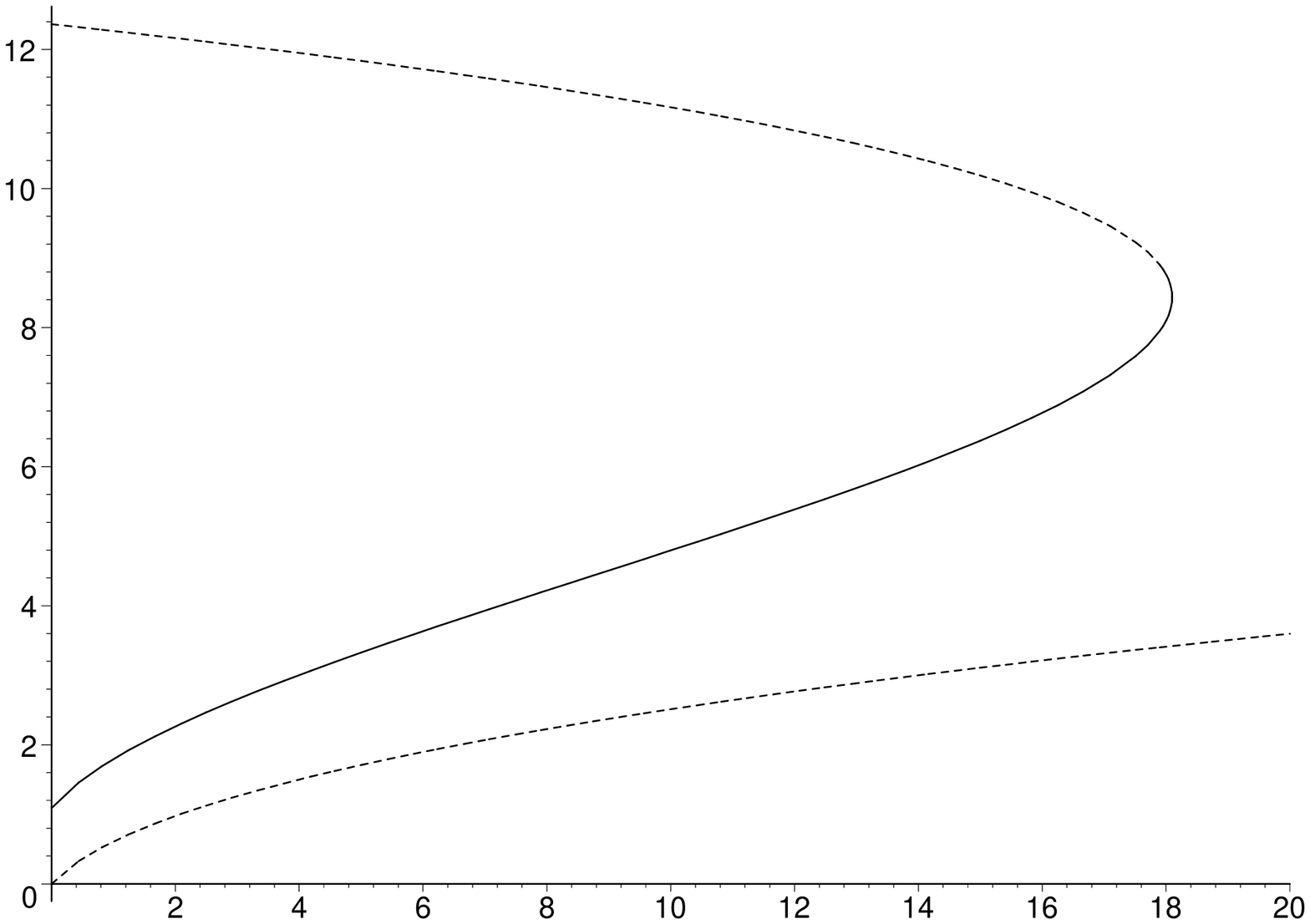}
\includegraphics[height=3in,width=2.75in,angle=270]{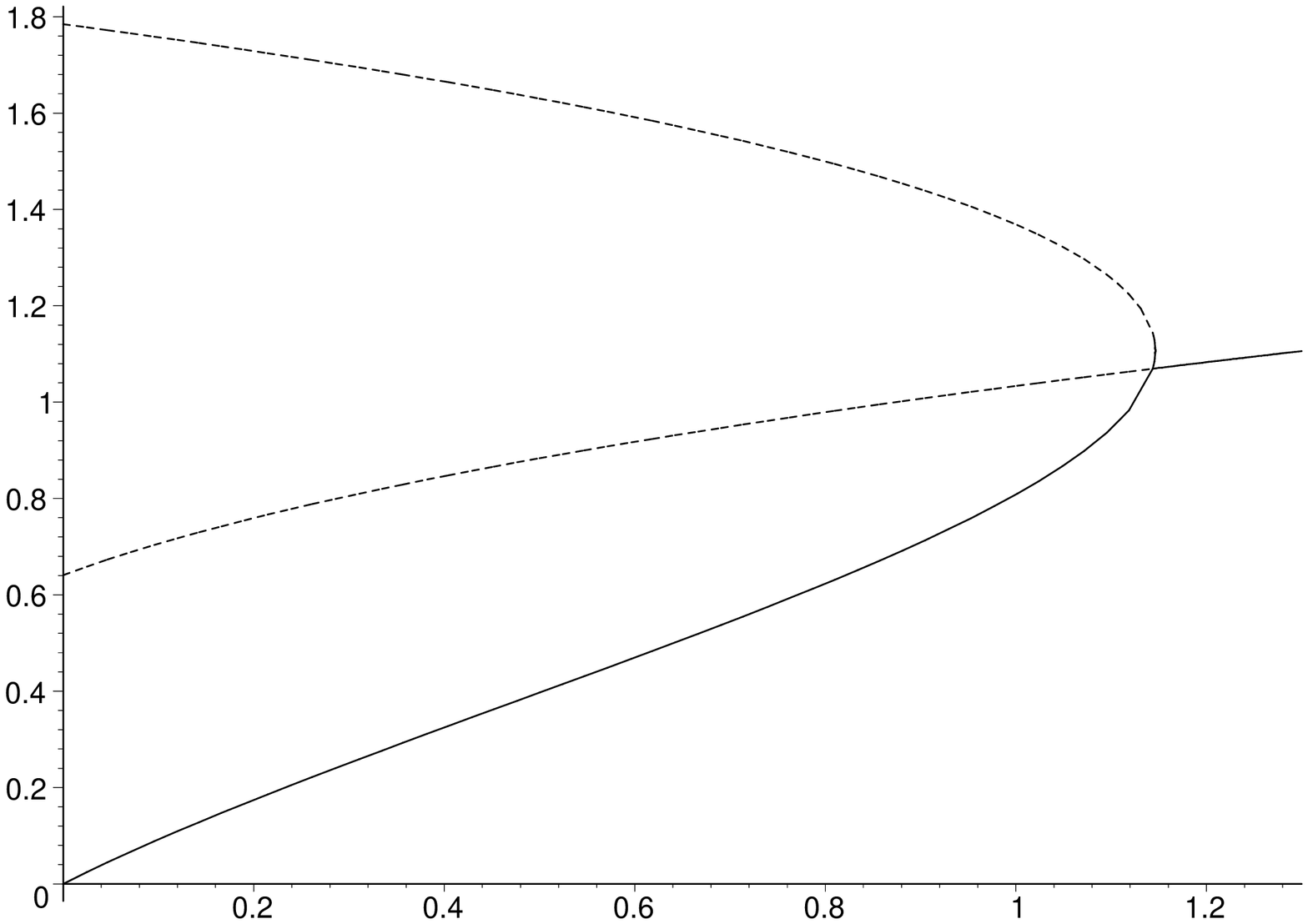}
\rput(3.8,7.4){$z+1$}\rput(11.2,7.4){$z+1$}
\rput(0.5,11.4){$E$}\rput(8.1,11.2){$E$}\rput(0.5,3.8){$h$}
\rput(8.1,3.8){$h$}\rput(4.1,0.5){$\mu$}\rput(11.7,0.5){$\mu$}
\caption{The left plots use the plus branch of Eq.~(\ref{ac}), the
right plots use the negative branch. Values used are $\Omega_{\rm
r_c}=0.1,~\Omega_{\rm m }=1/4,~\Omega_{\phi }=0$ and
$\Omega_{\lambda }=0$, so that $\Omega_{\alpha_+}=0.372$ and
$\Omega_{\alpha_-}=-4.372$. The top plots show the Hubble rates
against redshift. The bottom two plots are the equivalent plots in
$h$, $\mu$.}\label{zet}
\end{figure}

With non-zero $\Phi$ and $\lambda$, $E_{\rm i}$ is given by:

\begin{equation}\label{Eie2}
E^2_{\rm i}=\frac{1}{9\Omega^2_{\alpha}\Omega_{\rm
r_c}}\left\{2-6\Omega_{\alpha}\Omega_{\rm
r_c}+9\Omega^2_{\alpha}\Omega_{\rm
r_c}\Omega_{\phi}\pm2\sqrt{1-6\Omega_{\alpha}\Omega_{\rm
r_c}-9\Omega^2_{\alpha}\Omega_{\rm r_c}\Omega_{\phi}}\right\}.
\end{equation}
By considering $dz/dE=0$ in the Friedmann equation we get:

\begin{equation}\label{tp}
z_{\rm i}=\left\{\frac{\orc\left(4\oa\op+3\oa^2\op^2-4\right)-2\ob+
2\left(1-4\orc\oa\right)E^2_{\rm i}-3\orc\oa^2E_{\rm
i}^4}{2\om}\right\}^{1/3}-1.
\end{equation}

We will only consider the case of $\Phi=0=\lambda$, for which
$E_{\rm i}$ and $z_{\rm i}$ reduce to:
\begin{equation}\label{Eie22}
E^2_{\rm i}=\frac{1}{9\Omega^2_{\alpha}\Omega_{\rm
r_c}}\left\{2-6\Omega_{\alpha}\Omega_{\rm
r_c}\pm2\sqrt{1-6\Omega_{\alpha}\Omega_{\rm r_c}}\right\},
\end{equation}
and:
\begin{equation}\label{tp2}
z_{\rm i}=\left\{\frac{2\left(1-4\orc\oa\right)E^2_{\rm
i}-3\orc\oa^2E_{\rm i}^4-4\orc}{2\om}\right\}^{1/3}-1,
\end{equation}
with:
\begin{equation}\label{ac2}
\Omega_{\alpha_{\pm}}=\pm\frac{(1-\Omega_{\rm m})}{\sqrt{\Omega_{\rm
r_c}}}-2.
\end{equation}

The DGP limit, $\Omega_{\alpha}\rightarrow0$, gives $E_{\rm
i}=\infty$ due to the presence of $\Omega_{\alpha}$ in the
denominator, and thus of $z_{\rm i}=\infty$.

We see, from Eq.~(\ref{ac2}), that when $\Omega_{\alpha_{\pm}}=0$:

\begin{equation}\label{ac0}
\Omega_{m}=1\pm2\sqrt{\orc}.
\end{equation}
Therefore for $\Omega_{\alpha_{\pm}}$ to be positive we require:
\begin{eqnarray}
\om &<& 1-2\sqrt{\orc},~~~~~~
{\rm for}~\Omega_{\alpha_+}, \label{mc1}\\
\om &>& 1+2\sqrt{\orc},~~~~~~{\rm for}
~\Omega_{\alpha_-}.\label{mc2}
\end{eqnarray}
Since $\Omega_{\alpha_{\pm}}>0$ implies $\alpha>0$, Eqs.~(\ref{mc1})
and (\ref{mc2}) describe the GBIG 1 model. When
$\Omega_{\alpha_{\pm}}<0$, we have $\alpha<0$, and this corresponds
to the non-self-accelerating GB modifications of the DGP(--) model,
as discussed in section \ref{sec}.

For $\Omega_{\alpha_{\pm}}$ to be negative we require:
\begin{eqnarray}
\om &>& 1-2\sqrt{\orc},~~~~~~
{\rm for}~ \Omega_{\alpha_+}, \\
\om &<& 1+2\sqrt{\orc},~~~~~~{\rm for}~ \Omega_{\alpha_-}.
\end{eqnarray}
We only consider $0<\om<1$, therefore $\Omega_{\alpha_-}$ is always
negative and the solutions do not self-accelerate.
$\Omega_{\alpha_+}$ has two regions in the $\om,~\orc$ plane. One
has solutions that self-accelerate ($\oa>0$), one that does not
($\oa<0$). If we included a non-zero $\Omega_{\lambda}$ or
$\Omega_{\phi}$, self-acceleration would be possible on both
branches. Fig.~\ref{zet} illustrates the two branches for
$\Omega_{\lambda}=0=\Omega_{\phi}$.

\begin{figure}
\includegraphics[height=3in,width=2.75in,angle=270]{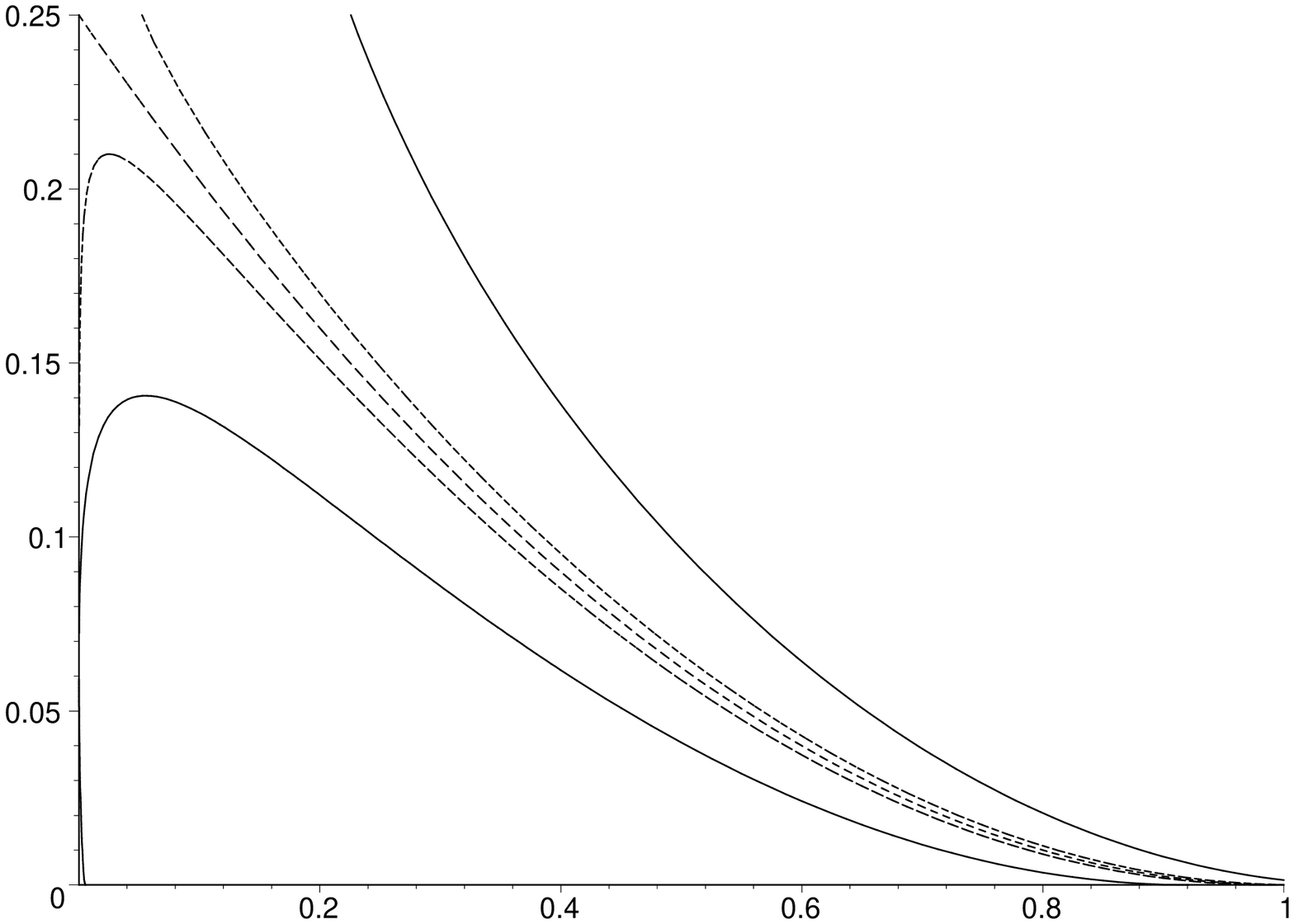}
\includegraphics[height=3in,width=2.75in,angle=270]{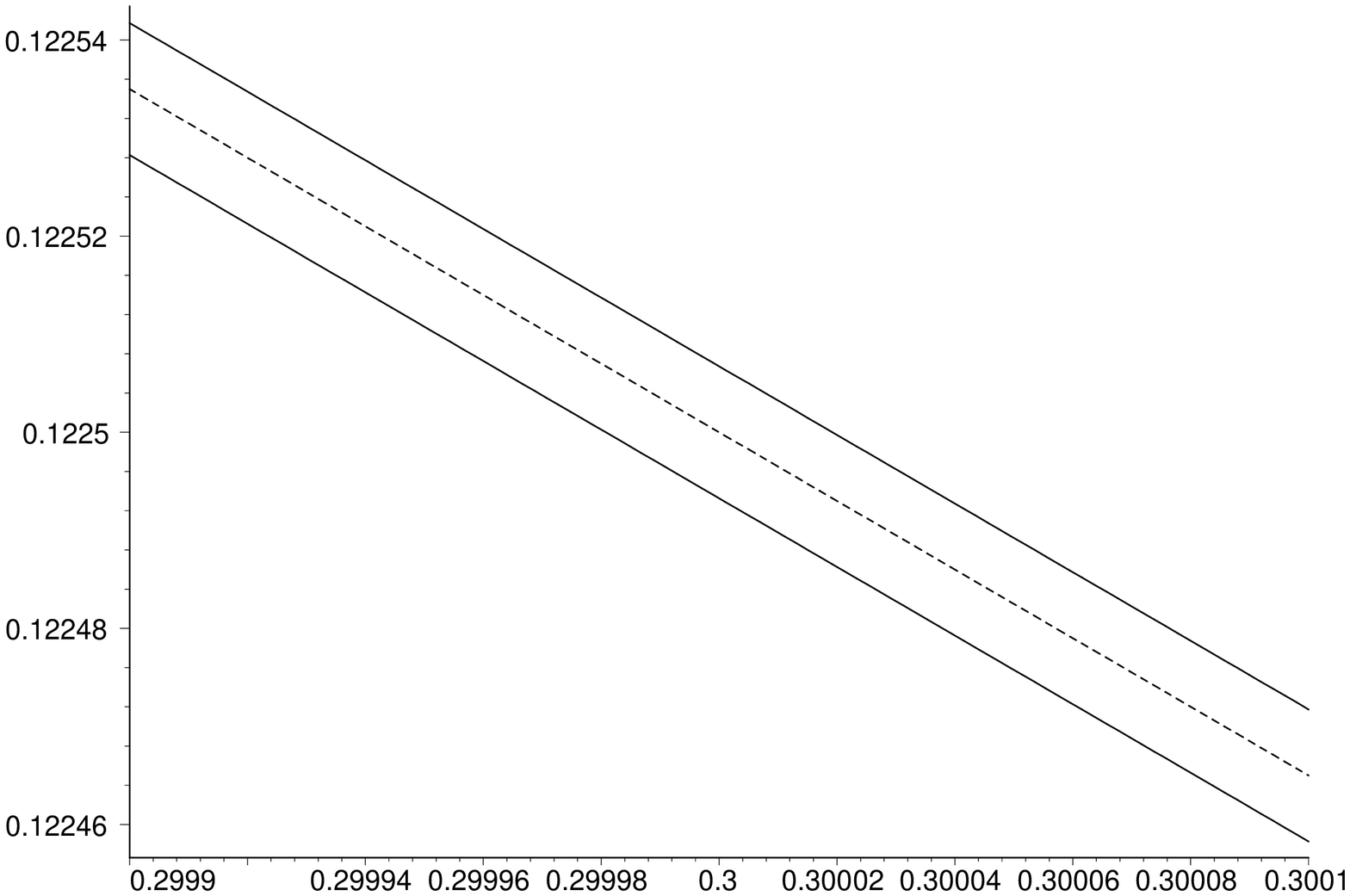}
\rput(0.5,3.6){$\orc$}\rput(3.7,0.4){$\om$}
\rput(8.2,4){$\orc$}\rput(11.4,0.3){$\om$}
 \caption{The $\om$,~$\orc$ plane for GBIG 1.
The central dashed line is for $\Omega_{\alpha}=0$. On  the left the
dashed lines are $z_{\rm i}=10$ and the solid lines are $z_{\rm
i}=2$. On the right the solid lines are $z_{\rm i}=1100$, the
decoupling redshift. The lines above $\Omega_{\alpha}=0$ have
$\Omega_{\alpha}<0$, while those below $\Omega_{\alpha}=0$
correspond to $\Omega_{\alpha}>0$. The magnitude of
$\Omega_{\alpha}$ increases as you move away from the
$\Omega_{\alpha}=0$ line. }\label{Zi2101}
\end{figure}

The initial redshift of the universe must be at least large enough
to accommodate nucleosynthesis, i.e. $z_{\rm i}>10^{10}$. This
enforces an extremely small value for $\Omega_{\alpha}$
($\Omega_{\alpha}\approx 5.5\mbox{x}10^{-5}$), as illustrated in
Fig.~\ref{Zi2101}.

\begin{figure}
\begin{center}
\includegraphics[height=3in,width=2.75in,angle=270]{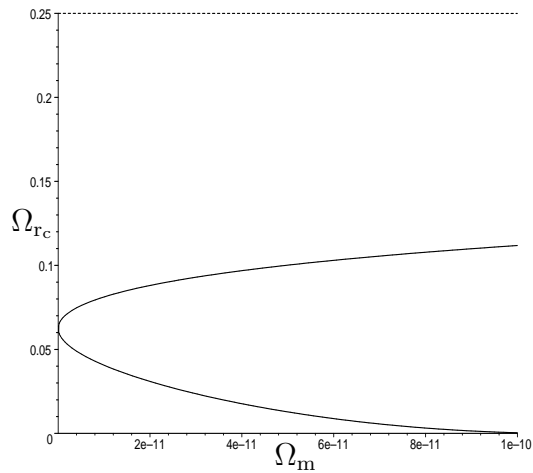}
\rput(-7.3,-3.5){$\orc$}\rput(-3.8,-6.6){$\om$}
 \caption{The $\om$,~$\orc$ plane for $\Omega_{\alpha_{+}}$ and $z_{\rm
i}=1100$. The dotted line is for
$\Omega_{\alpha}=0$.}\label{Zi11002}
\end{center}
\end{figure}

For each solution of $z_{\rm i}$ there is a region of the
$\om$,~$\orc$ plane that is significantly different from the DGP
model. This region, with small $\om$ and $\orc$, is shown for the
$z_{\rm i}=1100$ case in Fig.~\ref{Zi11002}.
This region, which is disallowed by observations, is present due to
the form of the denominator in Eqs.~(\ref{tp2}) and~(\ref{Eie2}).
$E_{\rm i}$ becomes infinite when $\orc=0$ and $z_{\rm
i}\rightarrow\infty$ for $\om=0$.

\begin{figure}
\begin{center}
\includegraphics[scale=0.4]{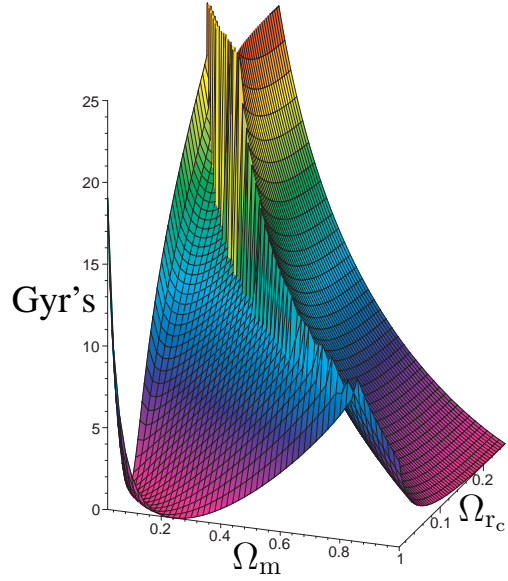}
\rput(-3.4,0.2){\large $\om$}\rput(-0.4,0.8){\large
$\orc$}\rput(-6.1,3.6){\large Gyr's}\caption{The age of the universe
for $\Omega_{\alpha +}$.}\label{pf21}
\end{center}
\end{figure}

\begin{figure}
\begin{center}
\includegraphics[scale=0.4]{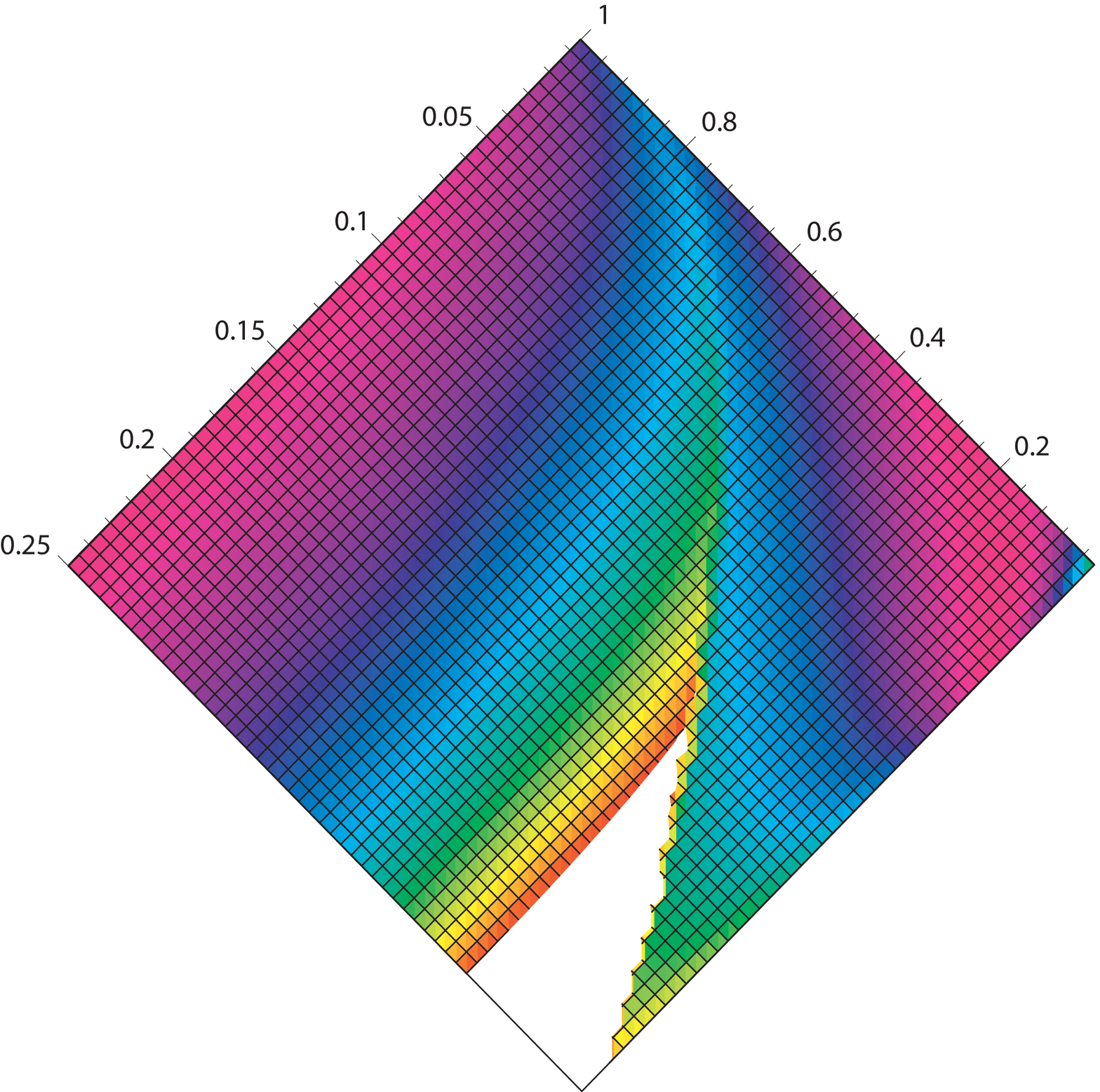}
\rput(-1.3,5.9){\large $\om$}\rput(-5.9,5.9){\large
$\orc$}\caption{As in Fig.~\ref{pf21}, but viewed from
above.}\label{pf23}
\end{center}
\end{figure}

We can work out the age (look-back time) of the universe by
evaluating the following integral:

\begin{equation}\label{int}
t_0-t_{\rm i}=H^{-1}_0\int^{z_{\rm i}}_0\frac{dz}{(1+z)E(z)}.
\end{equation}
Using $H_0=73{\rm kms^{-1}Mpc^{-1}}$ ($H_0=73\pm3{\rm
kms^{-1}Mpc^{-1}}$ is the WMAP 3 year data
result~\cite{Spergel:2006hy}) we can get the results in years,
Figs.~\ref{pf21} and~\ref{pf23}.
In Fig.~\ref{pf21} we see a sharp transition along the $\oa=0$
ridge. This is because the $\oa>0$ and $\oa<0$ branches are
disjoint. In the region where $\oa>0$ (left hand side) $z_{\rm i}$
is for the self-accelerating branch. On the right hand side of the
plot $\oa<0$, so $z_{\rm i}$ is for the non-self-accelerating
branch.
From Fig.~\ref{pf23} we see that in general, solutions that lie in
the $\oa<0$ region will give us larger ages. If we considered
solutions with a matter density less than that observed ($\om\approx
0.3$) it is possible to get a sufficiently large ages with $\orc$
measurably different from that of the DGP ($\oa=0$) model. If we
want to have self-acceleration and the finite density big bang
($\oa>0$) and have enough time for the evolution of the universe we
are restricted to solutions that lie along (or at least extremely
close to) the DGP limit. This means that observationally the GBIG
model will be indistinguishable from the DGP model even if the early
universe is dramatically different.

\begin{figure}
\begin{center}
\includegraphics[scale=0.4]{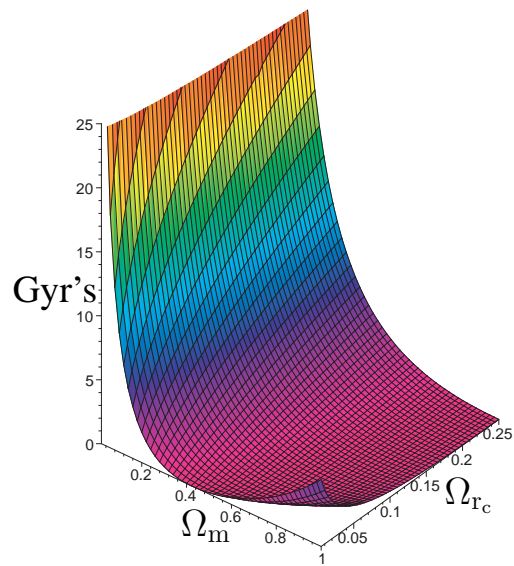}
\rput(-4.1,0.5){\large $\om$}\rput(-0.6,0.9){\large
$\orc$}\rput(-6.1,3.6){\large Gyr's}\caption{The age of the universe
for $\Omega_{\alpha -}$.}\label{pf221}
\end{center}
\end{figure}

\begin{figure}
\begin{center}
\includegraphics[scale=0.4]{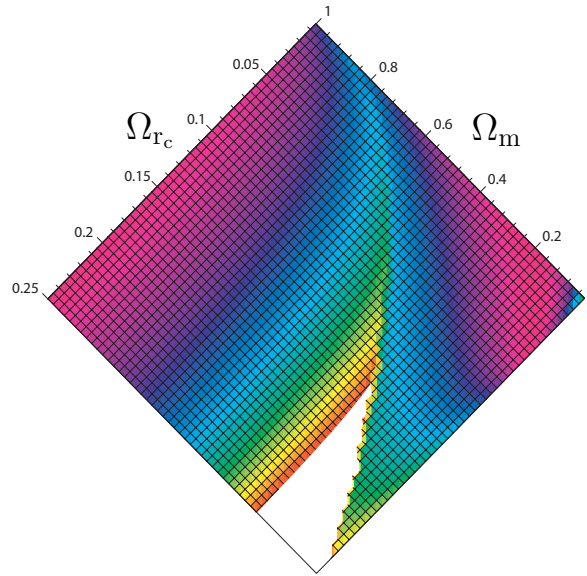}
\rput(-1.4,5.8){\large $\om$}\rput(-5.9,5.8){\large
$\orc$}\caption{As in Fig.~\ref{pf221}, but viewed from
above.}\label{pf232}
\end{center}
\end{figure}

In Figs.~\ref{pf221} and~\ref{pf232} we have the age results for the
$\Omega_{\alpha -}$ case. As $\Omega_{\alpha -}<0$ for all values
for $\om<1$ all the results are for the non-self-accelerating
solution.


\section{Conclusions}

We have seen that by including the Gauss-Bonnet bulk term with the
Induced gravity brane term in the gravitational action, we get some
striking new features. The ultra-violet correction from the
Gauss-Bonnet term in combination with the infra-red correction from
the induced gravity term gives rise to a solution that starts with a
finite density and ends with self-acceleration. We saw that the
infra-red induced gravity term on its own can give a
self-accelerating solution which is still present in the GBIG model.
The GB term modifies the early era but but does not produce the
finite density big bang on its own. It is only the combination that
does. There is no threshold for the finite density big bang to
occur-the contributions from the GB and IG terms only need to be
non-zero.

The density, pressure and temperature are finite but the big bang
does have a curvature singularity at the beginning. This deserves
further investigation.

We have seen that the UV-IR bootstrap confines $\gamma$ to be very
small i.e.~the GBIG model is very close to the DGP(+) model for most
of the history of the universe. This was also seen in the analysis
of the age of the universe. In order for there to sufficient time
for events such as nucleosynthesis to have taken place $\gamma$ must
be almost indistinguishable from the DGP(+) model ($\gamma=0$).

%% file: CHGeneralGBIGBranes.tex
\chapter{Generalised cosmologies with Induced Gravity and Gauss-Bonnet terms}\label{GGBIG}

\section{Introduction}

In this chapter we shall extend the work in Chapter~\ref{GBIGB} to
models with non-zero brane tension and bulk curvature. We keep the
assumptions that the bulk black hole mass is zero and the brane is
spatially flat. We shall see that this leads to many more possible
solutions, but not all of them are physically relevant.

\section{Field Equations}

The general form of the Friedmann equation in this case is:

\begin{equation}\label{Fried2}
\left[1+\frac{8}{3}\alpha\left(H^2+\frac{\Phi}{2}\right)\right]^2\left(H^2-\Phi\right)
=\left[rH^2-\frac{\kappa^2_5}{6}(\rho+\lambda)\right]^2,
\end{equation}
where $\Phi$ is a solution to:

\begin{equation}\label{Phi2}
\Phi+2\alpha\Phi^2=\frac{\Lambda_5}{6}.
\end{equation}
Due to the presence of the GB term in the bulk action, the bulk
cosmological constant is given by:

\begin{equation}\label{Lamb5}
\Lambda_5=-\frac{6}{\ell^2}+\frac{12\alpha}{\ell^4}.
\end{equation}
as seen in Chapter~\ref{GBIGB}. Equations~(\ref{Phi2}) and
(\ref{Lamb5}) give us two solutions for $\Phi$:

\begin{equation}
\Phi_1=-\frac{1}{\ell^2},~~\Phi_2=\frac{1}{\ell^2}-\frac{1}{2\alpha}.
\end{equation}
\newpage
\thispagestyle{myheadings}
\begin{center}
\markright{CHAPTER 4.~~~GENERALISED COSMOLOGIES WITH IG AND GB
TERMS}
\end{center}
\addvspace{-30pt}
$\Phi_1$ is the generalised form of the solution
considered in the previous chapter.

We work with the Friedmann equation in dimensionless form. We shall
use the same variables as defined in the previous chapter
(Eq.~\ref{DV}) with the addition of:
\begin{equation}\label{DV2}
\sigma =\frac{r\kappa^2_5}{6}\lambda,~\chi=\frac{r^2}{\ell^2},
~\phi=\Phi r^2.
\end{equation}
Using these variables we can write the two solutions for $\Phi$
as:

\begin{equation}\label{Phi3}
\phi_1=-\chi,~~\phi_2=\chi-\frac{4}{3\gamma}.
\end{equation}

\begin{figure}
\begin{center}
\includegraphics[height=3in,width=2.75in,angle=270]{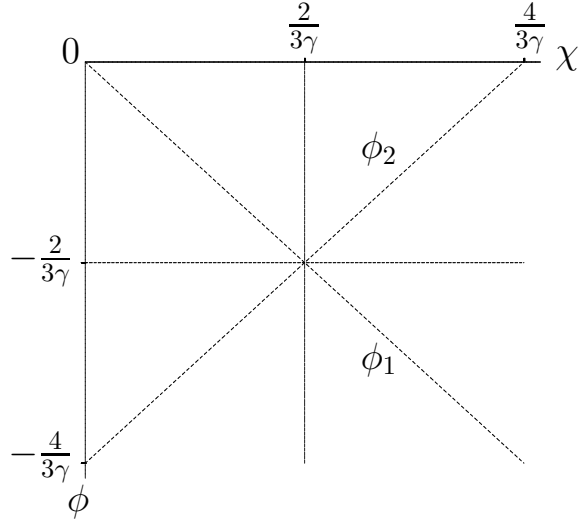}
\rput(-0.5,-0.7){\large$\chi$}
\rput(-4,-0.3){\large$\frac{2}{3\gamma}$}\rput(-1,-0.3){\large$\frac{4}{3\gamma}$}
\rput(-7,-6.6){\large $\phi$} \rput(-7.1,-0.6){\large$0$}
\rput(-7.5,-6){\large $-\frac{4}{3\gamma}$} \rput(-7.5,-3.4){\large
$-\frac{2}{3\gamma}$} \rput(-3,-1.9){\large $\phi_2$}
\rput(-3,-4.7){\large$\phi_1$} \caption{The effective bulk
cosmological constant $\phi$ as a function of $\chi$.} \label{bulkC}
\end{center}
\end{figure}

The bulk cosmological constant gives us an upper bound on the GB
coupling constant $\alpha$. Equation~(\ref{Lamb5}) gives us:

\begin{equation}\label{lcon}
\frac{1}{\ell^2}=\frac{1}{4\alpha}\left[1\pm\sqrt{1+\frac{4}{3}\alpha\Lambda_5}\right].
\end{equation}
For a RS ($\alpha\rightarrow0$) limit we take the minus branch.
$\Lambda_5>0$ would imply $\ell^2<0$ and thus we must have
$\Lambda_5\leq0$ and:

\begin{equation}\label{acon1}
\alpha\leq\frac{\ell^2}{4}.
\end{equation}
In dimensionless form this is given by:

\begin{equation}\label{alCon1}
\gamma\leq \frac{2}{3\chi}.
\end{equation}
Maintaining a RS limit would also rule out the $\phi_2$ branch. We
would be restricted to the solutions lying along the line in the top
left quadrant in Fig.~\ref{bulkC}. We are interested in the whole
range of the model so we include the plus branch in
Eq.~(\ref{lcon}). We assume $\Lambda_5\leq0$ therefore the
constraint on $\alpha$ is given by:

\begin{equation}\label{acon2}
\alpha\leq\frac{\ell^2}{2},~\Rightarrow~\gamma\leq
\frac{4}{3\chi}.
\end{equation}
If we take $\chi=0$ then we have no bound on $\gamma$ (apart from
being positive and real). This is the case considered in
Chapter~\ref{GBIGB}.

In Fig.~\ref{bulkC} we have the two $\phi$ solutions plotted as
functions of $\chi$. The two solutions with $\phi=0$ both live in a
Minkowski bulk, all the rest live in an AdS bulk. Note that one of
these AdS solutions ($\chi=0,\phi=-4/3\gamma$) has $\Lambda_5=0$ but
$\Phi=-1/2\alpha$ acting as an effective cosmological constant,
Chapter~\ref{GBIGB}. We see that for any allowed value of $\phi$ we
can be on either of the two branches. This means that we need not
consider the $\phi_1$ and $\phi_2$ solutions to the Friedmann
equation separately. We therefore consider Eq.~(\ref{acon2}) in
terms of $\phi$. We define the maximum value of $\gamma$, for a
particular value of $\phi$, that is allowed by the constraint
equation:

\begin{equation}\label{gammaM}
\gamma_{\rm M}=-\frac{4}{3\phi}.
\end{equation}

The dimensionless Friedmann equation is:

\begin{equation}\label{dFried}
\left[1+\gamma\left(h^2+\frac{\phi}{2}\right)\right]^2\left(h^2-\phi\right)
=\left[h^2-(\mu+\sigma)\right]^2.
\end{equation}
with the conservation equation now given by:
\begin{equation}\label{ecd}
\mu'+3h(1+w)\mu=0,
\end{equation}
where $'=d/d\tau$ and $h=a'/a$. The Raychaudhuri and acceleration
equations are given by:

\begin{equation}\label{Ray2}
h'=\frac{3\mu(1+w)[h^2-(\mu+\sigma)]}{(\gamma h^2+1)(3\gamma
h^2+1)-2[h^2-(\mu+\sigma)]-\phi\gamma\left(1+\frac{3}{4}\phi\gamma\right)},
\end{equation}
and:

\begin{equation}\label{accel}
\frac{a''}{a}=\frac{h^2(\gamma h^2+1)(3\gamma
h^2+1)-[h^2-(\mu+\sigma)][2h^2-3\mu(1+w)]-h^2\phi\gamma\left
(1+\frac{3}{4}\phi\gamma\right)}{(\gamma h^2+1)(3\gamma
h^2+1)-2[h^2-(\mu+\sigma)]-\phi\gamma\left(1+\frac{3}{4}
\phi\gamma\right)}.
\end{equation}

\section{Friedmann Equation Solutions}

\begin{figure}
\begin{center}
\includegraphics[height=3in,width=2.75in,angle=270]{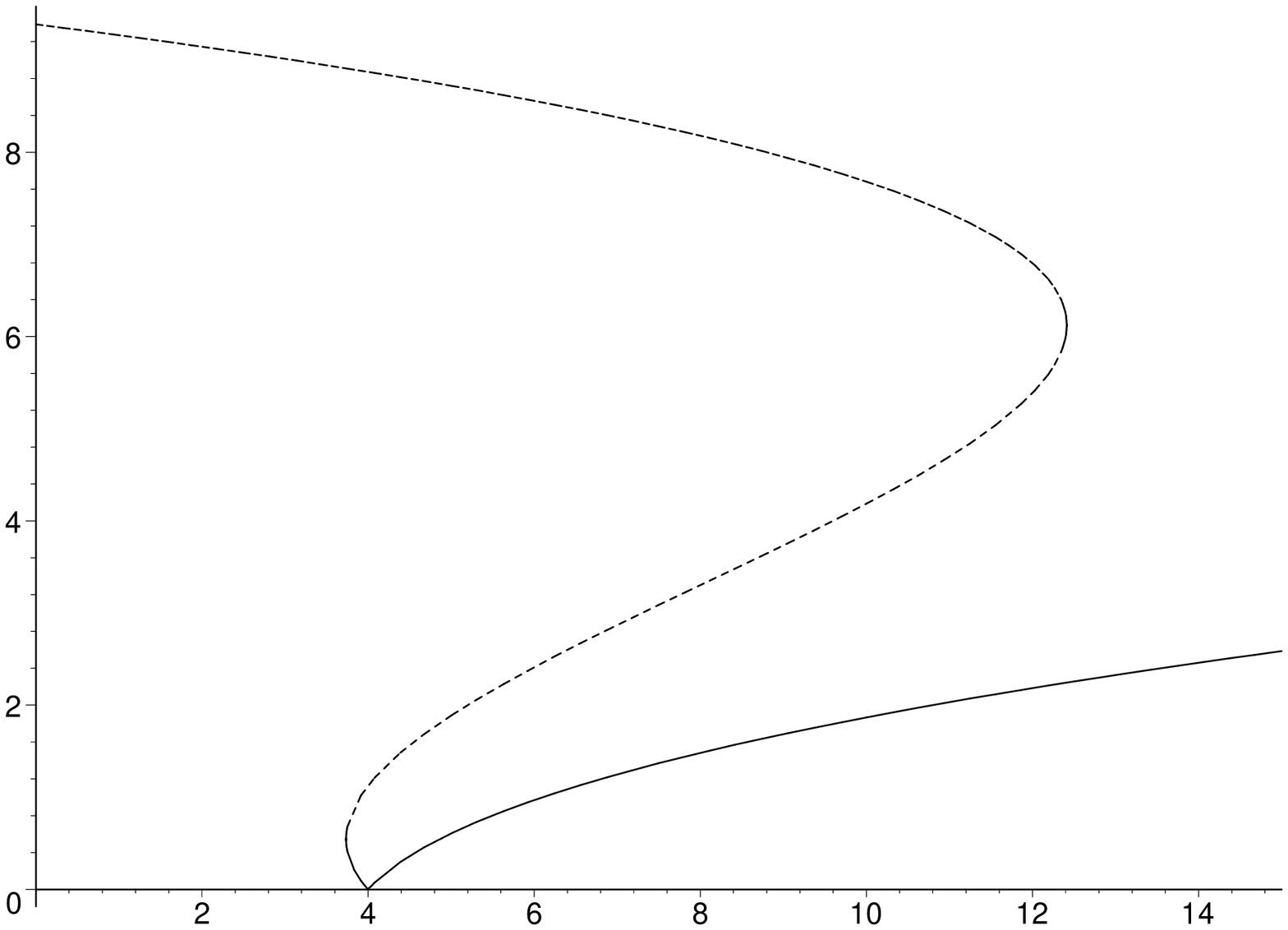}
\rput(-1.2,-2.6){\large$\mu_{\rm i},h_{\rm i}$}
\rput(-6,-6.6){\large$\mu_l$} \rput(-6.2,-5.6){\large $\mu_{\rm
e},h_{\rm e}$} \rput(-7.4,-0.8){\large$h_{\infty}$}
\rput{345}(-3.5,-1.5){\large $\leftarrow$}
\rput{25}(-3.5,-4.3){\large $\leftarrow$}
\rput{11}(-1.2,-4.7){\large $\leftarrow$}
\rput{300}(-5.5,-6){\tiny$\leftarrow$}\rput(-3.9,-6.6){\large
$\mu$} \rput(-7.4,-3.2){\large
$h$}\psline{->}(-5.8,-6.6)(-5.3,-6.2) \rput(-2,-4){\large GBIG 1}
\rput(-2,-1.6){\large GBIG 2}\rput(-1,-5.1){\large GBIG 3}
 \caption{Solutions of
the Friedmann equation ($h$ vs $\mu$) with negative brane tension
($\sigma=-4$) in a Minkowski bulk ($\phi=0$) with $\gamma=1/10$.
The curves are independent of the equation of state $w$. The
arrows indicate the direction of proper time on the brane.}
\label{Friedpl1}
\end{center}
\end{figure}

\begin{figure}
\begin{center}
\includegraphics[height=3in,width=2.75in,angle=270]{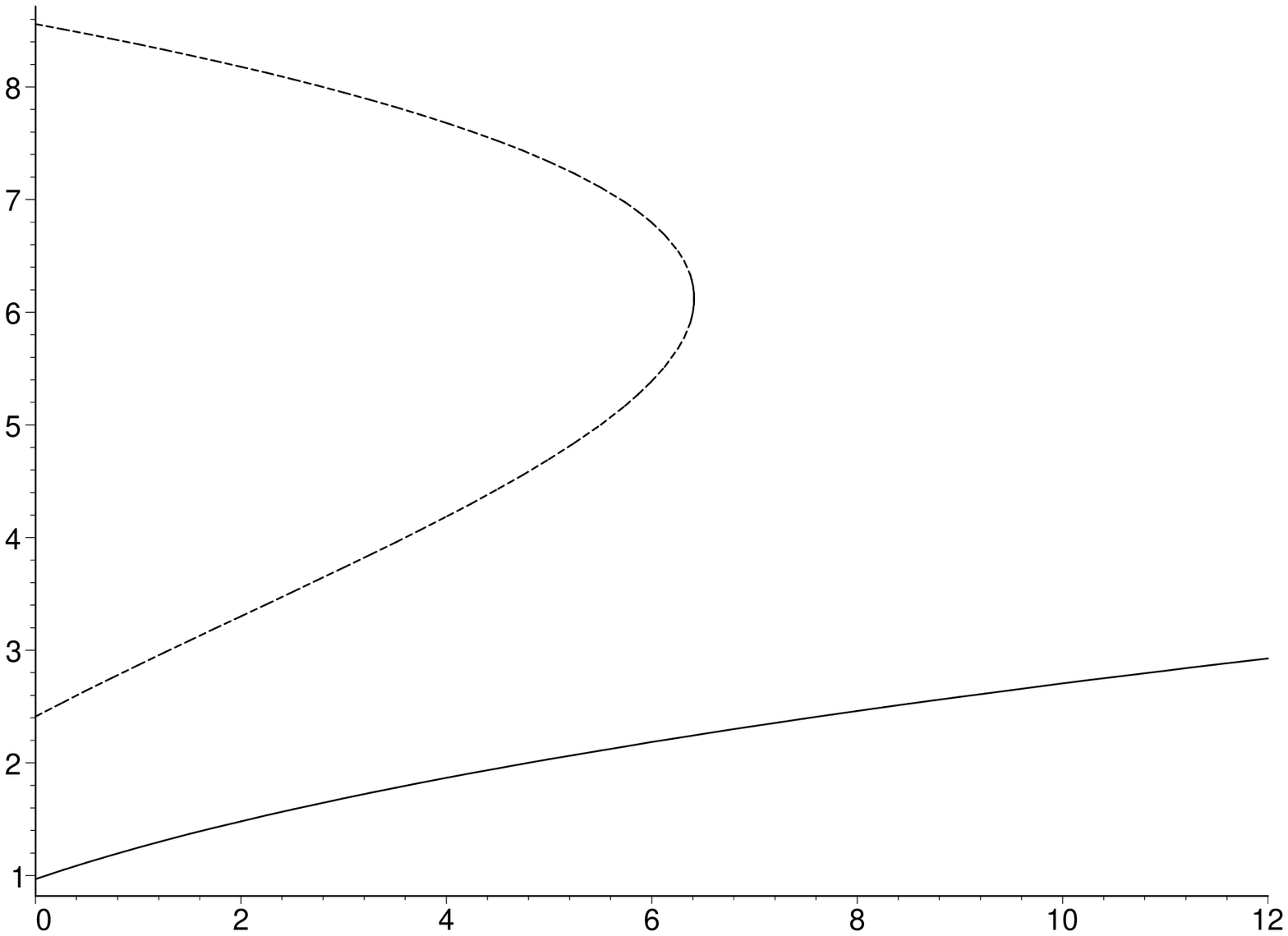}
\rput(-3,-2.5){\large$\mu_{\rm i},h_{\rm i}$}
\rput(-7.4,-0.8){\large $h_{\infty}$} \rput(-7.4,-5.2){\large
$h_{\infty}$}\rput(-7.4,-6.2){\large $h_{\infty}$}
\rput{347}(-5,-1.3){\large $\leftarrow$} \rput{20}(-5,-4.2){\large
$\leftarrow$} \rput{8}(-1.2,-4.7){\large
$\leftarrow$}\rput(-3.9,-6.8){\large $\mu$}
\rput(-7.4,-3.4){\large $h$}\rput(-3.7,-3.8){\large GBIG 1}
\rput(-3.2,-1.6){\large GBIG 2}\rput(-1,-5.2){\large GBIG 3}
 \caption{Solutions of
the Friedmann equation with positive brane tension ($\sigma=2$) in
a Minkowski bulk ($\phi=0$) with $\gamma=1/10$.} \label{Friedpl2}
\end{center}
\end{figure}

The model we considered in Chapter~\ref{GBIGB} is the Minkowski bulk
limit ($\chi=0$) of the $\phi_1$ case, which we now see to be
equivalent to $\chi=\frac{4}{3\gamma}$ in the $\phi_2$ case. The
results we present below are in terms of $\phi$ and thus include
both $\phi_1$ and $\phi_2$. We will consider the effect of including
brane tension in a Minkowski bulk before we look at the AdS bulk
cases. The Minkowski bulk case is given by $\phi=0$, since
Eqs.~(\ref{Lamb5}) and (\ref{Phi2}) imply $\Lambda_5=0$.
\subsection{Minkowski bulk ($\phi=0$) with brane tension}

In Figs.~\ref{Friedpl1} and~\ref{Friedpl2} we can see the solutions
to the Friedmann equation in a Minkowski bulk for both positive and
negative brane tensions. By introducing brane tension into the model
we are effectively adding a brane cosmological constant. These
solutions are therefore less desirable than the zero brane tension,
self-accelerating ones. For both negative and positive brane
tensions there are three solutions (this is not always true as we
will show later) denoted GBIG 1-3. There are four points of interest
in Figs.~\ref{Friedpl1} and~\ref{Friedpl2}, these are:
\begin{itemize}
\item ($\mu_{\rm i},h_{\rm i}$) and ($\mu_{\rm e},h_{\rm i}$): The initial
density for GBIG 1 and GBIG 2 ($\mu_{\rm i}$) and the final
density for GBIG 1 and 3 ($\mu_{\rm e}$), found by considering
$d\mu/d(h^2)=0$, are given by:

\begin{equation}\label{mui}
\mu_{\rm
i,e}=\frac{2-9\gamma\pm2(1-3\gamma)^{3/2}}{27\gamma^2}-\sigma,
\end{equation}
where the plus sign is for $\mu_{\rm i}$ and the negative sign is
for $\mu_{\rm e}$. The Hubble rates for these two densities are
given by:

\begin{equation}\label{Hie}
h_{\rm i,e}=\frac{1\pm\sqrt{1-3\gamma}}{3\gamma},
\end{equation}
where the sign convention is the same as above. The points
($\mu_{\rm i,e},h_{\rm i,e}$) have $h'=|\infty|$. Therefore
cosmologies that evolve to $\mu_{\rm e},~h_{\rm e}$ end in a
``quiescent'' (finite density) future singularity. This
singularity is of type 2 in the notation of
Ref.~\cite{Shtanov:2002ek}.

\item ($\mu_{ l},0$): This is the density at which GBIG 3 ``loiters''. The
point was found by considering $h=0$ and is given by :

\begin{equation}\label{muL}
\mu_{l}=-\sigma.
\end{equation}
In the case considered in Chapter~\ref{GBIGB} we had $\sigma=0$ so
$\mu_l=0$. We can show that for $\sigma<0$ GBIG 3 will not collapse
but will loiter at a density of $\mu=-\sigma$ before evolving
towards ($\mu_{\rm e},h_{\rm e}$), by considering $h'_{ l}$. At
$\mu=-\sigma$, $h=0$ we get $h'_{ l}=0$ from Eq.~(\ref{Ray2}). In a
standard expanding or collapsing cosmology $h'<0$ at all times. The
evolution for radiation and dust dominated universes can be seen in
Fig.~\ref{Phi1GBIG3Mink1}. A dust dominated universe loiters for
longer than the radiation dominated universe. The length of time a
universe will loiter for is also dependent on $\gamma$ in a
non-trivial way. This dependence is a matter for further
investigation.
\begin{figure}
\includegraphics[height=3in,width=2.75in,angle=270]{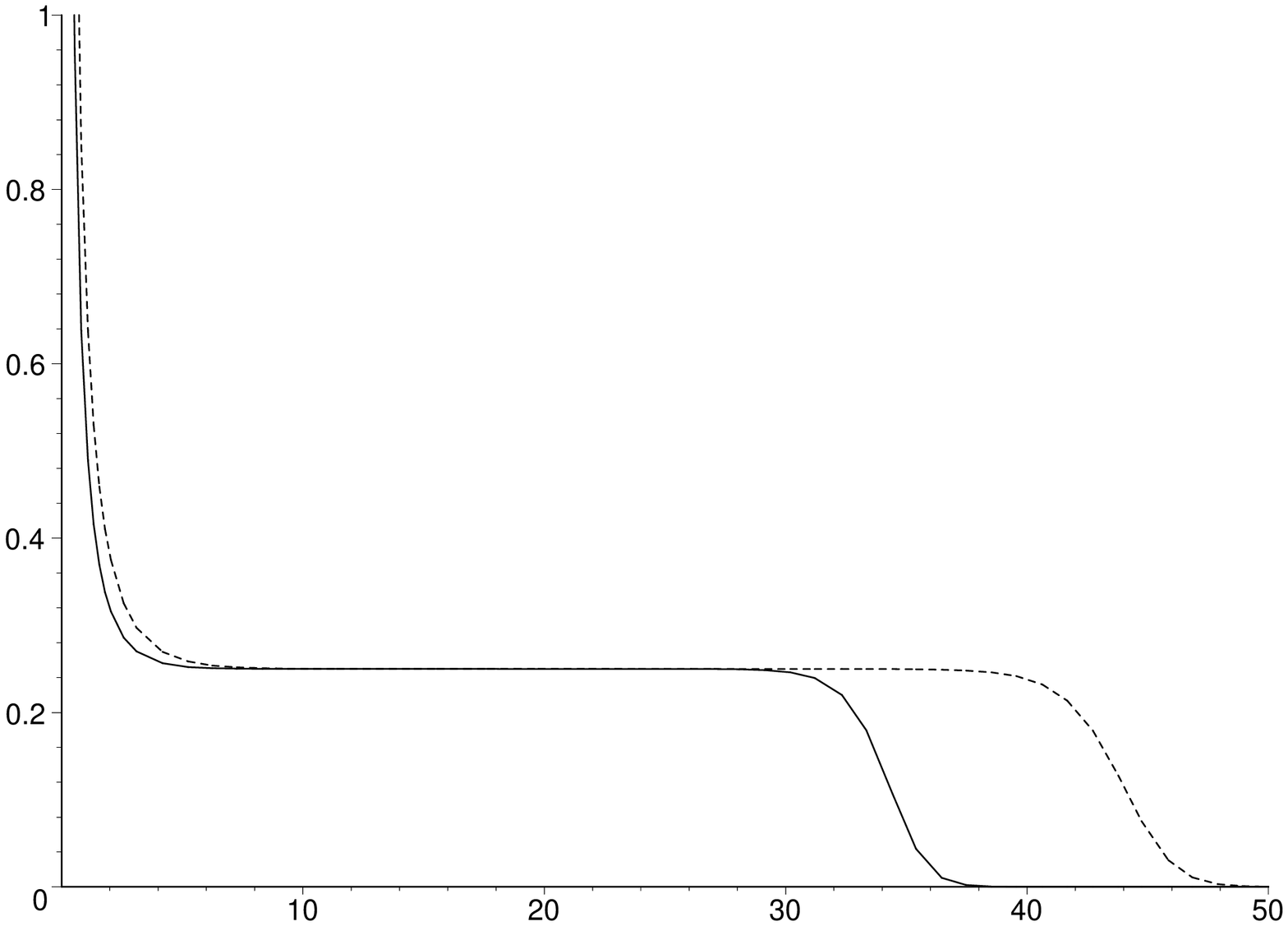}
\includegraphics[height=3in,width=2.75in,angle=270]{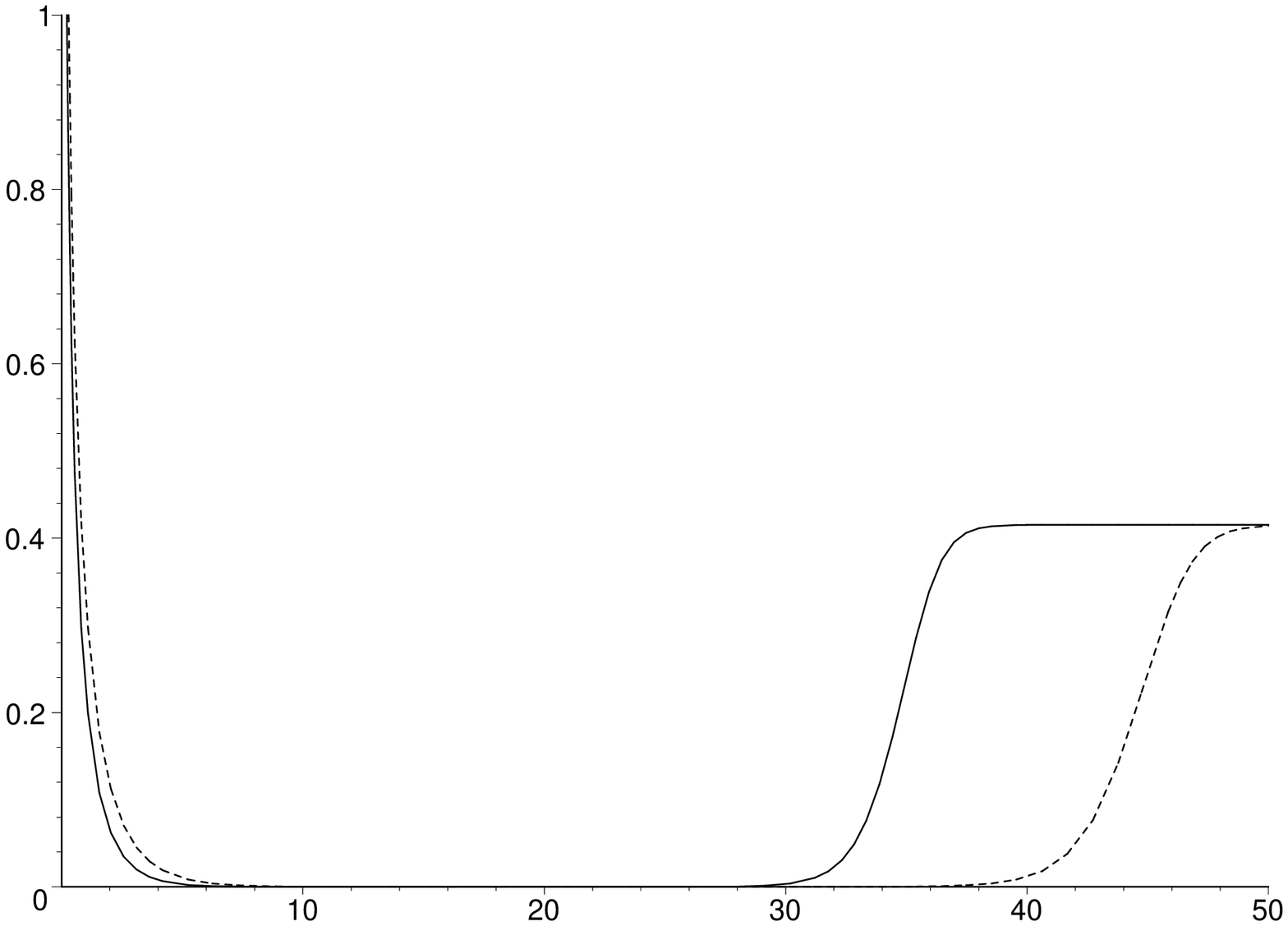}
\rput(0.4,3.5){\large $\mu$} \rput(3.8,0.2){\large
$\tau$}\rput(8,3.5){\large $h$}\rput(11.6,0.2){\large $\tau$}
 \caption{Plots of $\mu$ vs $\tau$ and
$h$ vs $\tau$ for GBIG 3 in a Minkowski bulk ($\phi=0$) with
negative brane tension ($\sigma=-1/4$). We see that GBIG 3 loiters
around $\mu=\mu_l=-\sigma$, $h=0$ before evolving towards a vacuum
de Sitter solution ($\mu_{\rm e}<0$). (Here $\gamma=1/10$. The solid
line is $w=1/3$. The dotted line is $w=0$.)} \label{Phi1GBIG3Mink1}
\end{figure}
In Ref.~\cite{Sahni:2004fb} they consider a loitering braneworld
model. An important difference between the two models is that in
Ref.~\cite{Sahni:2004fb} they require negative dark radiation,
i.e.~a naked singularity in the bulk or a de Sitter bulk. Also in
the model the Hubble rate at the loitering phase is exactly zero.
The time spent at this point is only dependent on $w$ and the
density of the loitering phase is only dependent on the brane
tension.

\item ($0,h_{\infty}$):  This is the asymptotic value of the Hubble rate
as $\mu\rightarrow 0$. For the value of the parameters used in
Fig.~\ref{Friedpl1}, GBIG 2 is the only case with this limit.  In
general any of the models can end in a similar state
(Fig.~\ref{Friedpl2}). The different values of $h_{\infty}$ are
given by the solutions to the cubic:
\begin{equation}\label{Phi1hin}
h_{\infty}^6+\frac{(2\gamma-1)}{\gamma^2}h_{\infty}^4+\frac{2\sigma}{\gamma^2}h_{\infty}^2
-\frac{\sigma^2}{\gamma^2}=0.
\end{equation}
In the case we considered in Chapter~\ref{GBIGB} ($\sigma=0$) the
last term in the above cubic is zero. Therefore we have
$h_{\infty}=0$ for GBIG 3 while the solutions for GBIG 1 and GBIG 2
are given by:

\begin{equation}\label{Phi100}
h_\infty=\frac{1\pm\sqrt{1-4\gamma}}{2\gamma},
\end{equation}
where the minus sign corresponds to GBIG 1 and the plus sign to
GBIG 2. This is the only case where we can write simple analytic
solutions as in all other cases we have to solve the cubic.
\end{itemize}

\begin{figure}
\begin{center}
\includegraphics[height=3in,width=2.75in,angle=270]{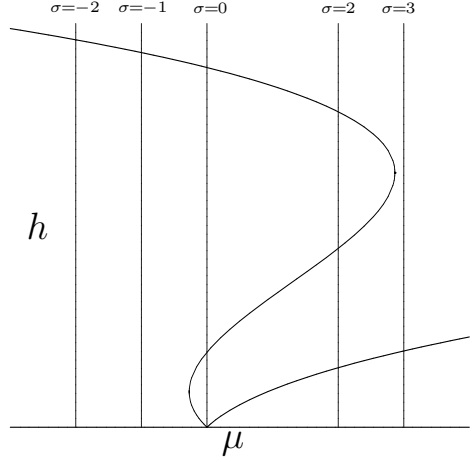}
\rput(-4,-6.4){\large $\mu$} \rput(-6.6,-3.5){\large $h$}
\rput(-6.1,-0.6){\tiny{$\sigma\!\!=\!\!-2$}}\rput(-5.2,-0.6){\tiny{$\sigma\!\!=\!\!-1$}}
\rput(-4.3,-0.6){\tiny{$\sigma\!\!=\!\!0$}}\rput(-2.6,-0.6){\tiny$\sigma\!\!=\!\!2$}
\rput(-1.8,-0.6){\tiny$\sigma\!\!=\!\!3$} \caption{Solutions of
the Friedmann equation ($h$ vs $\mu$) in a Minkowski bulk
($\phi=0$) with $\gamma=1/7$. The vertical lines represent the
$\mu=0$ axis for the labeled brane tensions.} \label{Phi1nszc}
\end{center}
\end{figure}
The two Hubble rates, $h_{\rm i}$ and $h_{\rm e}$, are independent
of the brane tension which simply shifts the $\mu=0$ axis.
Therefore $\mu_{\rm e}$ only corresponds to a physical (positive)
energy density when $\sigma<\sigma_{\rm e}<0$, see
Eq.~(\ref{sil}). The effect of the brane tension on the $\mu=0$
axis is illustrated in Fig.~\ref{Phi1nszc}.

\begin{figure}
\begin{center}
\includegraphics[height=3in,width=2.75in,angle=270]{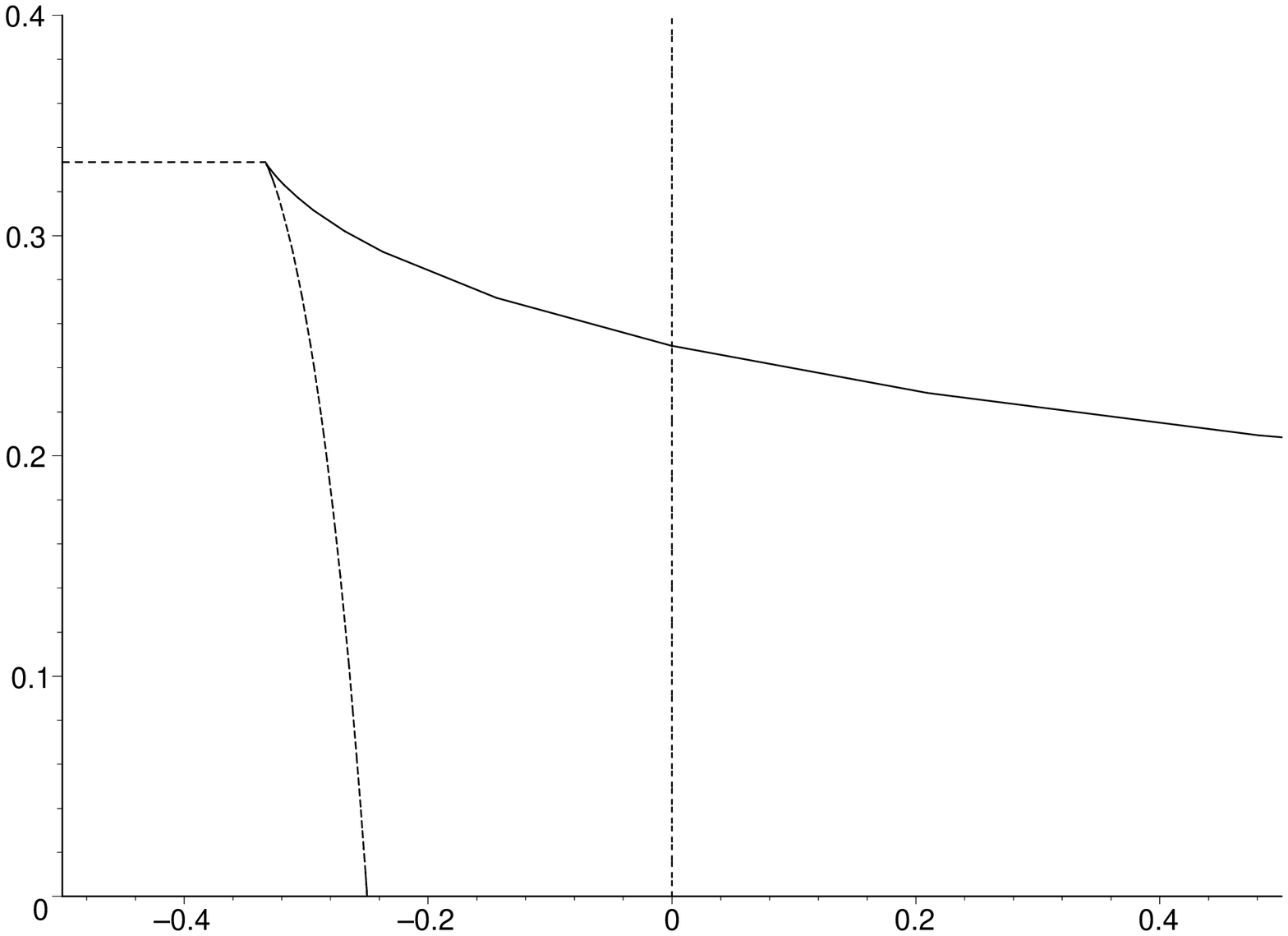}
\rput(-4.8,-1.25){\large I}\rput(-1.8,-1.25){\large II}
\rput(-6.2,-4.5){\large III} \rput(-4.8,-4.5){\large IV}
\rput(-1.8,-4.5){\large V}
 \rput(-6.3,-3){\large $\sigma_{\rm
e}(\gamma)$} \rput(-1,-3){\large $\gamma_{\rm
i}(\sigma)$}\rput(-3.7,-6){\large $\sigma_l$}
\rput(-6.4,-1.4){\large $\gamma_{\rm m}$}\rput(-7.4,-3.5){\large
$\gamma$}\rput(-3.9,-6.7){\large $\sigma$} \caption{The
$(\sigma,\gamma)$ plane for solutions in a Minkowski ($\phi=0$)
bulk. The short dotted horizontal line is $\gamma_{\rm m}=1/3$.
GBIG 1-2 exist with positive energy density in regions III, IV and
V.} \label{gs0}
\end{center}
\end{figure}

We showed in Chapter~\ref{GBIGB} that for $\phi=0=\sigma$ we require
$\gamma\leq1/4$ for GBIG 1 and GBIG 2 solutions to exist within the
positive energy density region (if $\gamma=1/4$ GBIG 1 and GBIG 2
reduce to the same vacuum de Sitter universe). If we have some
non-zero brane tension this constraint is modified. The maximum
value of $\gamma$, for GBIG 1 and GBIG 2 to exist with positive
energy density, as a function of $\sigma$ can be seen in
Fig.~\ref{gs0}. We have defined new quantities, $\sigma_{\rm e,i}$
and $\sigma_l$, for which $\mu_{\rm e,i,l}=0$. We can see from
Eq.~(\ref{muL}) that $\sigma_l=0$. $\sigma_{\rm e,i}$ are given in
terms of $\gamma$ by:

\begin{equation}\label{sil}
\sigma_{\rm
i,e}=\frac{2-9\gamma\pm2(1-3\gamma)^{3/2}}{27\gamma^2}.
\end{equation}

The maximum value of $\gamma$ ($\gamma_{\rm m}$) for GBIG 1 and GBIG
2 to exist (i.e. for Eqs.~(\ref{mui}) and (\ref{Hie}) to have real
solutions) is $\gamma_{\rm m}=1/3$ . Actually for $\gamma=1/3$, GBIG
2 exists but GBIG 1 is lost. This is since $h_{\rm i}=h_{\rm e}$ at
$\gamma=1/3$. The point $h_{\rm i}=h_{\rm e}$ is now a point of
inflection and Eq.~(\ref{Hie}) is still valid and therefore
$|h'|=\infty$. For values of $\gamma>1/3$, $h_{\rm i,e}$ and
$\mu_{\rm i,e}$ become complex, it is no longer a point of
inflection. Therefore $|h'|\neq\infty$ and GBIG 3 can continue its
evolution through this point to $h_{\infty}$.
\begin{figure}
\includegraphics[height=3in,width=2.75in,angle=270]{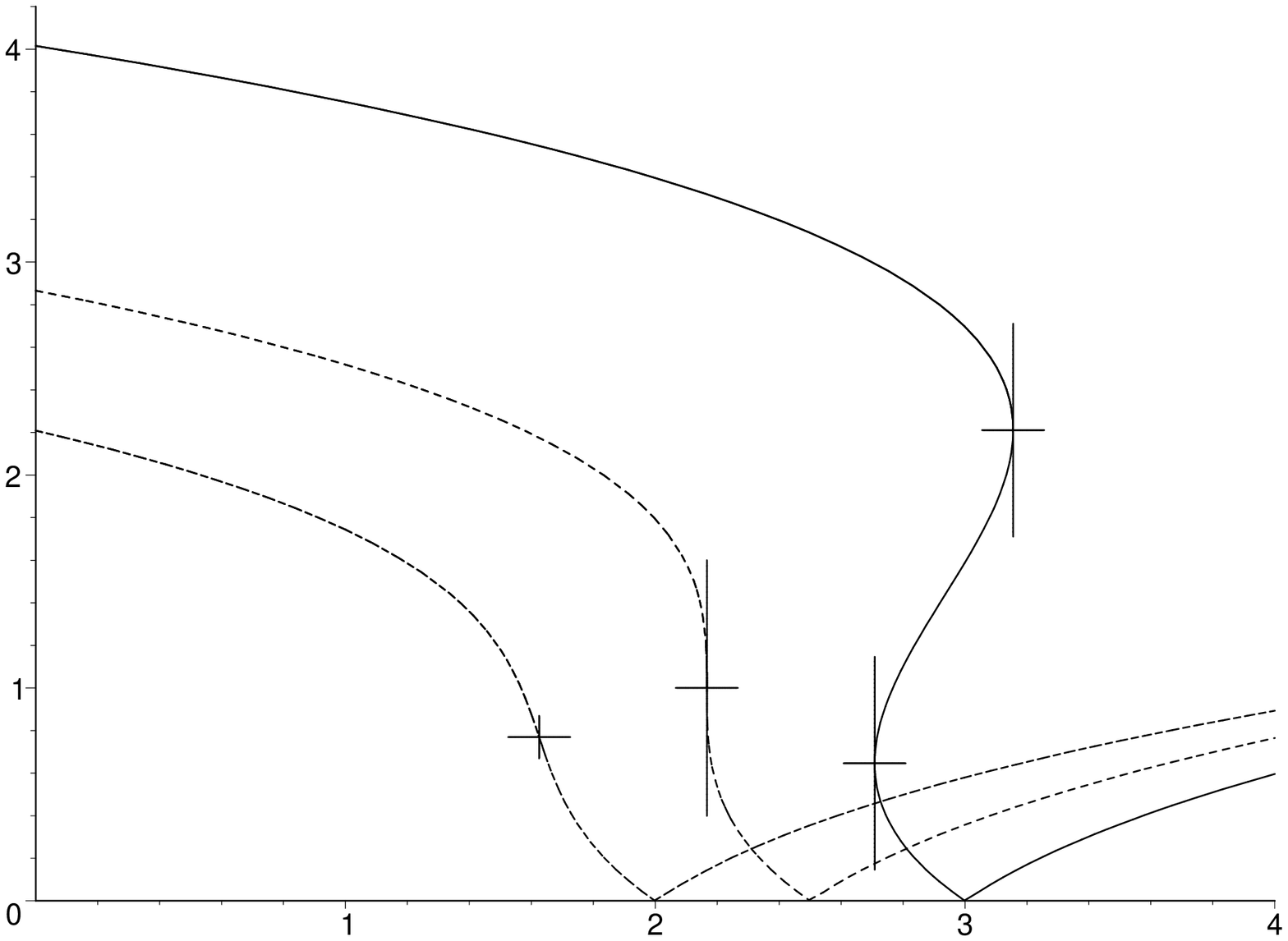}
\includegraphics[height=3in,width=2.75in,angle=270]{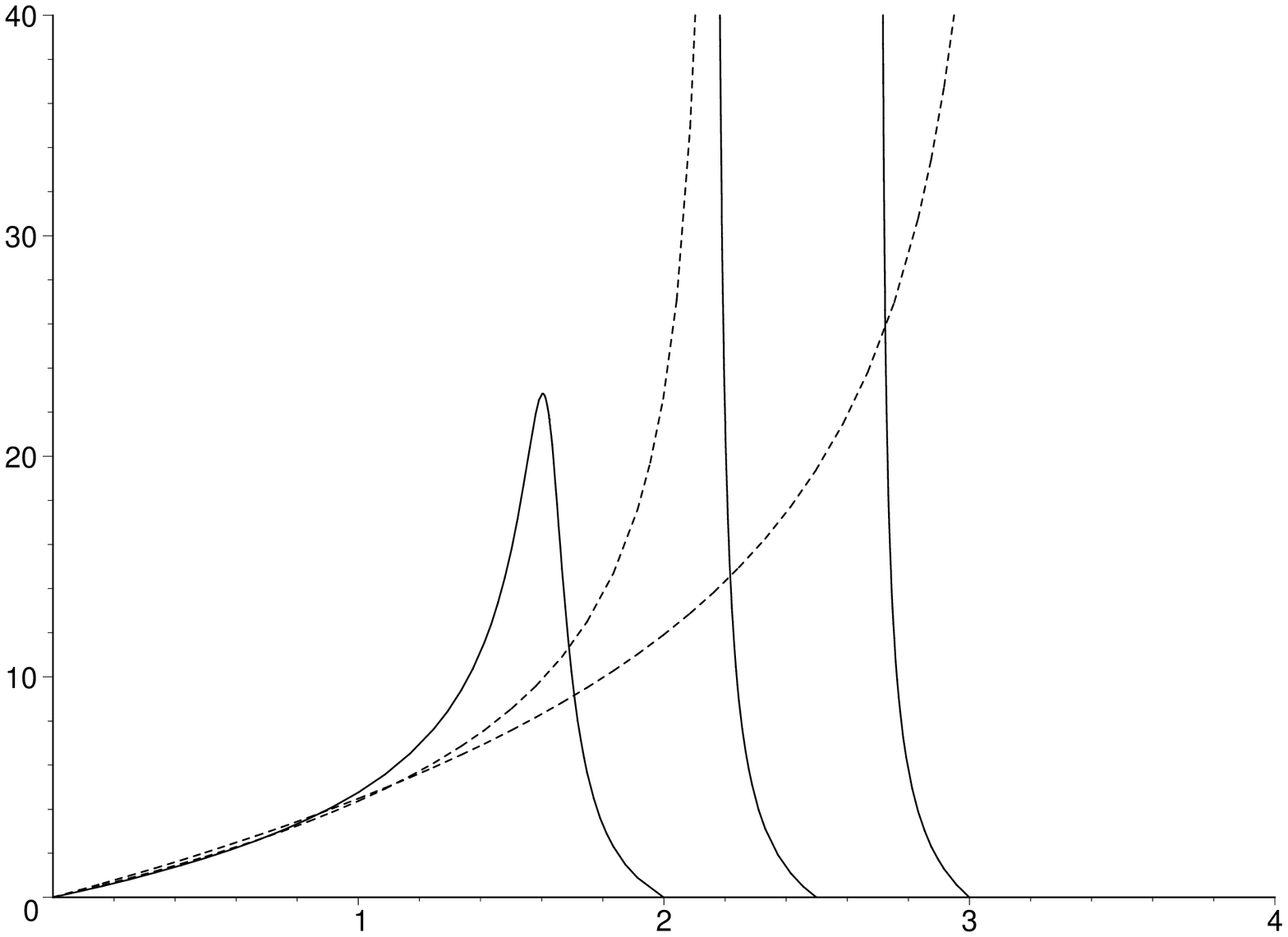}
\rput(12.8,6.6){\large $1$}\rput(11.8,6.6){\large $2$}
\rput(10.9,4.2){\large $3$} \rput(3,2.7){\large $3$}
\rput(3,4){\large $2$}
 \rput(3,5.8){\large $1$} \rput(0.4,4){\large $h$}
 \rput(8,4){\large $h'$}
 \rput(3.9,0.3){\large $\mu$}\rput(11.8,0.3){\large $\mu$}
\caption{The plot on the left is $h$ vs $\mu$ for three different
solutions, all with $\phi=0$. The plot on the right shows $h'$ vs
$\mu$ for the same solutions. ($1$): $\gamma=1/3-0.1,~\sigma=-3$.
($2$): $\gamma=1/3,~\sigma=-2.5$. ($3$):
$\gamma=1/3+0.1,~\sigma=-2$. Different brane tensions are used
purely for clarity.} \label{mhg12}
\end{figure}
In Fig.~\ref{mhg12} we show results for $h$ and $h'$ for three
values of $\gamma$, $\gamma=1/3-0.01,~1/3,~1/3+0.01$. In the top
plot we see that the solution denoted (1) has GBIG 1-3 present as
$\gamma<1/3$, we have two real and different values for $h_{\rm i}$
and $h_{\rm e}$. This solution in the bottom plot has three parts
(GBIG 1 has negative values not shown in Fig.~\ref{mhg12}), GBIG 2
is the dotted solution and the solid is GBIG 3. As ($1$) approaches
$h_{\rm i,e}$ $|h'|\rightarrow\infty$. Solution (2) has
$\gamma=1/3$, in the top plot we see that GBIG 1 now no longer
exists as $h_{\rm i}=h_{\rm e}$. In the bottom plot GBIG 3 (solid)
and GBIG 2 (dotted) solutions both go to $h'=\infty$ at the point
$h_{\rm i}=h_{\rm e}$. Solution (3) has $\gamma>1/3$. Now $h_{\rm
i,e}$ does not represent $d\mu/dh=0$, therefore $h'$ stays well
behaved throughout the evolution (bottom plot). GBIG 3 moves
seamlessly onto GBIG 2 making a single solution. As GBIG 2
super-accelerates this new solution will show late time phantom-like
behaviour ($h'>0\rightarrow w<-1$).

Solutions that lie along the $\sigma_l$ line in Fig.~\ref{gs0} are
those considered in Chapter~\ref{GBIGB}. Regions I and II in
Fig.~\ref{gs0} extend up to $\gamma_{\rm M}=\infty$. Solutions in
each region of Fig.~\ref{gs0} have the following properties:

\begin{itemize}
\item $I:~\sigma_{\rm e}(\gamma_m)<\sigma<0,~\gamma>\gamma_{\rm i}$ and
$\sigma<\sigma_{\rm e}(\gamma_m),~\gamma>\gamma_{\rm m}$. GBIG 1
and GBIG 2 do not exist. GBIG 3 loiters at $\mu_l$ before evolving
to a vacuum de Sitter universe.

\item $II:~\sigma>0,~\gamma>\gamma_{\rm i}$. GBIG 1 and GBIG 2 do not exist.
GBIG 3 evolves to a vacuum de Sitter universe.

\item $III:~\sigma\leq\sigma_{\rm e},~\gamma\leq\gamma_{\rm m}$. GBIG 1 will evolve to
 ($\mu_{\rm e},h_{\rm e}$). GBIG 2 evolves to ($0,h_{\infty}$). GBIG 3 loiters at
$\mu_l$ before evolving towards ($\mu_{\rm e},h_{\rm e}$). When
$\sigma=\sigma_{\rm e}$ GBIG 1 and 3 evolve to ($0,h_{\rm e}$). When
$\gamma=\gamma_{\rm m}$, GBIG 1 ceases to exist ($h_{\rm i}=h_{\rm
e}$).

\item $IV:~\sigma_{\rm e}<\sigma<0,~\gamma\leq\gamma_{\rm i}$.
GBIG 1 and GBIG 2 both evolve to vacuum de Sitter states. GBIG 3
loiters before evolving to a vacuum de Sitter universe. Each
vacuum de Sitter state has a different value of $h_{\infty}$. When
$\gamma=\gamma_{\rm i}$ GBIG 1 and GBIG 2 live at ($0,h_{\rm i}$).

\item $V:~\sigma\geq0,~\gamma\leq\gamma_{\rm i}$. GBIG 1-3 all
end in vacuum de Sitter states. With $\gamma=\gamma_{\rm i}$ GBIG
1-2 live at ($0,h_{\rm i}$). When $\sigma=0$ GBIG 3 ends in a
Minkowski state.
\end{itemize}

\begin{figure}
\begin{center}
\includegraphics[height=3in,width=2.75in,angle=270]{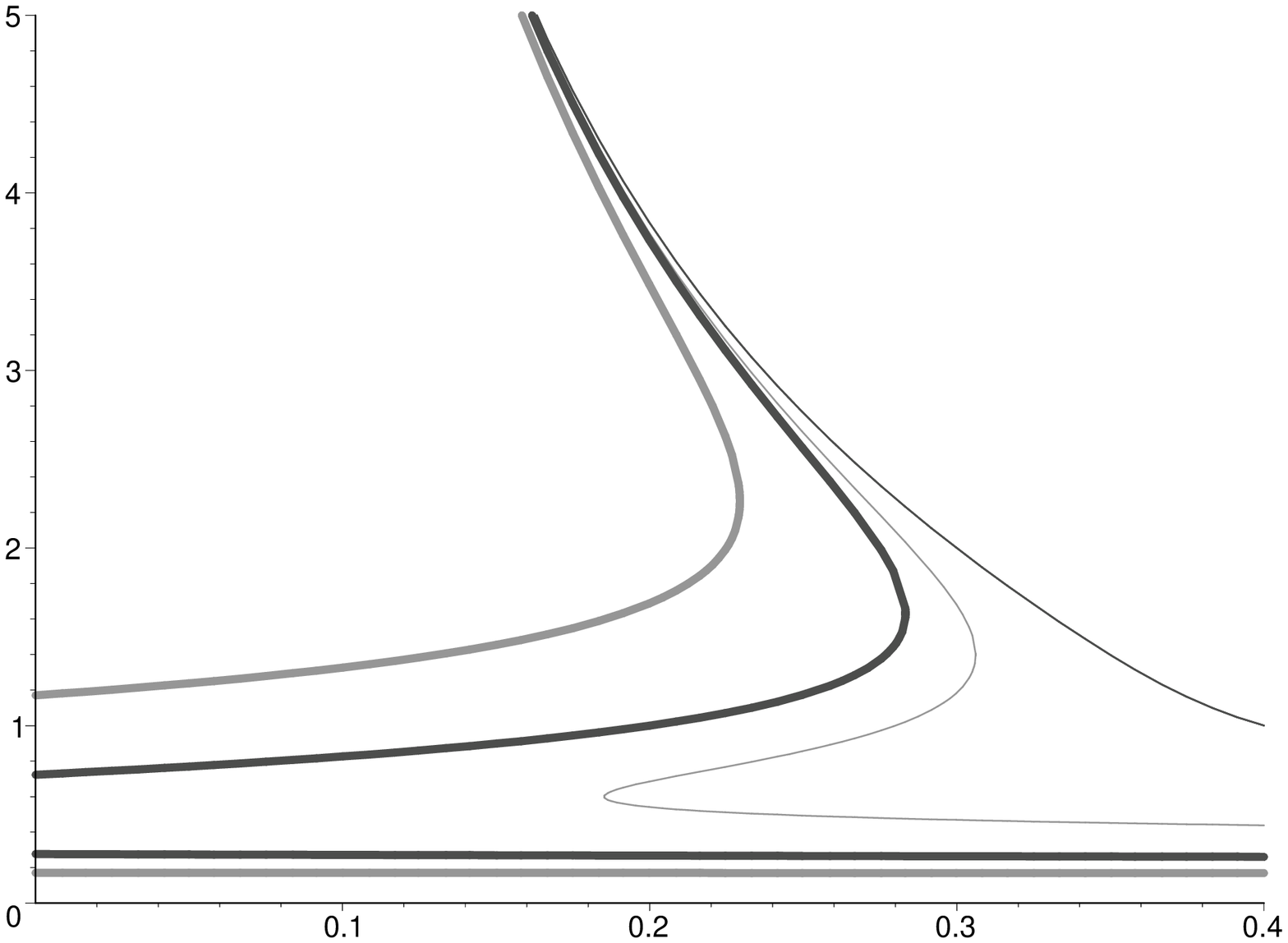}
\rput(-7.4,-3.5){\large $h_{\infty}$}\rput(-3.9,-6.7){\large
$\gamma$}
 \caption{$h_{\infty}$ for solutions in a Minkowski ($\phi=0$) bulk.
 The thin-dark line has $\sigma=-0.4$, thin-light line has $\sigma=-0.28$,
 thick-dark lines have $\sigma=-0.2$ and the thick-light lines have $\sigma=0.2$.}
\label{hinp0} \end{center}
\end{figure}
In Fig.~\ref{hinp0} we can see how solutions in each of the
regions mentioned above affect $h_{\infty}$. The thin-dark line
($\sigma=-0.4$) lies in III and I in Fig.~\ref{gs0}. Only GBIG 2
exists for $0\leq\gamma<1/3$. For $\gamma>1/3$ we've seen that
GBIG 3 and GBIG 2 connect to make a single solution, which is why
the line is continuous through $\gamma=1/3$. The thin-light line
($\sigma=-0.28$) has an interesting feature due to the solutions
lying on a line that cuts through both region III and IV as well
as I. Both the thick lines have GBIG 1-3 ending at $h_{\infty}$.
The thick-dark solution loiters before evolving to $h_{\infty}$,
but this has no effect upon the results in Fig.~\ref{hinp0}.

\subsection{AdS bulk ($\phi\neq0$) with brane tension}

When $\phi=0$, we have $\Lambda_5\neq0$, with one exception: the
$\phi_2$ solution with $\chi=0$, i.e. $\phi_2=-4/3\gamma$, has
$\Lambda_5=0$, but the bulk is AdS. For $\chi>0$, $\Lambda_5\neq0$.
Thus in all cases, $\phi\neq0$ implies an AdS bulk.

\begin{figure}
\begin{center}
\includegraphics[height=3in,width=2.75in,angle=270]{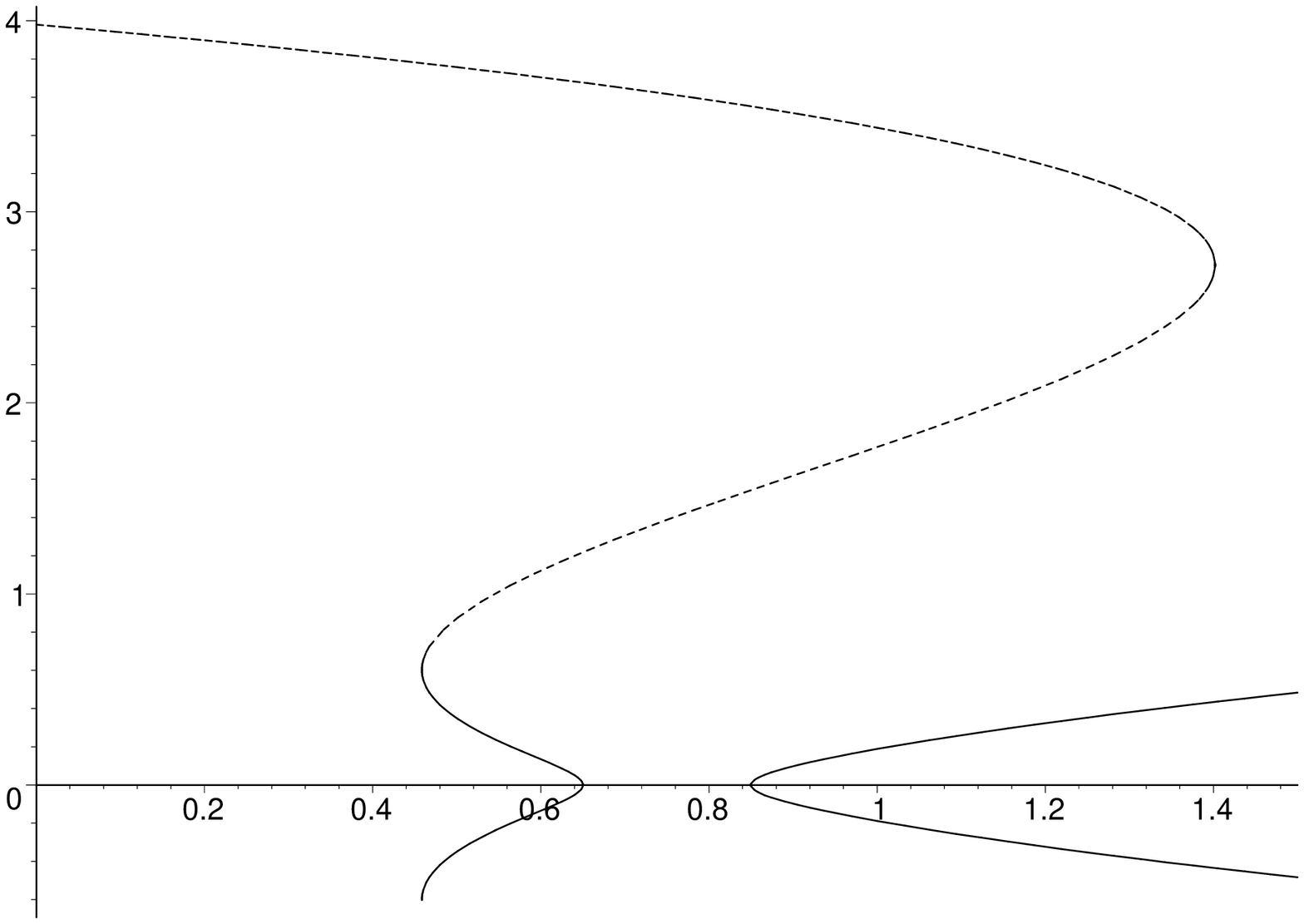}
\rput(-0.6,-2.2){\large$\mu_{\rm i},h_{\rm i}$}
\rput(-4.2,-5){\large$\mu_{\rm b}$} \rput(-3.6,-5){\large$\mu_{\rm
c}$}
 \rput(-5.7,-4.7){\large
$\mu_{\rm e},h_{\rm e}$} \rput(-7.4,-0.5){\large$h_{\infty}$}
\rput(-7.4,-3){\large $h$}\rput(-3.8,-5.9){\large $\mu$}
\rput{350}(-3.5,-1.2){\large $\leftarrow$}
\rput{20}(-3.5,-3.8){\large $\leftarrow$}
\rput{11}(-1.2,-4.9){\large $\leftarrow$}
\rput{325}(-5,-5.2){\large$\leftarrow$}
\rput{343}(-1.2,-6.1){\large $\rightarrow$}
\rput{33}(-5,-5.9){\large$\rightarrow$} \rput(-2,-3.6){\large GBIG
1} \rput(-2,-1.2){\large GBIG 2}\rput(-1,-4.5){\large GBIG
3}\rput(-5,-6.4){\large GBIG 4}\caption{Solutions of the Friedmann
equation ($h$ vs $\mu$) with negative brane tension
($\sigma=-3/4$) in an AdS bulk ($\phi=-0.01$) with $\gamma=1/5$.
The curves are independent of the equation of state $w$. The
arrows indicate the direction of proper time on the brane.}
\label{Friedpl1ds}
\end{center}
\end{figure}

When we allow the bulk to be warped ($\phi\neq0$) we open up
another possible solution, denoted GBIG 4. There is a maximum
value of $\phi$ for which GBIG 4 can exist as we shall see later.
The nature of GBIG 3 is also changed. These solutions, for a
negative brane tension, can be seen in Fig.~\ref{Friedpl1ds}.
Brane tension affects the solutions in the same manner as in
$\phi=0$ case.
The qualitative effect of the warped bulk is to make GBIG 3
collapse and to introduce the new bouncing branch GBIG 4. This is
due to $h=0$ now giving two solutions:

\begin{equation}\label{mucP1}
\mu_{\rm
c,b}=\pm\sqrt{-\phi}\left(1+\frac{\gamma\phi}{2}\right)-\sigma,
\end{equation}
where the plus sign is for the GBIG 3 collapse ($\mu_{\rm c}$) and
the minus for the GBIG 4 bounce ($\mu_{\rm b}$). Effectively the
loitering point in the Minkowski bulk is split into the max/min
densities of the bouncing/collapsing cosmologies of GBIG 3-4. The
other points in Fig.~\ref{Friedpl1ds} are modified by the warped
bulk; they are now given by:

\begin{equation}\label{muim}
\mu_{\rm
i,e}=\frac{4-18\gamma+27\gamma^2\phi\pm\sqrt{2}(2-6\gamma-9\gamma^2\phi)^{3/2}}{54\gamma^2}
-\sigma,
\end{equation}
with the plus sign for $\mu_{\rm i}$ and the negative sign for
$\mu_{\rm e}$. The Hubble rates $h_{\rm i,e}$ are given by:

\begin{equation}\label{Hiem}
h_{\rm
i,e}=\frac{\sqrt{2}}{6\gamma}\sqrt{4-6\gamma+9\gamma^2\phi\pm2\sqrt{4-12\gamma-18\gamma^2\phi}},
\end{equation}
with the same sign convention. $h_{\infty}$ is given by a
similarly modified equation:

\begin{eqnarray}\label{Phi1hinm}
h_{\infty}^6&+&\frac{(2\gamma-1)}{\gamma^2}h_{\infty}^4+\frac{\left[1+2\sigma-
\phi\gamma\left(1+\frac{3}{4}\phi\gamma\right)\right]}{\gamma^2}h_{\infty}^2\nonumber\\
&-&\frac{\left[\phi\left(1+\frac{\phi\gamma}{2}\right)^2+\sigma^2\right]}{\gamma^2}=0.
\end{eqnarray}

\begin{figure}
\begin{center}
\includegraphics[height=3in,width=2.75in,angle=270]{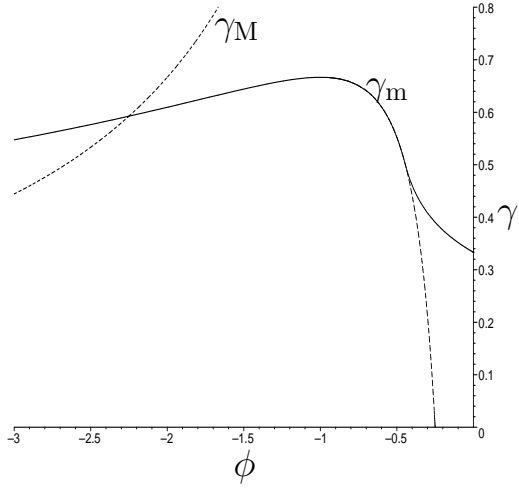}
\rput(-3.9,-6.8){\large $\phi$}\rput(-0.4,-3.5){\large $\gamma$}
\rput(-4,-1){\large $\gamma_{\rm M}$} \rput(-2,-1.8){\large
$\gamma_{\rm m}$}
 \caption{In order for $h_{\rm i}$ to be real $\gamma$
must lie beneath the solid line ($\gamma_{\rm m}$). The region to
the right of the vertical dotted line is where GBIG 4 ($h_{\rm
e}>0$) is allowed (except where $\phi=0$). }\label{gpbound}
\end{center}
\end{figure}

With a warped bulk the constraint from $h_{\rm i}$ ($\gamma\leq1/3$
in the Minkowski case) is modified. In the equation for $h_{\rm i}$
we effectively have two bounds from the two square root terms. In
the $\phi=0$ case only the inner term is of any consequence (giving
rise to the bound $\gamma\leq1/3$). When $\phi\neq0$ there are two
bounds which are applicable in different regimes. If we consider the
bound from the inner square root we get:

\begin{equation}\label{gbound}
\gamma\leq\frac{\sqrt{1+2\phi}-1}{3\phi}.
\end{equation}
This is valid for $\phi\geq-1/2$. Considering the outer square
root term we get the bound:

\begin{equation}\label{gbound2}
\gamma\leq\frac{2\left(1-2\sqrt{-\phi}\right)}{3\phi}.
\end{equation}
This bound becomes negative when $\phi>-1/4$ which is disallowed as
this prevents self-acceleration. The bound on $\gamma$ is given by
the lower of the two constraints when $-1/2<\phi<-1/4$ i.e. when in
the range where both exists. Therefore the bound on $\gamma$ for
$h_{\rm i}$ to be real is given by (see Fig.~\ref{gpbound}):

\begin{equation}\label{gboundT}
\gamma\leq\left\{ \begin{array}{lc}
 \frac{\sqrt{1+2\phi}-1}{3\phi}  &-\frac{4}{9}\leq\phi\leq0\\\\
 \frac{2\left(1-2\sqrt{-\phi}\right)}{3\phi}  &\phi\leq-\frac{4}{9}\\
  \end{array}
 \right.
\end{equation}
For a particular value of $\phi$ the maximum value of $\gamma$
allowed by this constraint is denoted $\gamma_{\rm m}$. When
$\phi<-9/4$ the constraint in Eq.~(\ref{acon2}) is tighter than
that in Eq.~(\ref{gbound2}) ($\gamma_{\rm M}<\gamma_{\rm m}$).
This means that for $\phi\leq-9/4$, $h_{\rm i}$ is always real for
allowed values of $\gamma$.

There is a bound for GBIG 4 to exist, found by considering $h_{\rm
e }=0$. This bound is given by:

\begin{equation}\label{chib}
\phi>-\frac{2}{9\gamma^2}\left\{4-3\gamma-2\sqrt{4-6\gamma}\right\}.
\end{equation}
This is a solution to the quadratic obtained from $h_{\rm e}=0$, the
other root of the quadratic does not obey the constraint $\phi\leq
-4/3\gamma$ so it is ignored. The minimum value for GBIG 4 to exist
will be denoted $\phi_{GBIG4Lim}$. Using the constraints we can
split the $\gamma,~\phi$ plane into three sections, see
Fig.~\ref{gpbound}. The solid line is constructed from the bounds in
Eq.~(\ref{gboundT}). For $h_{\rm i}$ to be real we must choose
values below this line. The area to the right of the vertical dotted
line (but excluding $\phi=0$) allows GBIG 4. Points to the left of
this line have GBIG 1 collapsing after a minimum energy density
($\mu_{\rm b}$) is reached. The dotted curve on the left comes from
initial constraint $\gamma\leq -4/3\phi$.

In order to consider the $\sigma,~\gamma$ plane as we did in the
$\phi=0$ Minkowski case, we need the modified $\sigma_{\rm i,e}$
equations:

\begin{equation}\label{silm}
\sigma_{\rm
i,e}=\frac{4-18\gamma+27\gamma^2\phi\pm\sqrt{2}(2-6\gamma-9\gamma^2\phi)^{3/2}}{54\gamma^2}
.
\end{equation}
We now have two more formulas for the collapse density of GBIG 3
and the bounce density of GBIG 4:

\begin{equation}\label{sicP1m}
\sigma_{\rm
c,b}=\pm\sqrt{-\phi}\left(1+\frac{\gamma\phi}{2}\right),
\end{equation}
with the plus sign corresponding to the collapse and the minus to
the bounce.

We shall consider the $\sigma,~\gamma$ plane for three different
values of $\phi$ corresponding to three distinct regions in
Fig.~\ref{gpbound}. We shall first consider $\phi=-0.1$.

\subsubsection{Typical example $\phi=-0.1$}

\begin{figure}
\begin{center}
\includegraphics[height=3in,width=2.75in,angle=270]{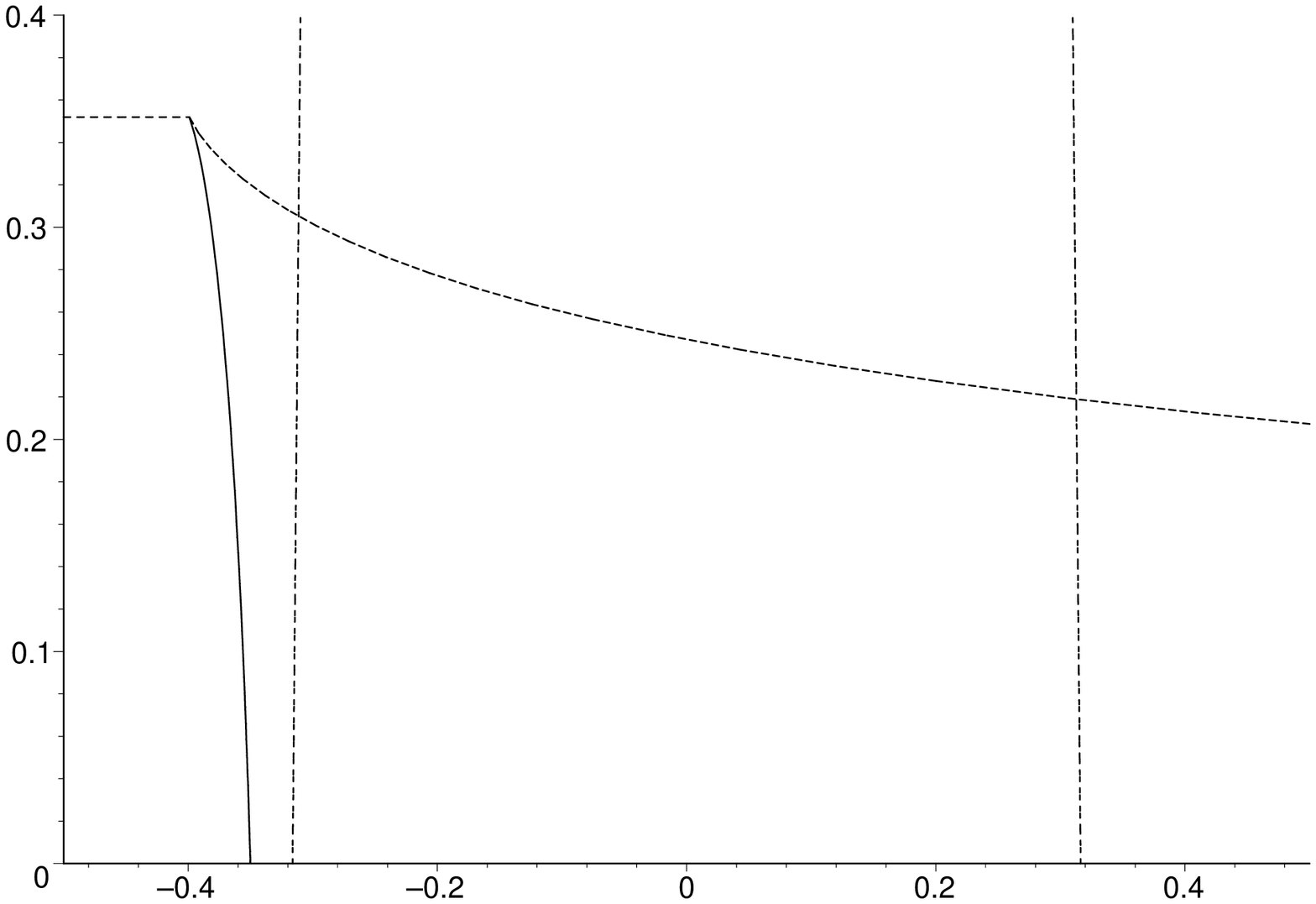}
\rput(-7.4,-3.6){\large $\gamma$}\rput(-3.9,-6.7){\large $\sigma$}
\psline{->}(-7.9,-2.8)(-6.3,-3)\rput(-6.5,-1.25){\large
I}\rput(-4,-1.25){\large II} \rput(-1,-1.25){\large III}
\rput(-6.6,-4.5){\large IV} \rput(-6,-2.5){\small
V}\rput(-4,-4.5){\large VI}\rput(-1,-4.5){\large VII}
 \rput(-8,-2.5){\large $\sigma_{\rm
e}(\gamma)$} \rput(-1,-3.2){\large $\gamma_{\rm
i}(\sigma)$}\rput(-2.55,-6){\large $\sigma_{\rm
c}(\gamma)$}\rput(-5.2,-6){\large $\sigma_{\rm
b}(\gamma)$}\rput(-7.3,-1.6){\large $\gamma_{\rm m}$} \caption{The
$(\sigma,\gamma)$ plane for solutions in a AdS ($\phi=-0.1$) bulk.
The short dotted horizontal line is the maximum value of $\gamma$
as obtained from Eq.~(\ref{acon2}).} \label{gs12}
\end{center}
\end{figure}

If we take $\phi=-0.1$, we are in the region where GBIG 4 is
allowed. The $\sigma,~\gamma$ plane can be seen in
Fig.~\ref{gs12}. The regions I, II and III in Fig.~\ref{gs12}
extend up to $\gamma_{\rm M}=40/3$.

In each region we have the following cosmologies.
\begin{itemize}
\item
$I:~\sigma_{\rm e}(\gamma_m)<\sigma<0,~\gamma>\gamma_{\rm i}$ and
$\sigma\leq\sigma_{\rm e}(\gamma_m),~\gamma>\gamma_{\rm m}$. GBIG
1-2 do not exist. GBIG 3 reaches a minimum energy density ($\mu_{\rm
c}$) and then evolves back to $\mu=\infty$. GBIG 4 starts at
($0,-h_{\infty}$) then bounces at ($\mu_{\rm b},0$) before evolving
to ($0,+h_{\infty}$). When $\sigma=\sigma_{\rm b}$, GBIG 4 exists as
a Minkowski universe.

\item $II:~\sigma_{\rm b}\leq\sigma<\sigma_{\rm c},~ \gamma>\gamma_{\rm
i}$. Only GBIG 3 exists. GBIG 3 reaches a minimum energy density
($\mu_{\rm c}$) and then evolves back to $\mu=\infty$. When
$\sigma=\sigma_{\rm c}$ GBIG 3 ends in a Minkowski universe.

\item $III:~\sigma>\sigma_{\rm c},~ \gamma>\gamma_{\rm
i}$.  Only GBIG 3 exists and evolves to a vacuum de Sitter universe.

\item $IV:~\sigma\leq\sigma_{\rm e}<0,~\gamma\leq\gamma_{\rm m}$. GBIG 1
evolves to ($\mu_{\rm e},h_{\rm e}$). GBIG 2 evolves to
$h_{\infty}$.
 GBIG 3 reaches a minimum energy density ($\mu_{\rm c}$) and then
evolves back to $\mu=\infty$. GBIG 4 starts at ($\mu_{\rm
e},-h_{\rm e}$), bounces at ($\mu_{\rm b},0$) before evolving to
($\mu_{\rm e},+h_{\rm e}$). When $\sigma=\sigma_{\rm e}$ GBIG 1
and 3 evolve to ($0,h_{\rm e}$). GBIG 4 starts at ($0,-h_{\rm e}$)
and bounces before evolving to ($0,h_{\rm e}$). When
$\gamma=\gamma_{\rm m}$ GBIG 1 ceases to exist ($h_{\rm i}=h_{\rm
e}$).

\item $V:~\sigma_{\rm e}<\sigma\leq\sigma_{\rm b},~\gamma\leq\gamma_{\rm i}$.
 GBIG 1 and GBIG 2 both end in vacuum de Sitter universes with different
values of $h_{\infty}$. GBIG 3 reaches a minimum energy density
($\mu_{\rm c}$) and then evolves back to $\mu=\infty$. GBIG 4
starts at ($0,-h_{\infty}$) then bounces at ($\mu_{\rm b},0$)
before evolving back to ($0,+h_{\infty}$). When
$\sigma=\sigma_{\rm b}$ GBIG 4 exists as a Minkowski universe
($0,0$).

\item $VI:~\sigma_{\rm b}<\sigma\leq\sigma_{\rm c},~\gamma\leq\gamma_{\rm
i}$. GBIG 1 and GBIG 2 both end in vacuum de Sitter universes with
different values of $h_{\infty}$. GBIG 3 reaches a minimum energy
density ($\mu_{\rm c}$) and then evolves back to $\mu=\infty$.
GBIG 4 does not exist. When $\sigma=\sigma_{\rm c}$ GBIG 3 ends in
a Minkowski universe. When $\gamma=\gamma_{\rm i}$ GBIG 1 and GBIG
2 live at ($0,h_{\rm i}$).

\item $VII:~\sigma>\sigma_{\rm c},~\gamma\leq\gamma_{\rm i}$. GBIG 1-3 all
end in vacuum de Sitter universes with different values of
$h_{\infty}$. GBIG 4 does not exist. When $\gamma=\gamma_{\rm i}$
GBIG 1 and GBIG 2 live at ($0,h_{\rm i}$).
\end{itemize}

\begin{figure}
\includegraphics[height=3in,width=2.75in,angle=270]{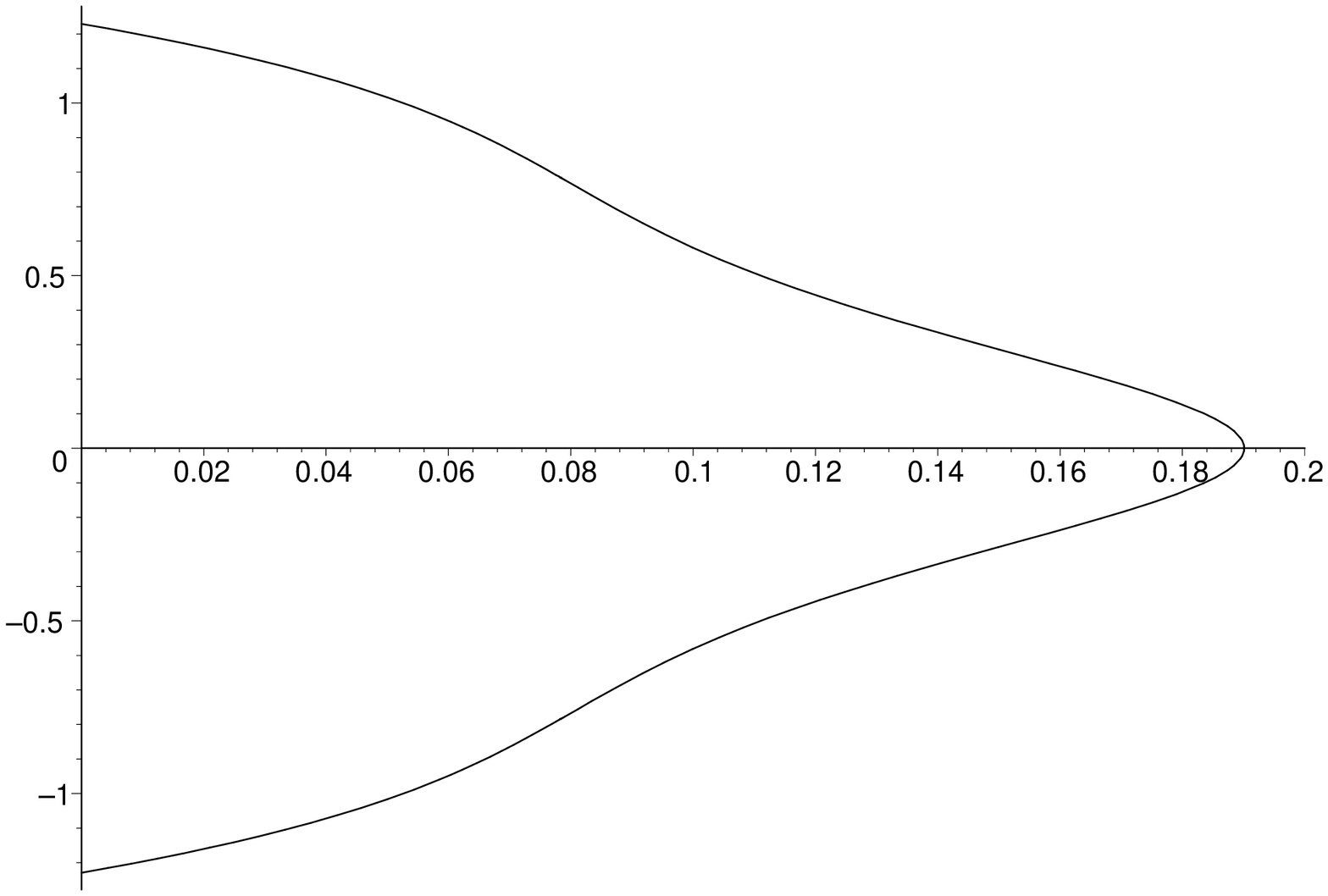}
\includegraphics[height=3in,width=2.75in,angle=270]{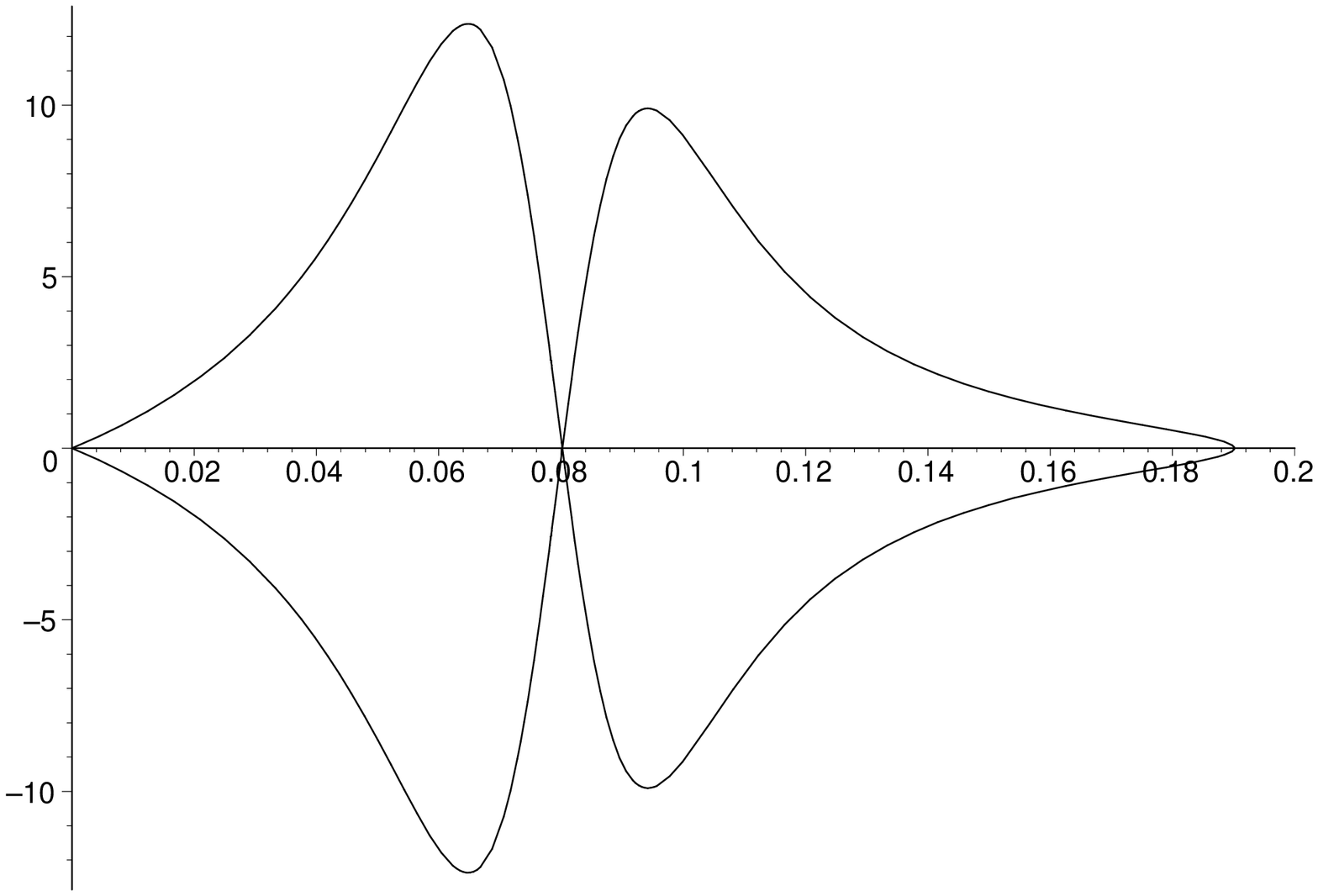}
\rput(4,3){\large $\mu$}\rput(11.6,3){\large $\mu$}
\rput{335}(4,4.8){\large $\leftarrow$} \rput{25}(4,2.2){\large
$\rightarrow$} \rput{45}(9.5,4.5){\large $\leftarrow$}
\rput{315}(9.5,2.4){\large
$\rightarrow$}\rput{45}(12.4,2.6){\large $\leftarrow$}
\rput{325}(12.4,4.4){\large $\rightarrow$}
 \rput(8,3.5){\large $h''$}\rput(0.4,3.5){\large $ h$} \caption{The GBIG 4 solution in
region I for $\phi=-0.1,\gamma=0.4>\gamma_{\rm m}$ and
$\sigma=-1/2$. The plot on the right is $ h''$ vs $\mu$ for this
solution. This shows the differences between this solution and the
similar case when $\phi=-1$, where GBIG 4 is no longer allowed.
Arrows denote proper time.} \label{uh12I}
\end{figure}

In region I there is a GBIG 4 solution, which is modified in the
same way as the GBIG 3 solution in the Minkowski bulk. As $h_{\rm
i,e}$ are no longer valid in region I GBIG 4 joins onto GBIG 2,
see Fig.~\ref{uh12I}.
The plot on the right in this figure shows $h''$, in order to
clearly distinguish this solution from the one in the $\phi=-1$ case
where GBIG 4 is no longer allowed.

\begin{figure}
\begin{center}
\includegraphics[height=3in,width=2.75in,angle=270]{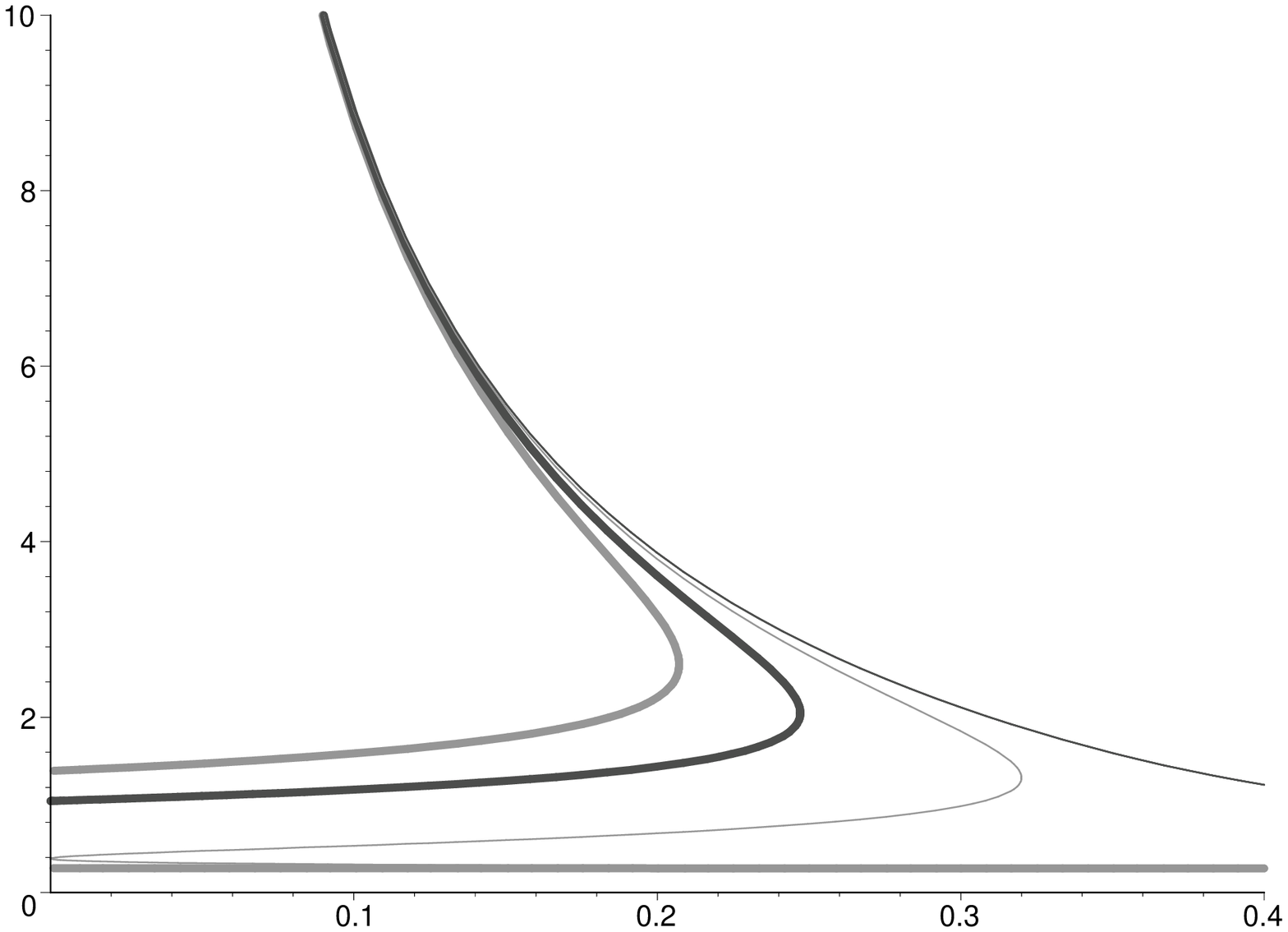}
\rput(-7.4,-3.5){\large $h_{\infty}$}\rput(-3.9,-6.7){\large
$\gamma$}
 \caption{$h_{\infty}$ for solutions in a AdS ($\phi=-0.1$) bulk.
 The thin-dark line has $\sigma=-1/2$, thin-light lines have $\sigma=-0.35$,
 thick-dark line has $\sigma=0$ and the thick-light lines have $\sigma=1/2$.}
\label{hinp12}
\end{center}
\end{figure}

In Fig.~\ref{hinp12} are the results for $h_{\infty}$ with
$\phi=-0.1$. The thin-dark line ($\sigma=-1/2$) lies in regions I
and IV. Therefore we have only one solution for $h_{\infty}$,
which corresponds to GBIG 2 for $\gamma\leq\gamma_{\rm m}$. For
$\gamma_{\rm m}<\gamma<\gamma_{\rm M}$ this root now corresponds
to the end point for GBIG 3 as in the $\phi=0$ case. The
thin-light lines ($\sigma=-0.35$) lie in regions I and V. There
are the two GBIG 1 and GBIG 2 solutions and the GBIG 4 solution
which converges with the light-thick line at the bottom. The
thick-dark line ($\sigma=0$) lies in regions II and VI, so only
has GBIG 1 and GBIG 2 present. The thick-light lines
($\sigma=1/2$) lie in regions III and VII, so has GBIG 1 and GBIG
2 and the GBIG 3 solution (the horizontal line).

\subsubsection{Typical example: $\phi=-1$}

\begin{figure}
\begin{center}
\includegraphics[height=3in,width=2.75in,angle=270]{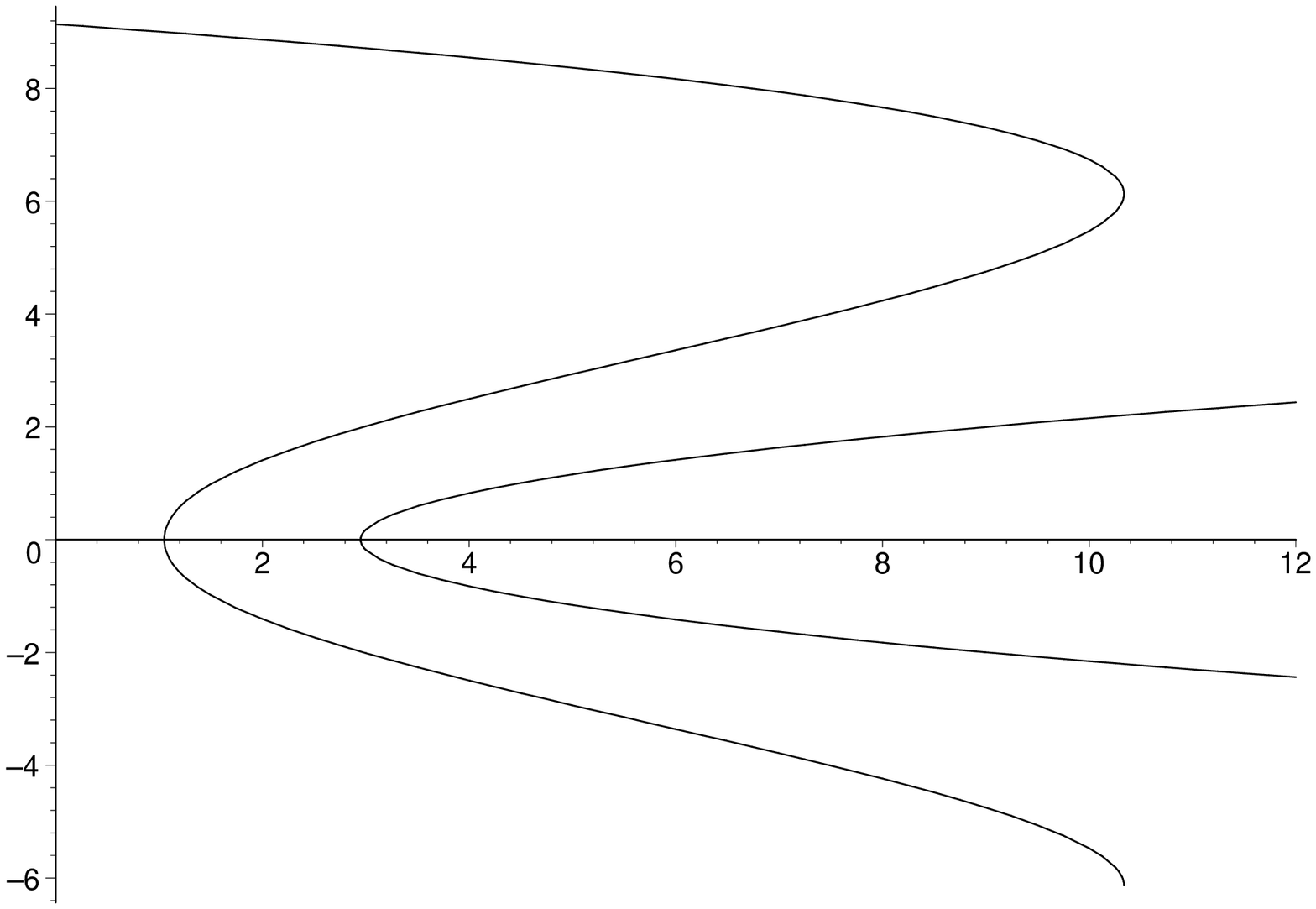}
\rput(-1.0,-6.2){\large$\mu_{\rm i},-h_{\rm
i}$}\rput(-1.1,-1.8){\large$\mu_{\rm i},h_{\rm i}$}
\rput(-6.7,-3.8){\large$\mu_{\rm b}$}
\rput(-5.6,-3.8){\large$\mu_{\rm c}$}
  \rput(-7.5,-0.8){\large$h_{\infty}$}
\rput(-7.4,-4){\large $h$}\rput(-3.6,-4.4){\large $\mu$}
\rput{350}(-5,-0.9){\large $\leftarrow$} \rput{20}(-5,-3.1){\large
$\leftarrow$} \rput{11}(-1,-3.1){\large $\leftarrow$}
\rput{345}(-1,-5){\large $\rightarrow$}
\rput{329}(-5,-5){\large$\rightarrow$}
 \rput(-2,-2.7){\large GBIG
1} \rput(-2,-1.1){\large GBIG 2}\rput(-1,-3.5){\large GBIG
3}\caption{Solutions of the Friedmann equation ($h$ vs $\mu$) with
negative brane tension ($\sigma=-2$) in an AdS bulk ($\phi=-1$)
with $\gamma=1/10$. For this value of $\phi$ GBIG 4 no longer
exists and GBIG 1 can collapse. The curves are independent of the
equation of state $w$. The arrows indicate the direction of proper
time on the brane.} \label{Phi12}
\end{center}
\end{figure}

\begin{figure}
\begin{center}
\includegraphics[height=3in,width=2.75in,angle=270]{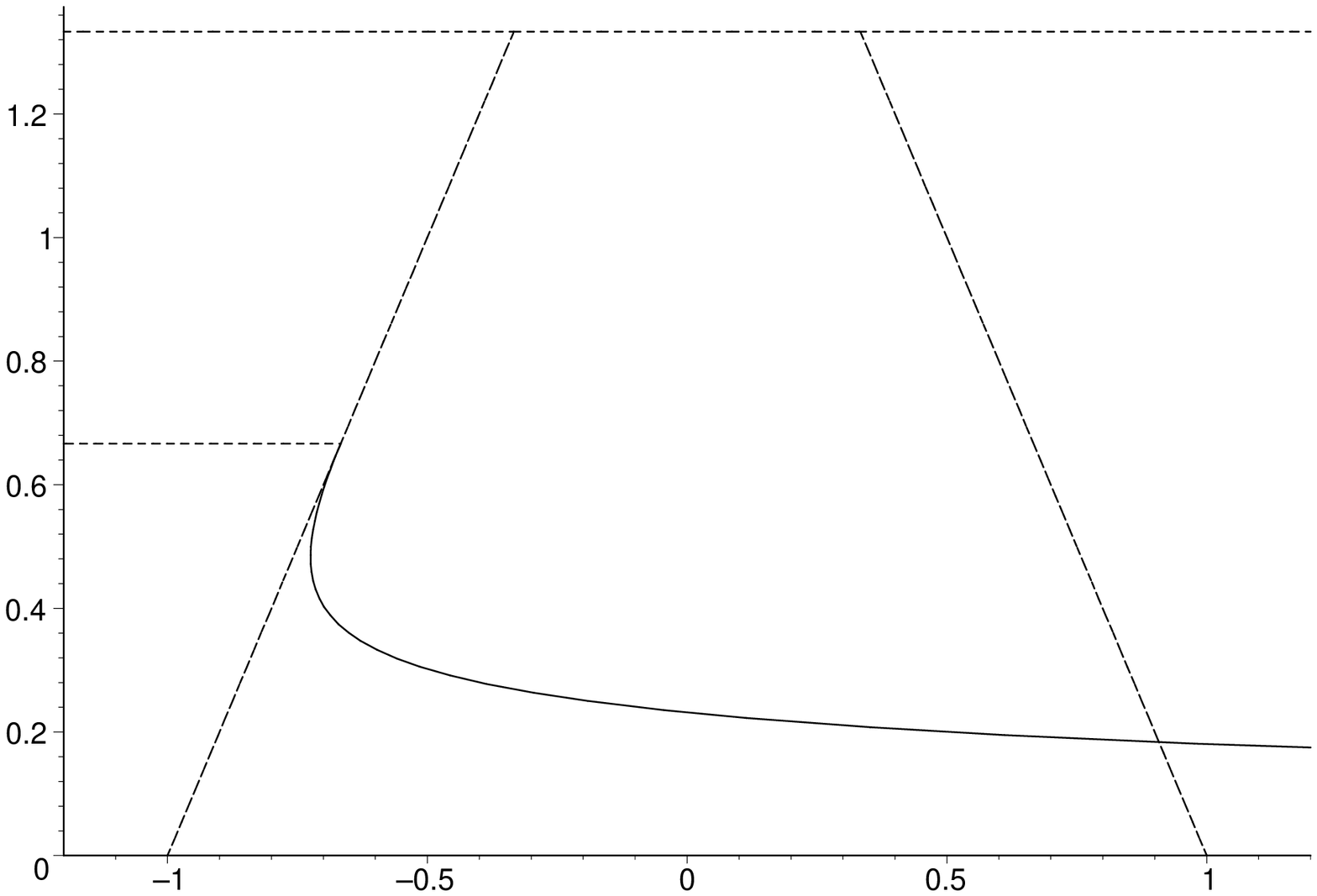}
\rput(-7.4,-3.6){\large $\gamma$}\rput(-3.9,-6.7){\large $\sigma$}
\rput(-6.2,-2.4){\large I}\rput(-3.9,-2.4){\large II}
\rput(-1,-2.4){\large III} \rput(-6.4,-4.5){\large IV}
\rput(-3.9,-6){\large V}\rput(-1,-6){\large VI}
 \rput(-4.8,-4.9){\large $\gamma_{\rm i}(\sigma)$}\rput(-1.2,-4.6){\large
$\sigma_{\rm c}(\gamma)$}\rput(-5.6,-5.8){\large $\sigma_{\rm
b}(\gamma)$}\rput(-6.5,-3.5){\large $\gamma_{\rm m}$}
\rput(-6.5,-0.9){\large $\gamma_{\rm M}$}\caption{The
$(\sigma,\gamma)$ plane for solutions in a AdS ($\phi=-1$) bulk.
The short dotted horizontal line is the maximum value of $\gamma$
as obtained from Eq.~(\ref{gbound}). The top horizontal line is
from the initial bound in Eq.~(\ref{gpbound}).} \label{gs4}
\end{center}
\end{figure}

We now consider $\phi=-1$ which is in the region where GBIG 4 no
longer exists (see Fig.~\ref{Phi12}), as $h_{\rm e}=0$.
Therefore $\mu_{\rm e}$ is not relevant and GBIG 1 bounces at
$\mu_{\rm b}$. This means that the ($\sigma,~\phi$) plane,
Fig.~\ref{gs4}, is simpler.
The regions in Fig.~\ref{gs4} are:

\begin{itemize}
\item$I:~\sigma\leq\sigma_{\rm
b},~\gamma_{\rm m}<\gamma\leq\gamma_{\rm M}$. GBIG 1 does not
exist. GBIG 2 starts at ($0,-h_{\infty}$), collapses to ($\mu_{\rm
b},0$) and then expands back to ($0,h_{\infty}$). GBIG 3 expands
to ($\mu_{\rm c},0$) and then collapses. When $\sigma=\sigma_{\rm
b}$ GBIG 2 exists as a Minkowski universe.

\item $II:~\sigma_{\rm b}<\sigma\leq\sigma_{\rm
c},~\gamma_{\rm i}<\gamma\leq\gamma_{\rm M}$. GBIG 1 and GBIG 2 do
not exist. GBIG 3 expands to $\mu_{\rm c}$ and then collapses.
When $\sigma=\sigma_{\rm c}$ GBIG 3 evolves to a Minkowski
universe.

\item $III:~\sigma>\sigma_{\rm c},~\gamma_{\rm i}<\gamma\leq\gamma_{\rm M}$.
 GBIG 1 and GBIG 2 do not exist. GBIG 3 evolves to $h_{\infty}$.

\item $IV:~\sigma\leq\sigma_{\rm b},~\gamma\leq\gamma_{\rm m}$. GBIG 1
evolves from $\mu_{\rm i}$ to $\mu_{\rm b}$ and then expands back
to $\mu_{\rm i}$. GBIG 2 evolves to $h_{\infty}$. GBIG 3 expands
to $\mu_{\rm c}$ and then collapses. When $\gamma=\gamma_{\rm m}$
GBIG 1 ceases to exist, GBIG 2 expands from ($\mu_{\rm b},0$).
When $\sigma=\sigma_{\rm b}$ GBIG 1 evolves to a Minkowski
universe.

\item $V:~\sigma_{\rm b}<\sigma\leq\sigma_{\rm
c},~\gamma\leq\gamma_{\rm i}$. GBIG 1 and GBIG 2 evolve to
$h_{\infty}$. GBIG 3 expands to $\mu_{\rm c}$ and then collapses.
When $\gamma=\gamma_{\rm i}$, GBIG 1 and GBIG 2 exist as the same de
Sitter universe with ($0,h_{\infty}$). When $\sigma=\sigma_{\rm c}$,
GBIG 3 ends as a Minkowski universe.

\item $VI:~\sigma>\sigma_{\rm c},~\gamma\leq\gamma_{\rm i}$.
GBIG 1-3 evolve to $h_{\infty}$. When $\gamma=\gamma_{\rm i}$, GBIG
1-2 live at ($0,h_{\infty}$).
\end{itemize}

\begin{figure}
\includegraphics[height=3in,width=2.75in,angle=270]{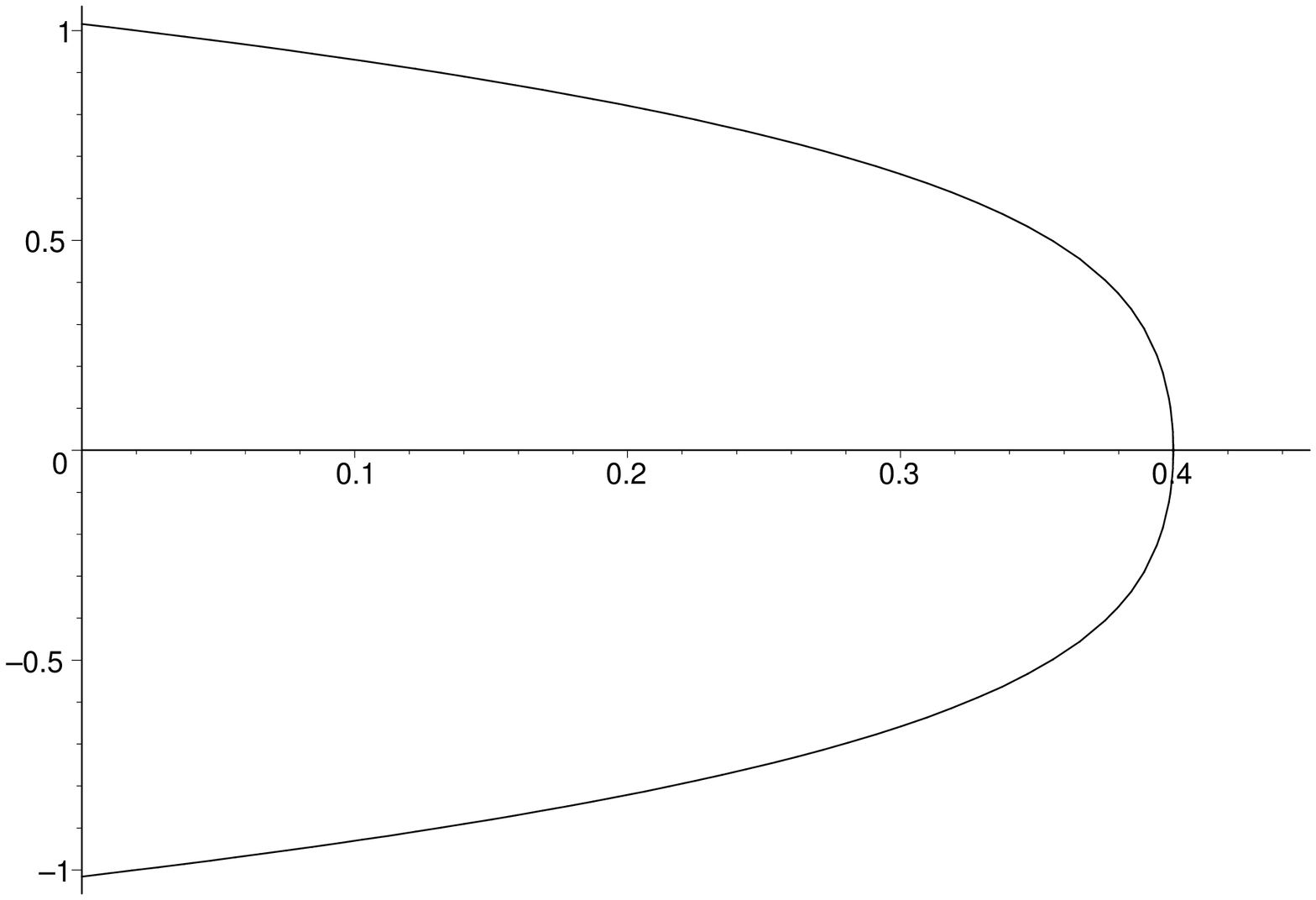}
\includegraphics[height=3in,width=2.75in,angle=270]{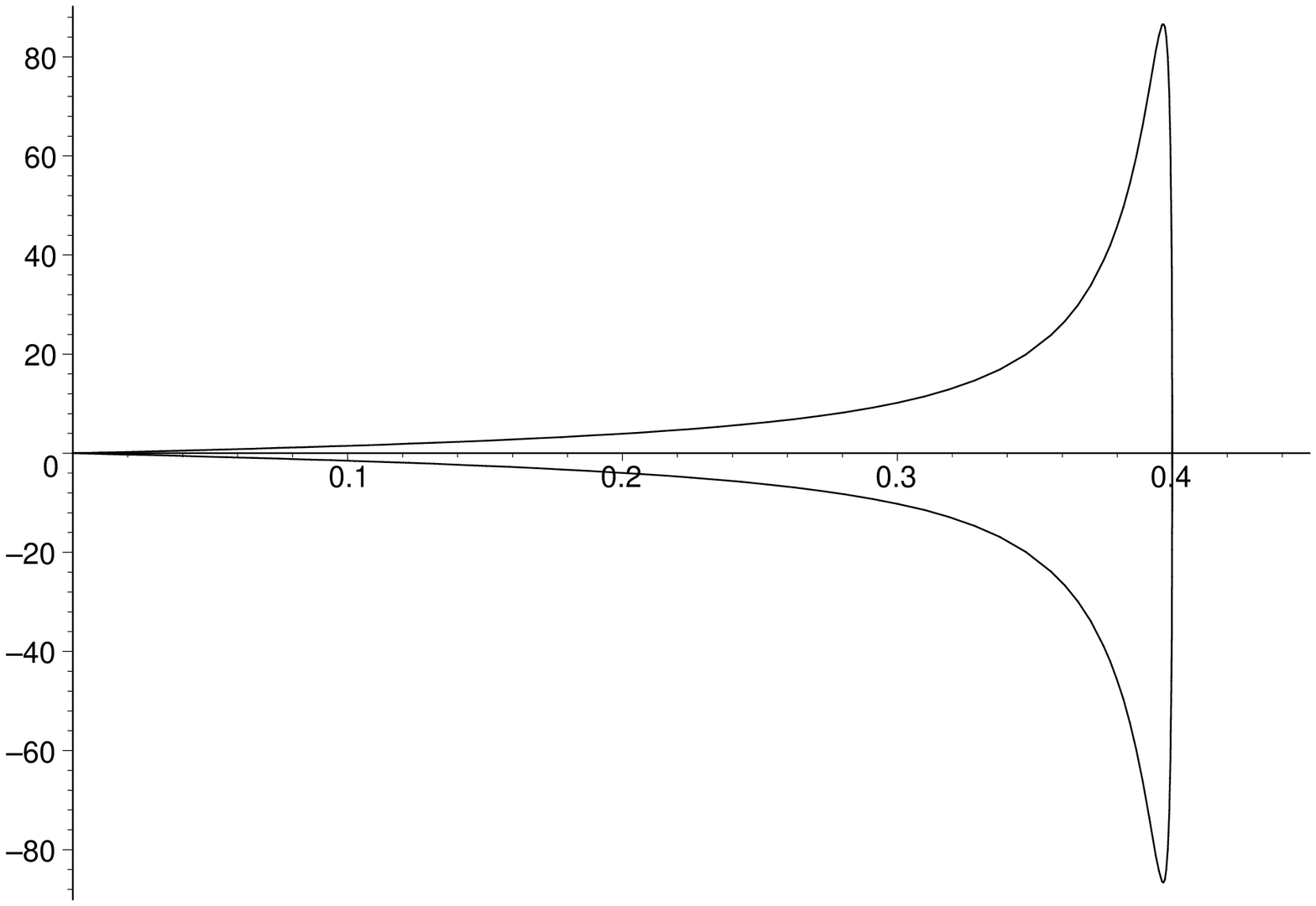}
\rput(4,3.2){\large $\mu$}\rput(10.9,3.2){\large $\mu$}
\rput{345}(4,5.6){\large $\leftarrow$} \rput{15}(4,1.35){\large
$\rightarrow$} \rput{45}(13,4.2){\large $\leftarrow$}
\rput{315}(13,2.8){\large $\rightarrow$} \rput(-7.3,-3.5){\large
$h''$}\rput(-7.3,3.5){\large $h$} \caption{The GBIG 4 solution in
region I for $\phi=-1,\gamma=0.8>\gamma_{\rm m}$ and $\sigma=-1$.
The plot on the right is $h''$ vs $\mu$ for this solution. This
shows the differences between this solution and the case when
$\phi=-0.1$. Arrows denote proper time.} \label{uh4I}
\end{figure}

\begin{figure}
\begin{center}
\includegraphics[height=3in,width=2.75in,angle=270]{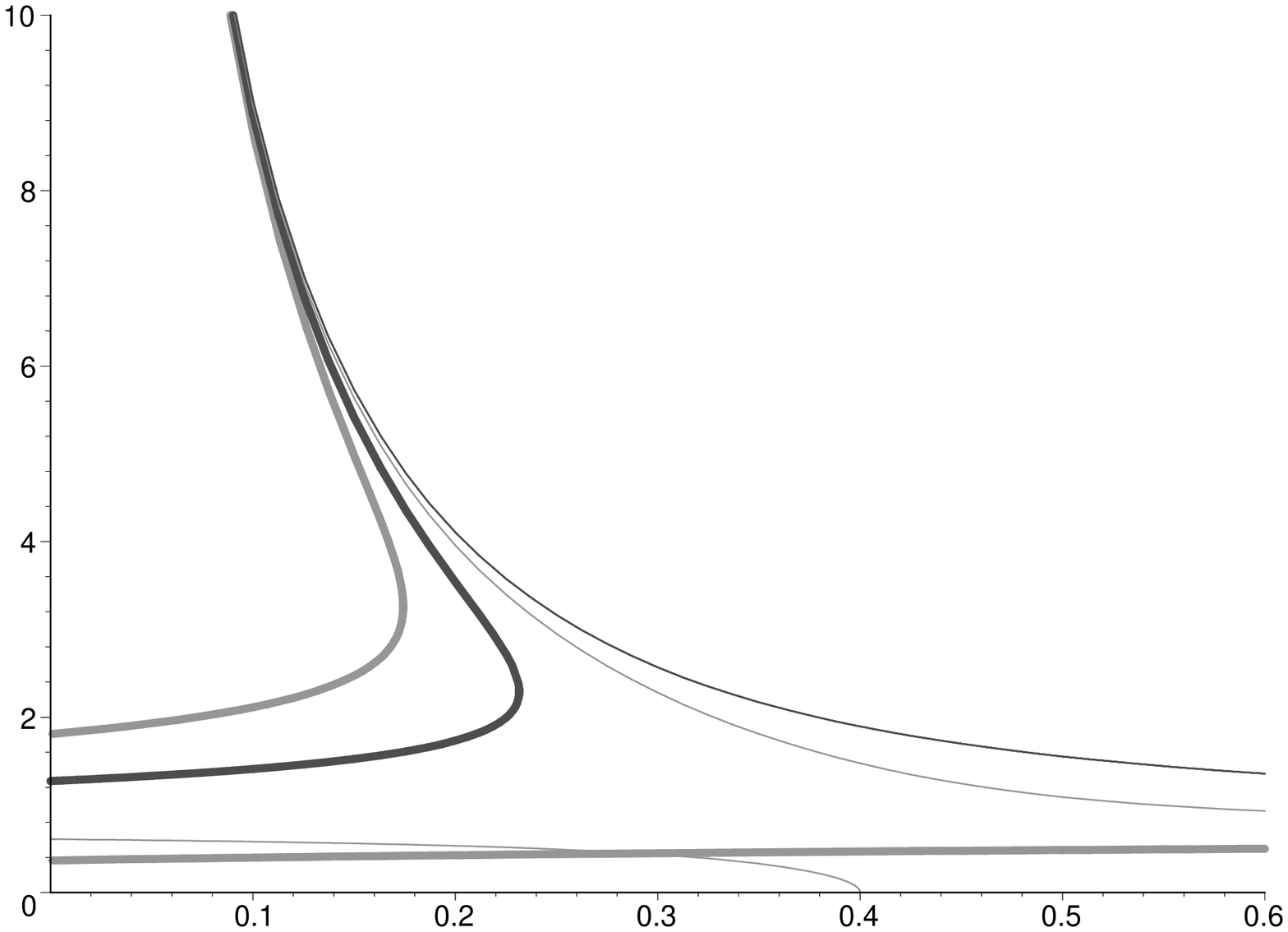}
\rput(-7.4,-3.5){\large $h_{\infty}$}\rput(-3.9,-6.7){\large
$\gamma$}
 \caption{$h_{\infty}$ for solutions in a AdS ($\phi=-1$) bulk.
 The thin-dark line have $\sigma=-1.2$, thin-light lines have $\sigma=0.8$,
 thick-dark line have $\sigma=0$ and the thick-light lines have $\sigma=1.2$.}
\label{hinp4}
\end{center}
\end{figure}

In region I we again have a combined solution as GBIG 1 has
vanished. As GBIG 4 is not allowed ($h_{\rm e}=0$) when we take
$\gamma>\gamma_{\rm m}$, which causes $h_{\rm i}=0$, GBIG 2
matches up with its negative counterpart. We see the bouncing GBIG
2 solution in Fig.~\ref{uh4I}. Note the nature of $h''$ is very
different to that of the GBIG 2, 4 bounce in the $\phi=-0.1$ case.
In Fig.~\ref{hinp4} we present results for $h_{\infty}$ for
$\phi=-1$.

\subsubsection{Typical example: $\phi=-5$}

\begin{figure}
\begin{center}
\includegraphics[height=3in,width=2.75in,angle=270]{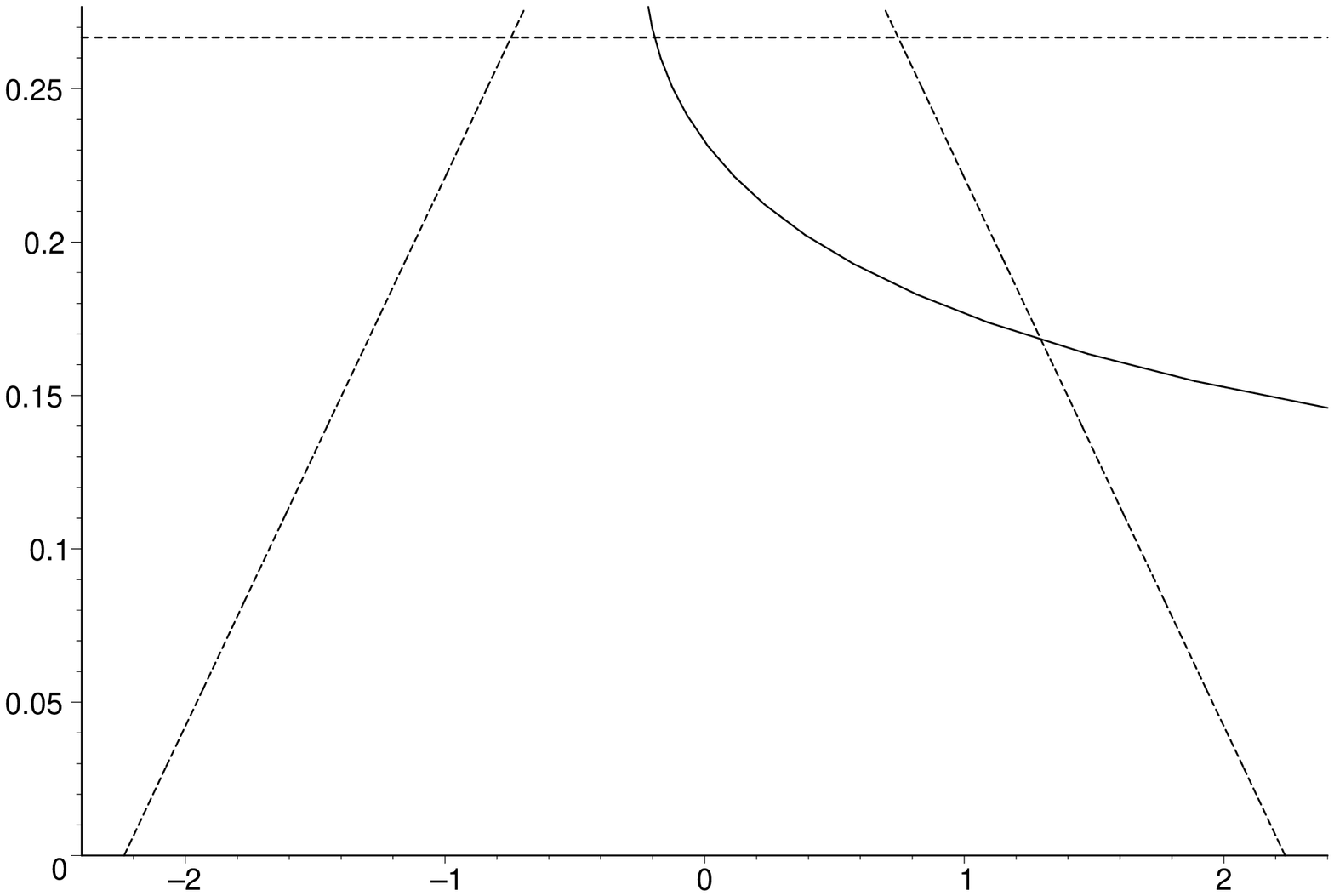}
\rput(-7.4,-3.8){\large $\gamma$}\rput(-3.9,-6.7){\large
$\sigma$}\rput(-6,-2){\large I}\rput(-3.2,-2){\large II}
\rput(-1.2,-2){\large III} \rput(-4,-4.5){\large
IV}\rput(-1.2,-4.5){\large V} \rput(-1,-3.1){\large $\gamma_{\rm
i}(\sigma)$}\rput(-1.8,-6){\large $\sigma_{\rm
c}(\gamma)$}\rput(-6,-6){\large $\sigma_{\rm
b}(\gamma)$}\rput(-6.5,-1){\large $\gamma_{\rm M}$} \caption{The
$(\sigma,\gamma)$ plane for solutions in a AdS ($\phi=-5$) bulk. The
top horizontal line is the bound from the initial bound in
Eq.~(\ref{gpbound}).} \label{gs20}
\end{center}
\end{figure}

Here we shall present results for $\phi=-5$ for completeness. This
value of $\phi$ lives in the region of Fig.~\ref{gpbound} where
$h_{\rm i}$ is always real. This means that the $\sigma,~\gamma$
plane is much simpler, Fig.~\ref{gs20}.

The regions in Fig.~\ref{gs20} are:

\begin{itemize}
\item$I:~\sigma\leq\sigma_{\rm
b},~\gamma\leq\gamma_{\rm M}$. GBIG 1 evolves from $\mu_{\rm i}$ to
$\mu_{\rm b}$ and back. GBIG 2 evolves to $h_{\infty}$. GBIG 3
expands to $\mu_{\rm c}$ and then collapses. When
$\sigma=\sigma_{\rm b}$, GBIG 1 ends in a Minkowski universe.

\item $II:~\sigma_{\rm i}<\sigma\leq\sigma_{\rm
c},~\gamma_{\rm i}<\gamma\leq\gamma_{\rm M}$. GBIG 1 and GBIG 2 do
not exist. GBIG 3 expands to $\mu_{\rm c}$ and then collapses.
When $\sigma=\sigma_{\rm c}$ GBIG 3 evolves to a Minkowski
universe.

\item $III:~\sigma>\sigma_{\rm c},~\gamma_{\rm i}<\gamma\leq\gamma_{\rm M}$.
GBIG 1 and GBIG 2 do not exist. GBIG 3 evolves to $h_{\infty}$.

\item $IV:~\sigma_{\rm b}<\sigma\leq\sigma_{\rm i}$ and $~\sigma_{\rm
b}<\sigma\leq\sigma_{\rm c}$; $\gamma\leq\gamma_{\rm M}$ and
$\gamma\leq\gamma_{\rm i}$. GBIG 1 and GBIG 2 do not exist. GBIG 3
expands to $\mu_{\rm c}$ and then collapses. When
$\gamma=\gamma_{\rm i}$ (and for $\sigma=\sigma_{\rm i}$) GBIG 1
and GBIG 2 both live at ($0,h_{\infty}$). When $\sigma=\sigma_{\rm
c}$ GBIG 3 evolves to a Minkowski universe.

\item $V:~\sigma>\sigma_{\rm c},~\gamma\leq\gamma_{\rm i}$.
GBIG 1-3 evolve to $h_{\infty}$. When $\gamma=\gamma_{\rm i}$, GBIG
1-2 both live at ($0,h_{\infty}$).
\end{itemize}

As we decrease the value of $\phi$, region II in Fig.~\ref{gs20}
shrinks (the point $\sigma_{\rm i}(\gamma_{\rm M})$ becomes
increasingly positive). For $\phi\leq-64/9$ region II no longer
exists. The nature of the solutions in the other regions are
unaffected.

\begin{figure}
\begin{center}
\includegraphics[height=3in,width=2.75in,angle=270]{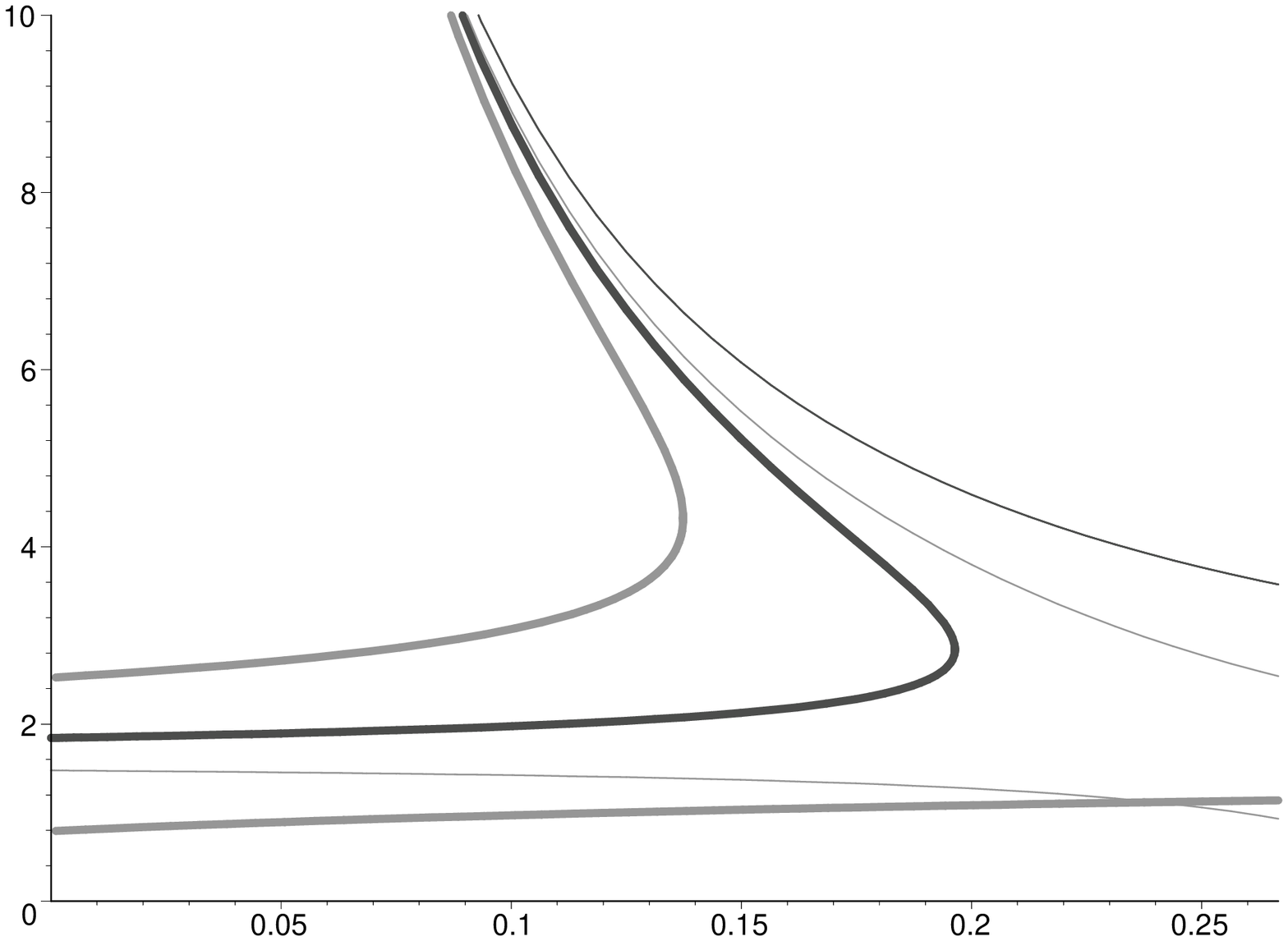}
\rput(-7.4,-3.5){\large $h_{\infty}$}\rput(-3.9,-6.7){\large
$\gamma$}
 \caption{$h_{\infty}$ for solutions in a AdS ($\phi=-5$) bulk.
 The thin-dark line has $\sigma=-3$, thin-light lines have $\sigma=-1/2$,
 thick-dark lines has $\sigma=1/2$ and the thick-light lines has $\sigma=3$.}
\label{hinp20}
\end{center}
\end{figure}

There are no bouncing solutions in this case (in fact for any case
with $\phi\leq-9/4$) as GBIG 4 is unallowed and $\gamma_{\rm
M}<\gamma_{\rm m}$ (which rules out the combined solutions). In
Fig.~\ref{hinp20} we show the $h_{\infty}$ results for $\phi=-5$.

\section{Conclusions}

In this chapter we have looked at the general GBIG model. We have
seen that there is a range of possible dynamics that can be achieved
depending on the parameters in the model. The main model of interest
is still GBIG 1 as this is the one that starts in a finite density
``quiescent'' singularity and evolves to a de Sitter universe. If we
warp the bulk enough, GBIG 1 can be allowed to collapse back to its
initial density, provided there is sufficient negative brane
tension. If the bulk is warped enough to allow GBIG 1 to collapse
this means that the bouncing cosmology GBIG 4 can no longer exist.
The exact nature of the late time dynamics can be changed by
including a non-zero brane tension. A negative brane tension can
reduce the Hubble rate at late time and a positive tension will
increase it. GBIG 1 can end in a future ``quiescent'' singularity
with a non-zero density if we have an appropriate (negative) brane
tension present. As the solution of interest is GBIG 1 and we want
it to provide the late time acceleration that we are experiencing it
needs to end as a vacuum de Sitter universe. Therefore if there is
some brane tension it must take values $\sigma_{\rm
i}>\sigma>\sigma_{\rm e}$.

GBIG 2 also starts in a ``quiescent'' singularity but it
super-accelerates so it is un-physical and of little interest.

GBIG 3 still starts in an infinite density big-bang but has a number
of possible late time dynamics. In a Minkowski bulk GBIG 3 can
either evolve to a Minkowski state, as shown in Chapter~\ref{GBIGB},
a vacuum de Sitter state or even loiter around $\mu=-\sigma$ before
ending in a vacuum de Sitter state or a ``quiescent'' singularity.
This loitering cosmology is different from that in
Ref.~\cite{Sahni:2004fb}, as we do not require a naked bulk
singularity or a de Sitter bulk. If we warp the bulk then GBIG 3
will generally collapse, unless there is sufficient (positive) brane
tension to allow the solution to end in a vacuum de Sitter state.

GBIG 4 can only ever exist in a mildly warped bulk with negative
brane tension. So it can be said that GBIG 4 is unphysical due to
the requirement that $\sigma<0$.

There are a number of bouncing cosmologies within this set-up, in
addition to the GBIG 4. There are also the solutions where GBIG 4
and 2 match up ($\gamma_{\rm m}<\gamma\leq\gamma_{\rm M}$ with
$\phi_{\rm GBIG 4 Lim}<\phi<0$) and the GBIG 2 bouncing cosmologies
($\gamma_{\rm m}<\gamma\leq\gamma_{\rm M}$ with
$-9<\phi\leq\phi_{\rm GBIG 4 Lim}$). Each of these have different
dynamics so would produce different evolutionary histories. The
solutions that spend time on the GBIG 2 branch will experience
phantom-like behaviour during this period.

%% file: Conclusions.tex
\chapter{Conclusions}

In this thesis we have looked at some brane world cosmological
models. The Randall-Sundrum is an interesting toy model of the
universe in which we are able to investigate some of the
phenomenological properties of string theory ideas. The idea that
everything apart from gravity is confined to a four dimensional
hypersurface leads to some striking new features.

In chapter \ref{RSB} we looked at the Kaluza-Klein modes of the
graviton in Randall-Sundrum models. We considered the nature of
these modes for both Minkowski and de Sitter branes in the
Randall-Sundrum one and two-brane models. For two Minkowski branes
there is a zero mode and then a series of massive modes. The
spacings of these modes are governed by Bessel functions, dependent
on the AdS length scale of the bulk and the inter-brane distance.
When we send the second brane out to infinity we obtain a continuum
of massive Kaluza-Klein modes. In the investigation of two de Sitter
branes we saw that there is a zero mode and a mass gap, which is a
function of the Hubble rate on the brane, to the first massive mode.
We derived a new expression for the mode spacing as a function of
associated Legendre functions, dependent on the AdS length scale and
the inter-brane distance but also on the Hubble rate on the brane.
Therefore they differ in the low energy and high energy limits.
Taking the one brane limit we again find a continuum of massive
modes.

In chapter \ref{RSB} we reviewed the 5D bulk equations and how these
are projected onto the brane in Randall-Sundrum models. It is the
projected equations that are used when we want to understand the
cosmological dynamics in these models.

In chapter \ref{GBIGB} we briefly reviewed the DGP and GB models
before combining the two into GBIG model. The DGP is a very
interesting brane world model that much work has been done on,
because of the late time acceleration in the DGP(+) branch. This is
a very interesting result as the observed acceleration of the
universe is usually interpreted via the introduction of a ``dark
energy'' field. The DGP model explains this phenomenon via modified
gravity. In the GBIG model we have included the Gauss-Bonnet term in
order to modify the high energy dynamics of the universe. The GB
term has been shown to arise naturally within a string theory
context. Therefore we investigated a model where both the induced
gravity on the brane and Gauss-Bonnet terms in the bulk are present.
Note, in the context of string theory terms of higher order than the
GB term will need to be incorporated at even higher energies, and
earlier times.
\begin{figure}
\begin{center}
\includegraphics[height=3in,width=2.75in,angle=270]{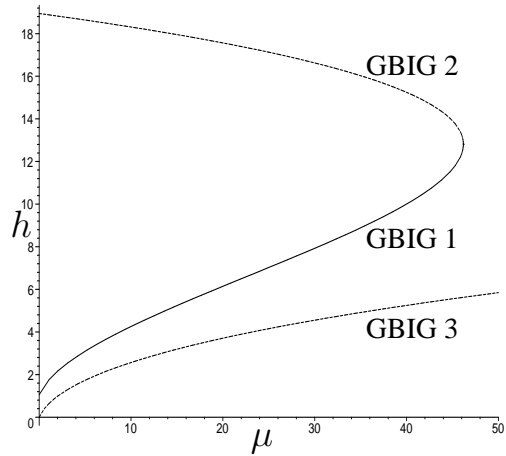}
 \rput(-7.2,-3.6){\large $h$}\rput(-4,-6.5){\large
$\mu$}\rput(-2,-1.5){\small GBIG 2}\rput(-2,-3.8){\small GBIG
1}\rput(-2,-5){\small GBIG 3}
 \caption{The three GBIG models with zero brane tension in a Minkowski bulk.}
\label{GBIG}
\end{center}
\end{figure}

We have shown that the GBIG model gives some intriguingly distinct
phenomenology from that of the DGP and GB models. With a non-zero
contribution from both the GB and IG terms we find we have three
solutions (in the zero brane tension case), as shown in
Fig.~\ref{GBIG}. We have a solution that starts in a finite density
big bang and then self accelerates in the future. This is the
solution of most interest. The DGP and GB terms do not remove the
initial singularity on their own. Only when both sets of terms are
present that this feature occurs. It has been shown that a 4D
heterotic string using a one loop corrected superstring effective
action with GB terms and dilaton and modulus fields can avoid the
initial singularity~\cite{Antoniadis:1993jc}.

There is a second GBIG solution which also starts with the finite
big bang but then super accelerates. This solution is therefore less
relevant. The third solution which starts with a standard big bang
singularity evolves to a Minkowski universe. The GBIG model has
finite density, pressure and temperature but it does have a
curvature singularity at the birth of the universe.

The GBIG model has a severe UV-IR bootstrap which forces $\gamma$ to
be very small in order to have high enough initial redshift. The
analysis of Big Bang nucleosynthesis gives us the constraint that
$\gamma$ must be very small. This also means that observations of
Big Bang nucleosynthesis are incapable of discriminating between the
GBIG and DGP(+) models. In order to discern between the two models
we need to consider the earlier universe where the GBIG model is
substantially different from the DGP(+). These differences will be
apparent in the analysis of inflation dynamics + perturbations
giving rise to formation of structure. The problem of structure
formation in the DGP model has been investigated in
Ref.~\cite{Koyama:2005kd}. Inflation with Gauss-Bonnet terms in the
bulk has been considered in Ref.~\cite{Dufaux:2004qs}. In both cases
there are substantial unresolved problems to be tackled. Combing the
two in the GBIG model will be a difficult but important future line
of research.

In chapter \ref{GGBIG} we looked at the GBIG model in more
generality by including both brane tension and a non zero effective
cosmological constant $\phi$. We saw that the effect of including
brane tension was to shift the solutions (of Fig.~\ref{GBIG}) along
the $\mu$ axis. The consequence of this is that by including
positive brane tension it is possible to make the GBIG 1-2 start at
lower densities and end with greater acceleration. It is also
possible to make the GBIG 1-2 branches inaccessible. GBIG 3 can now
accelerate at late time i.e. the brane tension acts as an effective
cosmological constant. Including a negative brane tension causes
GBIG 1-2 to start at higher densities and end with less
acceleration. If enough tension is added the GBIG 1 model ends in a
``quiescent'' singularity. GBIG 3 now loiters at a density given by
an absolute value of the brane tension. After loitering at this
density it will either evolve to a de Sitter solution or the
``quiescent'' singularity. The time GBIG 3 loiters for is a function
of the equation of state $w$ and $\gamma$. The exact solution for
this time requires further investigation. This loitering solution is
of interest as it does not require the presence of a naked bulk
singularity or a de Sitter bulk, as other loitering brane
cosmologies require.
\begin{figure}
\begin{center}
\includegraphics[height=3in,width=2.75in,angle=270]{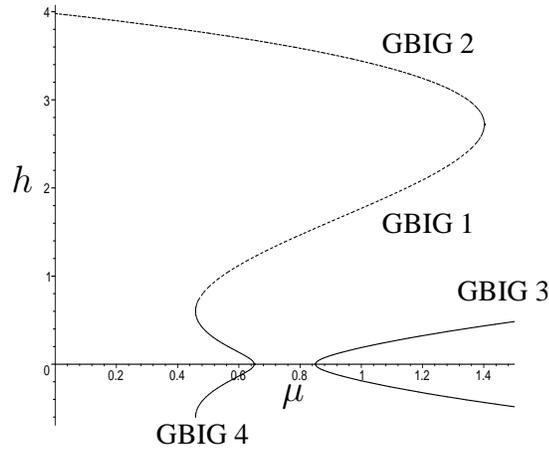}
\rput(-7.4,-3){\large $h$}\rput(-3.8,-5.9){\large $\mu$}
\rput(-2,-3.6){\small GBIG 1} \rput(-2,-1.2){\small GBIG
2}\rput(-1,-4.5){\small GBIG 3}\rput(-5,-6.4){\small GBIG
4}\caption{Generalised GBIG models with negative brane tension in an
AdS bulk.} \label{Friedpl1ds2}
\end{center}
\end{figure}

\begin{figure}
\begin{center}
\includegraphics[height=3in,width=2.75in,angle=270]{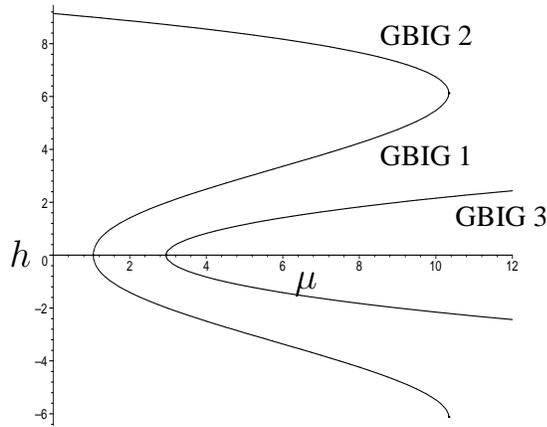}
\rput(-7.4,-4){\large $h$}\rput(-3.6,-4.4){\large $\mu$}
 \rput(-2,-2.7){\small GBIG
1} \rput(-2,-1.1){\small GBIG 2}\rput(-1,-3.5){\small GBIG
3}\caption{Another example of generalised GBIG models with negative
brane tension in an AdS bulk.} \label{Phi122}
\end{center}
\end{figure}

The effective cosmological constant $\phi$ modifies the solutions in
a more complex way. GBIG 1-2 are not altered in nature only in size.
GBIG 3 is the most affected. The GBIG 3 branch is now split in two,
as seen in Fig.~\ref{Friedpl1ds2}. GBIG 3 can now collapse after a
minimum density is reached, as long as sufficient brane tension is
present. GBIG 4 is a new solution which is only present for a
non-zero $\phi$ which is below a certain threshold. For values of
$\phi$ above this threshold GBIG 4 vanishes and GBIG 1 is allowed to
collapse, with sufficient brane tension, Fig.~\ref{Phi122}.
GBIG 4 is only accessible with negative brane tension present.

The are a number of constraints on $\gamma$. The most important is
imposed by the fact that $\Lambda_5\leq0$. Another constraint is
imposed by requiring GBIG 1 to exist. In the $\sigma=0$ cases we are
constrained by the density of the big bang being greater than zero.
With the presence of brane tension we can always have this density
greater than zero. Now the constraint on $\gamma$ comes from the
fact that as we increase $\gamma$, the GBIG 1 density range
decreases. At a certain point GBIG 1 disappears and GBIG 3 ends at a
``quiescent'' singularity of the same density as the GBIG 2 big bang
(this is in the $\phi=0$ cases; if $\phi\neq0$ then GBIG 2 meets up
with GBIG 4). Increasing $\gamma$ further allows evolution to
continue through this point causing GBIG 2 and 3 to become a single
solution.

We have seen that the GBIG model has many possible evolutionary
histories depending on the values of  the effective cosmological
constant $\phi$, the brane tension $\sigma$ and most importantly the
contribution of the IG and GB terms in $\gamma$. Not all the
solutions are physical. Our comprehensive analysis of the dynamics
is the starting point for further investigations into the GBIG
model, including inflation and structure formation.

The work on the GBIG model in chapter~\ref{GBIGB} has been cited in
a number of papers
including~\cite{Lidsey:2005nt,Papantonopoulos:2006uj,Copeland:2006wr,
Rizzo:2006wf,Apostolopoulos:2006si,deRham:2006pe,Panotopoulos:2006ui,
Fernandez-Jambrina:2006hj,Papantonopoulos:2006xi,Farakos:2006sr}.
The work in chapter~\ref{GGBIG} has been cited
in~\cite{Apostolopoulos:2006si,deRham:2006pe,Fernandez-Jambrina:2006hj,
Papantonopoulos:2006xi}.

%% file: ApCon.tex
\chapter{Conventions}

The conventions used in this thesis are as follows:

\begin{itemize}

\item Greek letters are used for 4D indices.

\item Roman letters are used for 5D indices.

\item 5D tensor are denoted by superscript $(5)$ with Roman indices.

\item The extra dimension is denoted by index $y$ when in a
Gaussian-normal coordinate system.

\item The extra dimension is denoted by index $z$ when in a
Poincar\'{e} coordinate system.

\item The metric signature is:

\begin{equation}\label{ms}
(-,+,+,+).
\end{equation}

\item The Riemann tensor is defined such that for AdS spacetime we have:

\begin{equation}\label{ries}
R^{ab}_{cd}=-\frac{1}{\ell^2}\left(\delta^a_c\delta^b_d-\delta^a_d\delta^b_c\right).
\end{equation}

\end{itemize}